\documentclass[11pt]{article}
\usepackage{latexsym,amsmath,amssymb,amsthm,amsfonts,graphicx}
\numberwithin{equation}{section}
\numberwithin{figure}{section}
\newtheorem{theorem}{Theorem}[section]
\newtheorem{lemma}[theorem]{Lemma}
\newtheorem{mainthm}[theorem]{`Main Theorem'}
\newtheorem{proposition}[theorem]{Proposition}

\newtheorem{remark}{Remark}

\newtheorem{definition}{Definition}

\numberwithin{equation}{section}
\newcommand{\nc}{\newcommand}
\def\R{\mathbb{R}}
\def\C{\mathbb{C}}
\nc{\diff}[2]{\frac{d #1}{d #2}}
\nc{\diffn}[3]{\frac{d^{#3} #1}{d {#2}^{#3}}}
\nc{\pdiff}[2]{\frac{\partial #1}{\partial #2}}
\nc{\pdiffn}[3]{\frac{\partial^{#3} #1}{\partial{#2}^{#3}}}
\nc{\abs}[1] {\lvert #1 \rvert}
\nc{\norm}[2] {{\lVert #1 \rVert}_{#2}}
\nc{\threeline}[1] {\lvert \lvert \lvert #1 \rvert\rvert\rvert}
\nc{\iamMc}{\frac{i\alpha-M}{c}}
\nc{\iapMc}{\frac{i\alpha+M}{c}}

\nc{\cF}{{\cal F}}
\nc{\cO}{{\cal O}}
\nc{\cQ}{{\cal Q}}
\nc{\cR}{{\cal R}}
\nc{\sqrtE}{\mu}
\nc{\cK}{{\cal K}}
\nc{\cL}{{\cal L}}
\nc{\cM}{{\cal M}}
\nc{\cN}{{\cal N}}
\nc{\cE}{{\cal E}}
\nc{\cH}{{\cal H}}
\nc{\cZ}{{\cal Z}}
\nc{\cT}{{\cal T}}
\nc{\rhoo}{{(z\cdot\xi)}}
\nc{\omegaa}{{(z\cdot\eta)}}

\nc{\order}{{\cal O}}
\nc{\ores}{{\omega_{\rm res}}}
\nc{\nit}{\noindent}
\nc{\Eplus}{E_+}
\nc{\Eminus}{E_-}
\nc{\Epm}{E_\pm}

\nc{\w}{\omega}
\nc{\eps}{\epsilon}
\nc{\e}{\varepsilon}
\nc{\g}{\gamma}
\nc{\z}{\zeta}
\nc{\G}{\Gamma}

\nc{\nn}{\nonumber}
\nc{\D}{\partial}
\nc{\pZ}{\partial_Z}
\nc{\pT}{\partial_T}
\nc{\pz}{\partial_z}
\nc{\pt}{\partial_t}
\nc{\vu}{\Vec u}
\nc{\vE}{\Vec {\cal E}}
\nc{\vr}{\Vec r}
\nc{\vrho}{\Vec \rho}
\nc{\Reps}{R^{\e}}
\nc{\Vreps}{\Vec \Reps}
\nc{\half}{\frac{1}{2}}
\nc{\bphi}{\bar{\phi}}
\nc{\efour}{{\Hat e}_4}
\nc{\marginnote}[1] {\marginpar{\tiny #1}}

\numberwithin{equation}{section}
\huge
\begin{document}
\title{Dynamics of Nonlinear Schr\"odinger \slash Gross-Pitaevskii Equations;\\ Mass Transfer in Systems with Solitons and Degenerate Neutral Modes}\author{Zhou Gang$^{\ast}$, \
Michael I. Weinstein$^{\dag}$} \maketitle
\centerline{\small{$^{\ast}$Department of Mathematics, Princeton University, 
Princeton, NJ, U.S.A.}} \centerline{\small{$^{\dag}$Department of
Applied Physics and Applied Mathematics, Columbia University, New York, NY, U.S.A.}}
\setlength{\leftmargin}{.1in}
\normalsize \vskip.1in \setcounter{page}{1}
\setlength{\leftmargin}{-.2in} \setlength{\rightmargin}{-.2in}
\section*{Abstract}
Nonlinear Schr\"odinger / Gross-Pitaevskii equations play a central role in the understanding of nonlinear optical and macroscopic quantum systems. The large time dynamics of such systems is governed by interactions of the  nonlinear ground state manifold, discrete neutral modes (``excited states'') and dispersive radiation. Systems with symmetry, in spatial  dimensions larger than  one, typically have degenerate neutral modes. Thus, we study the large time dynamics of  systems with degenerate neutral modes. This requires a new normal form ( nonlinear matrix Fermi Golden Rule) governing the system's large time asymptotic relaxation to the ground state (soliton) manifold. 
%
\vfil\eject
\tableofcontents
\section{Introduction}
Nonlinear Schr\"odinger / Gross-Pitaevskii (NLS/GP) equations are a class of dispersive Hamiltonian partial differential equations (PDEs) of the form:
\begin{equation}\label{eq:NLS}
\begin{array}{lll}
i\D_t\psi(x,t)&=&-\Delta\psi(x,t)\ +\ \left(\ V(x)\ -\ f(|\psi(x,t)|^{2})\ \right)\psi(x,t).
\end{array}
\end{equation}
Here, $\psi=\psi(x,t)$ is a scalar complex-valued function of position, $x\in\mathbb{R}^{d}$ and time, $t\in\mathbb{R}$. The function $V:\mathbb{R}^{d}\rightarrow \mathbb{R}$ denotes a linear  {\it potential} and  $f:\mathbb{R_+}\rightarrow \mathbb{R}$, a  nonlinear potential.
 For example, $V$ can be taken to be a smooth, non-positive potential well, with rapid decay as $|x|\to\infty$ and $f(|\psi|^2)=-g\ |\psi|^2,\ g\ {\rm constant}$. For $g>0$, the nonlinearity is called repulsive or defocusing. For $g<0$ it is called attractive or focusing. In this paper, we focus on spatial dimensions $d\ge3$. Precise hypotheses on $V$ and $f$ are given below. We are interested in the initial value problem (IVP) for (\ref{eq:NLS}) with finite energy data $\psi(x,0)$ and solution $\psi(x,t)$, which are sufficiently regular and  decaying to zero as $|x|\to\infty$. A precise well-posedness result is cited below; see Theorem ~\ref{thm:wellposedness}.
\\

NLS/GP equations play a central role in the understanding of nonlinear optical \cite{MoNe,Boyd,SulemSulem}
and macroscopic quantum systems; see, for example, \cite{EY:99}.
A striking and important feature of NLS/GP is that it can have localized standing waves or  {\it nonlinear bound state} solutions,  some of which are stable and play a central role in the general dynamics.  
 In particular, for a wide variety of potentials and nonlinearities there exists an
interval $\mathcal{I}\subset\mathbb{R}$ such that for any $\lambda\in
\mathcal{I}$, (\ref{eq:NLS}) has {\it nonlinear ground state} solutions. These are solutions of the form
\begin{equation}
\psi(x,t)=e^{i\lambda t}\phi^{\lambda}(x),
\nn\end{equation}
where
\begin{equation}
-\Delta\phi^\lambda+V\phi^\lambda-f(|\phi^\lambda|^2)\phi^\lambda=-\lambda\phi^\lambda
\label{bs-eqn}\end{equation}
with $\phi^{\lambda}\in H^1$ and $\phi^\lambda>0$.

The gauge (phase-translation) invariance of (\ref{eq:NLS}), 
\begin{equation}
\psi\mapsto e^{i\gamma}\psi,\ \gamma\in [0,2\pi),
\nn\end{equation}
 generates a  {\it nonlinear ground state} or  {\it ``soliton'' }
 \footnote{The term {\it soliton} sometimes refers, more specifically,  to particle-like solutions of completely integrable PDEs.}
   {\it manifold}:
\begin{equation}
\mathcal{M}_{\mathcal{I}}:=\{e^{i\gamma}\phi^{\lambda},\lambda\in
\mathcal{I},\ \gamma\in [0,2\pi)\}.
\label{phi-lambda-mfld}
\end{equation}
 If $V$ is identically zero, then NLS /GP admits a larger group of symmetries and the definition of soliton manifold (which exists in the focusing case, $g<0$) is naturally extended to incorporate these additional symmetries; see, for example, \cite{Wei86,GSS87}. \\ \\
 %
{\it Orbital Stability:}\ \  The soliton manifold $\mathcal{M}_{\mathcal{I}}$ is said to be {\it orbitally stable} if  any initial condition $\psi_{0}$, which is  close to $\mathcal{M}_{\mathcal{I}}$ in $H^{1}$, gives rise to a solution $\psi(t)$, which  is $H^1$ close for $t\ne0$.  There is an extensive literature on the {\it orbital stability} of the soliton manifold. For the case, $V\equiv0$,  orbital stability (stability modulo spatial and phase translations) of global energy minimizers was proved in \cite{CL79} by compactness arguments. In 
\cite{Wei85,Wei86}  it is shown that  positive solutions, which are  index one critical points (Hessian with one strictly negative eigenvalue)
and satisfying the slope condition
\footnote{$\mu=-\lambda$ is the typical definition of soliton frequency.  Therefore the slope condition, (\ref{eq:stab1}), often appears as a rate of change with respect to $\mu$ being negative.}:
\begin{equation}\label{eq:stab1}
\frac{d}{d\lambda}\ \int_{\R^d}\ |\phi^\lambda(x)|^2\ dx\ >\ 0,
\end{equation}
 are  $H^1$ orbitally stable. For $V\equiv0$ and $f(|\psi|^2)=-g|\psi|^2, g<0$ (focusing case), (\ref{eq:stab1}) is equivalent to $\sigma<2/d$. Orbital stability of solitary waves 
 of NLS/GP for a class of potentials, $V$, was studied in \cite{RoWe} and in a semi-classical setting by \cite{Oh88}. A general formulation of a stability / instability theory is presented in \cite{GSS87}.\\ \\
{\it Asymptotic stability:}\ \ 
 We say the soliton manifold, $\mathcal{M}_\mathcal{I}$  is asymptotically stable if $\psi_0$ close to $\mathcal{M}_\mathcal{I}$ in a suitable norm implies that $\psi(t)$ remains close to and {\it converges to} $\mathcal{M}_\mathcal{I}$ (in a possibly different norm),
  as $t$ tends to infinity.\\
   Are solitary waves  {\it asymptotically stable}? This is a local variant of the problem of {\it asymptotic resolution}\cite{Tao-08}, {\it i.e.} whether general initial conditions resolve into stable nonlinear bound states of the system plus dispersive radiation.
 A great deal of progress has been made on this problem in recent years. The study of asymptotic stability of solitary waves  was initiated in  \cite{SoWe:9092}; see also
 \cite{BuPe:93,PW97,GNT:04,Weder:00}.
  In the translation invariant case, asymptotic stability was then investigated by \cite{BuPe:95}.  
Asymptotic stability analysis requires  two new analytical features: one dynamical systems  and the other harmonic analysis / spectral theoretic. \\ \\
First, since we do not know in advance which nonlinear ground state in ${\cal M}_{\cal I}$ emerges in the large time limit, a decomposition with flexibility allowing for the asymptotic soliton to dynamically emerge is required\footnote{The case of  integrable systems, such as 1d NLS ($V=0,\ f(|\psi |^2)\psi=|\psi |^2\psi$) is an important class for which it {\it is} possible to determine the emerging coherent structures from the {\it scattering transform} of the initial data.}. To this end, the solution is decomposed in terms of a motion  along the soliton manifold and components  (symplectic or bi-) orthogonal to it. 
Dynamics along the soliton manifold, $\mathcal{M}_{\mathcal{I}}$,  are governed by {\it modulation equations}; see, for example,
  \cite{Wei85,FGJS04,HoZw07}. \\  \\
Secondly, in order to prove convergence to the soliton manifold, ${\cal M}_{\cal I}$, we need to show that the deviation of the solution from ${\cal M}_{\cal I}$ decays with advancing time. This requires time-decay estimates ($L^p(\R^d)$ or space-time norms) for the linearized (about the soliton) propagator on the subspace (symplectic or bi-) orthogonal to the discrete spectral subspace.  The discrete subspace is the union of a  {\it zero frequency mode subspace} spanned by infinitesimal generators of the NLS/GP symmetries (translation, gauge) acting on 
 $\phi^\lambda$,  and often a subspace of {\it neutral modes} (sometimes called internal modes) with non-zero frequencies. 
  
 Since a typical perturbation of the ground state solitary wave in $\mathcal{M}_{\mathcal{I}}$ excites all discrete spectral components, one must understand the mechanisms, due to which these do not interfere with the asymptotic convergence of $\psi(x,t)$ to $\mathcal{M}_{\mathcal{I}}$. In brief:  Concerning the zero modes, the choice of modulation equations ``quotients out'' the zero modes; perturbations exciting these, induce motion {\it along} the soliton manifold. And concerning  the
 non-zero frequency neutral modes, these are shown to damp to zero, as $t\to\infty$, due resonant nonlinear coupling of discrete to radiation modes. Related to this is a further dynamical systems aspect of the 
 analysis. The neutral mode amplitudes are governed by nonlinear oscillator equations, coupled to a dispersive wave field. Near-identity changes of variables are used to put the system in an appropriate normal form, wherein the mechanism of energy transfer from the neutral modes to the evolving soliton and propagating radiation,
 is made explicit. Energy transfer shows up as an explicit (nonlinear) damping term in the normal form; see the discussion below. The positive damping coefficient (matrix, in the present work) is a nonlinear variant of Fermi's Golden Rule \cite{CT:92}.    See  \cite{BuPe:95}, regarding the dynamics near solitary waves of the translation invariant NLS equations and  \cite{SoWe99} for ``breathers'' of a class of nonlinear wave equations.  In 
 \cite{SoWe04} this mechanism was proved to be responsible for  {\it ground state selection} in NLS/GP equations; see also  \cite{Wei05}.  Experimental verification of the prediction in  \cite{SoWe04,SoWe-PRL:05} is reported  in \cite{Silberberg:05}. Related work on resonant radiation damping appears in~\cite{TsaiYau02,BuSu03,TsaiYau02-2,Tsai03,CuKiPe,CuMi}. The role of the Fermi Golden rule in the non-persistence of coherent structures for nonlinear wave equations was first demonstrated, via Floquet analysis, in ~\cite{sigal}. There is a close relation to the perturbation theory of embedded eigenvalues for linear problems \cite{RS3-4,SW-GAFA:98,CPV:05}. 
 
The above works on nonlinear resonance required that the neutral modes frequencies (a) lie  sufficiently close to the essential spectrum and (b) be of geometric multiplicity one. For example: for the cubic nonlinearity, $f(|\psi|^2)=-g|\psi|^2$, {\it close} means:   coupling of 
 to radiation modes occurs at order $|g|^2$.   The situation where simple neutral modes with a  large spectral gap has been studied in \cite{GS2,GS3,G1,CuMi, Cu2008}.
  Here,   coupling  of the discrete to continuum modes occurs at some high order in $g$. Thus, the normal form expansion gives a damping term at some even order: $|g|^{2k},\ \ k\ge2$.
 \\ \\ \\
 \nit{\bf Results of this paper\ -
  systems with degenerate neutral modes:}\\ 
 An important situation, not covered by previous results, is the dynamics in the presence of degenerate neutral modes. This case arises naturally in systems of spatial dimension $d\ge2$ with symmetry. For example, if the potential is spherically symmetric, $V=V(|x|)$, then the first and higher excited states are degenerate, the degree of degeneracy related to the order of the associated spherical harmonics. Another interesting class of examples is a class of double-well potentials; see Appendix ~\ref{subsec:example}.\\ \\
In this paper we prove the asymptotic stability of the ground state / soliton manifold, ${\cal M}_{\cal I}$, of NLS/GP when the linearized  spectrum has degenerate neutral modes. We show that the solution has three {\it interacting} parts:\\
 (i)\  a  modulating soliton, parametrized by motion along ${\cal M}_{\cal I}$\\
 (ii)\ oscillatory, spatially localized, neutral modes which decay with time and\\
 (iii)\ a dispersive part, which decays in a local energy norm.\\
  The neutral modes and dispersive waves decay via transfer their  mass  to the  soliton manifold or to  spatially infinity. Additionally, degenerate neutral modes are coupled and exchange mass among themselves  in addition  to the soliton and radiation. These degenerate modes cannot be viewed as very weakly coupled ``oscillators''~\cite{Tsai03}. We require instead a new normal form expansion. This is related to  ideas developed in ~\cite{KW01}, where a parametrically forced linear Hamiltonian PDE was considered, and a normal form, uniform in discrete eigenvalue spacing, was required.  
  
  We outline the perspective we take and give a rough form of the main theorem, Theorem \ref{THM:MainTheorem}. Consider NLS/GP,
  where $-\Delta+V$ has a ground state, $\xi_0(x)>0$, $e_0<0$ and a {\it degenerate} excited state, whose energy, $e_1, \ e_0<e_1<0$, is assumed sufficiently near the zero.  Typical solutions of the linear 
  Schr\"odinger equation, evolving from localized initial data 
  $\psi_0,\ \ \psi(t)=\exp[-i(-\Delta+V)t]\psi_0$, will evolve a (typically) time quasi-periodic superposition of spatially localized ground  and excited time-periodic states, plus a part which disperses to zero, {\it i.e.} tends to zero as $t$ advances in $L^2_{loc}$. This picture emerges from the spectral decomposition of $-\Delta+V$ in $L^2$, with respect to which the bound state projections of the solution evolve as independent {\it oscillators} and the continuous spectral part of the solution has a character, qualitatively like a solution to the free Schr\"odinger equation. 
  
  For NLS/GP, {\it e.g.} $-g|\psi|^2\psi,\ \ g\ne0$, the dynamics of  discrete and continuum modes are coupled.
   We consider an appropriate open set of initial conditions near the soliton manifold. {\bf In contrast to the linear Schr\"odinger equation} we show that the solution converges to a nonlinear ground state.
  To see this, we view NLS/GP as a infinite dimensional Hamiltonian system comprised of two subsystems: (i) a finite dimensional system governing dynamics along the soliton manifold, ${\cal M}_{\cal I}$,  parametrized by $(\lambda(t),\gamma(t))$, the zero modes amplitudes $(a_1,a_2)$, and the neutral mode amplitudes, $z=(z_{1},z_{2},\cdot\cdot\cdot,z_{n})^{T}$, and (ii) an infinite dimensional dispersive Schr\"odinger wave equation. A very detailed analysis of this coupled system (the bulk of this paper) yields the following (rough) form for the asymptotic behavior of small amplitude solutions of NLS/GP:
\begin{mainthm}
Consider the initial value problem for NLS/GP. Assume arbitrary localized initial data, which are  sufficiently near a small amplitude nonlinear bound state, $\phi^{\lambda_0}$. Then the solution of NLS/GP evolves as a modulated soliton plus decaying error having the following form:
\begin{align}
\psi(t)\ =\ e^{i\int_{0}^{t}\lambda(s)ds}
 e^{i\left[\gamma(t)+a_2(z(t),\bar{z}(t))\right]}\ \times[\ \ \phi^{\lambda(t)+a_1(z(t),\bar{z}(t))}\  +\ {\cal O}(|z(t)|)\ +\ R(t)\ \ ],\label{rough-exp}\end{align}
 Here, $\lambda(t)\to\lambda_\infty$.
 ${\cal O}(|z(t)|)$  represents a localized, nonspreading but decaying part, satisfying $|z(t)|\le C\langle t\rangle^{-1/2}$. Also, 
  $a_j=a_j(z,\bar{z})={\cal O}(|z|^2)$.  $R(t)$ represents a spreading, dispersively decaying part, and  tends to zero as $t\to\infty$ in $L^2_{loc}$; more precisely, $\|\langle x\rangle^{-\nu}R(t)\|_2\to0,\ \ \nu>0$.
\end{mainthm}
\nit For the precise statement, see Theorem \ref{THM:MainTheorem}.\\ \\

A key part of the proof of Theorem \ref{THM:MainTheorem} is to show that $|z(t)|$ tends to zero and that $\lambda(t)$ has a limiting value $\lambda_\infty\in\mathcal{I}$ as $t$ tends to infinity.
 We prove the latter, by showing $\D_t\lambda(t)\in L^1(\R^+)$.  
 We have two comments on the approach of this article to these issues.\\ \\
 {\bf (1)\ New normal form:}\ We show that there exist a {\underline{non-negative}} {\underline{ symmetric}} {\underline{matrix}} $\Gamma(z,\bar{z})=O(|z|^{2})$ and a skew symmetric matrix $\Lambda(z,\bar{z})=O(|z|^{2})$ (see (\ref{eq:detailedDescription1}) below) such that
\begin{equation}
{\partial_t} z =-iE(\lambda) z -\Gamma( z ,\bar{ z }) z +\Lambda( z ,\bar{ z })
z\ +\ \cO\left((1+t)^{-\frac{3}{2}-\delta}\right),\ \ \delta>0\label{new-nf}
\end{equation}
The matrix $\Gamma$ is defined in terms of the spectral decomposition of the $L(\lambda)=JH(\lambda)$, the generator of the linearized flow about the nonlinear bound state, $\phi^\lambda$; see section \ref{SEC:Operator}. Our analysis requires that $\Gamma=\Gamma(z,\bar{z};\lambda)$ is positive definite for an open interval of $\lambda-$ values. A variant of this hypothesis appears in previous work ~\cite{SoWe04,TsaiYau02,BuSu03,TsaiYau02-2,Tsai03,GS2,GS3,CuKiPe,CuMi}
 It is expected to hold, in some sense, generically. In section ~\ref{SEC:FGR}  we state hypotheses under which positive definiteness holds for class of potentials of double-well type, constructed in section ~\ref{subsec:example}.%
%
%
This  hypothesis, denoted FGR\ ( see (\ref{Gammadef}-\ref{eq:FGR})\ ), is 
 a nonlinear variant of the Fermi Golden Rule \cite{CT:92,RS3-4,SW-GAFA:98}.
 We note that for finite dimensional Hamiltonian systems a
  damping term is absent
  ; it would violate phase-volume conservation. This term arises due to nonlinearity induced coupling between discrete and continuous spectral (radiation) modes, a phenomenon associated with continuous spectra, arising in PDEs on spatially infinite domains; see \cite{SoWe99,Wei05}.
We show that  (\ref{new-nf}) and FGR (see (\ref{Gammadef}-\ref{eq:FGR})\ )\ , imply
 the bound $|z(t)|=\cO(t^{-\frac{1}{2}}).$ 
For the case of multiple simple bound states with well-separated frequencies, a system of type (\ref{new-nf}) holds with $\Gamma$, a {\it diagonal} matrix~\cite{Tsai03}.  Equation (\ref{new-nf}) can be viewed as a new normal form, a special case of one valid uniformly in neutral mode eigenfrequency-separation
\\ \\
{\bf (2)\ Choice of basis for the neutral mode subspace:}\ We prove that  $\lambda(t)$ approaches some $\lambda_\infty$ as $t\rightarrow \infty$, by proving that $\D_t\lambda(t)$ is integrable. If there are $n$ simple well-separated neutral modes,  one initially finds
\begin{equation}
\partial_{t}\lambda(t)=\sum_{m=1}^{n}a_{m}|z_{m}|^{2}+O(t^{-\frac{3}{2}}).
\nn\end{equation}
 Since we expect $|z_{m}|=O(t^{-\frac{1}{2}})$ we can not  conclude integrability of $\D_t\lambda(t)$. However, it can be shown that after near identity change of variables: $z\mapsto z+\cO(|z|^2)$, we can take $a_{m}=0$; see the normal form expansion in ~\cite{GS2,GS3,SoWe04}.
 In the degenerate (similarly not well-separated) $\lambda(t)$ satisfies the equation 
 \begin{equation}
 \partial_{t}\lambda(t)=\sum_{m,k}a_{m,k}z_{m}\bar{z}_{k}+O(t^{-\frac{3}{2}}).
 \nn\end{equation}
 In the present paper we show, very generally, by appropriate choice of neutral subspace basis we can take $a_{m,k}=0$.\\ \\
  Finally, we 
expect that our techniques can be extended to more complicated situtations, {\it e.g.} where coupling of some neutral modes occurs at higher order in the nonlinearity.
%
\\ \\
\nit {\bf Outline of the paper:}\ \ 
The paper is organized as follows. In Section ~\ref{notation} display notation, which is often used. Section ~\ref{HaGWP} is a brief section outlining structural properties of NLS / GP  and gives a statement of a basic well-posedness result. Section \ref{ExSt} introduces solitary waves (solitons) in the regime of weak nonlinearity. Section \ref{SEC:Operator} has a detailed discussion of the spectral properties of $L(\lambda)=JH(\lambda)$, the generator of the linearized dynamics about the soliton: zero energy subspace, degenerate neutral subspaces and continuous spectral subspaces. Projections associated with theses subspaces are defined and decay estimates of the linearized evolution on the continuous spectral subspace are recalled. 
In section \ref{SEC:FGR} the Fermi Golden Rule matrix, $\Gamma$,  is introduced explicitly, (\ref{Gamma-kl}). The detailed calculations, proving symmetry and non-negativity are given in the appendix;
section \ref{subsec:Gamma-general}. 
 The main theorem, requires positive definiteness of $\Gamma$. Proposition \ref{prop:example-fgr} is a result reducing the required positive definiteness to a condition involving the spectral properties of $-\Delta +V$. 
Section \ref{MainTHM} contains a statement of the main theorem, Theorem 
\ref{THM:MainTheorem}.  In section 
~\ref{SEC:ReformMain} we give a more precise formulation of 
 Theorem  \ref{THM:MainTheorem}. This formulation makes  explicit the dynamical (modulation) equations for the solitary wave parameters, the neutral mode amplitudes  and  the dispersive part. These are  proved via  normal form methods  in sections ~\ref{SEC:effective} and ~\ref{SEC:NormalForm}. In section ~\ref{ProveMain} we prove the  main Theorem~\ref{THM:MainTheorem} in the setting of reformulated Theorem~\ref{GOLD:maintheorem}. 
%
Section \ref{sec:appendix}  contains some important calculations used in the body of the paper. Of particular interest is the appendix of section \ref{subsec:example}, where a class of double-well three-dimensional potentials is constructed, to which we apply Theorem~\ref{THM:MainTheorem}.
\section*{Acknowledgments} Zhou Gang was supported, in part, by a Natural Sciences and Engineering Research Council of Canada (NSERC)  Postdoctoral Fellowship.
 Michael I. Weinstein was supported, in part, by  U.S.
NSF Grants DMS-04-12305 and DMS-07-07850. Part of this research was completed, while the Z.G. was a visitor of the Department of Applied Physics and Applied Mathematics (APAM) of Columbia University. Z.G. wishes to thank the APAM for its hospitality.
\vfil\eject
\section{Notation}\label{notation}
\begin{itemize}
\item[(1)]\ 
$
\alpha_+ = \max\{\alpha,0\},\ \ [\tau]=\max_{\tilde\tau\in Z}\ \{\tilde\tau\le\tau\}
$
\item[(2)]\ $\Re z$ = real part of $z$,\ \ $\Im z$ = imaginary part of $z$
\item[(3)] Multi-indices
\begin{align}
&w\ =\ (w_1,\dots, w_N)\ \in \mathbb{C}^N,\ \bar{w}=(\bar{w}_1,\dots, \bar{w}_N)\\
&a\in \mathbb{N}^N,\ z^a=z_1^{a_1}\cdot\cdot\cdot z_N^{a_N}\nonumber\\
&|a|\ Ä=\ |a_1|\ +\ \dots\ +\ |a_N|
\nonumber
\end{align}
$z$ denotes the vector of {\it neutral mode} amplitudes\\
$\xi$ denotes the vector, whose $j^{th}$ entry is the $j^{th}$ neutral vector-mode of $JL(\lambda)$, $\xi_j$.
\item[(4)] $Q_{m,n}$ denotes an expression of the form
\begin{equation}
Q_{m,n}\ = \sum_{|a|=m,\ |b|=n}\ q_{a,b}\ z^a\ \bar{z}^b\   
          = \sum_{|a|=m,\ |b|=n}\ q_{a,b}\prod_{k=1}^{N}\ z_k^{a_k}\bar{z_k}^{b_k} \nonumber
\end{equation}
\item[(5)]  \begin{equation}
J\ =\ \left(\begin{array}{cc} 0 & 1\\ -1 & 0\end{array}\right),
\ \ H\ =\ \left(\begin{array}{cc} L_+ & 0\\ 0 & L_-\end{array}\right),\ \ 
 L=JH=\left(\begin{array}{cc} 0 & L_-\\ -L_+ & 0\end{array}\right)
\nonumber\end{equation}
\item[(6)] $\sigma_{ess}(L)=\sigma_c(L)$ is the essential (continuous) spectrum of $L$,\\ $\sigma_{disc}(L)$ is the discrete spectrum of $L$.
\item[(7)]\ Riesz projections: $P_{disc}(L)$ and $P_c(L)=I-P_{disc}(L)$\\
 $P_{disc}(L)$ projects onto the discrete spectral part of $L$\\
$P_c(L)$ projects onto the continuous spectral part of $L$
\item[(8)]\ $$\langle f,g\rangle = \int\ f(x)\ {\overline{g(x)}}\ dx $$
\item[(9)]\ $$\| f\|_p^p=\ \int_{\R^d}\ |f(x)|^p\ dx,\  \ 1\le p\le\infty$$
\item[(10)]\ $$\| f\|_{H^{s,\nu}}^2\ =\  \int_{\R^d}\ \left|\langle x\rangle^\nu\ (I-\Delta)^{\frac{s}{2}}f(x)\right|^2\ dx$$
\end{itemize}
\vfil\eject
\section{Hamiltonian Structure}\label{HaGWP}

NLS/GP  can be expressed as a Hamiltonian system:
\begin{equation}
i\D_t\psi\ =\ \frac{\delta\cE[\psi,\bar{\psi}]}{\delta\bar{\psi}}
\nn\end{equation}
where the Hamiltonian energy, $\cE[\cdot]$, is defined by
\begin{align}
\cE[\psi]\ =\ \mathcal{E}[\psi,\bar{\psi}] &=\int
\left(\frac{1}{2} \nabla\psi\cdot\nabla\bar{\psi}\ +\ 
\frac{1}{2}V(x) \psi\ \bar{\psi}\ -\ F(\psi\ \bar{\psi})\ \right)\ dx,\nn\\
F(u)&=\frac{1}{2}\int_{0}^{u}\ f(\xi)\ d\xi.
\nn\end{align} 

Equation (~\ref{eq:NLS}) is a Hamiltonian system on Sobolev space
$H^{1}(\mathbb{R}^{d},\mathbb{C})$ viewed as a real space
$H^{1}(\mathbb{R}^{d},\mathbb{R})\oplus
H^{1}(\mathbb{R}^{d},\mathbb{R})$, i.e. $H^1(\mathbb{R}^d,\mathbb{C})\ni f\leftrightarrow (Ref, Imf)\in H^{1}(\mathbb{R}^{d},\mathbb{R})\oplus
H^{1}(\mathbb{R}^{d},\mathbb{R}),$ 
with the
symplectic form
\begin{equation}
\omega(\psi,\phi)=Im\int_{\mathbb{R}^{d}}\ \psi\ \bar{\phi}.
\nn\end{equation}
Equation (~\ref{eq:NLS})  is invariant under time-translation ($t\mapsto t+t_0$) and gauge (phase)-translation $\phi\mapsto\ \phi e^{i\gamma},\ \gamma\in\R$  yielding, by Noether's theorem, the conservation laws
\begin{enumerate}
 \item[{}] Conservation of energy:\ \ \ 
  $\mathcal{E}[\psi(t)]=\mathcal{E}[\psi(0)];$
 \item[{}]
 Conservation of particle number (optical power):\ \ \ $\mathcal{N}[\psi(t)]=\mathcal{N}[\psi(0)],$ where\ \ \ \
  \begin{equation}
  \mathcal{N}[\psi]=\int
 |\psi|^{2}\ dx.\nn\end{equation}
\end{enumerate}
\nit We make the following\\
\nit{\bf Assumptions on the potential  $V$ and nonlinearity $f$:}
\begin{enumerate}
 \item[{\bf (fA)}]  $f(\tau)$ is a smooth function satisfying $f(\tau)=\cO(\tau)$ for $|x|$ is small.  Thus, the nonlinearity in NLS is cubic at small amplitudes,
  {\it i.e.} $f(|\psi|^2)\psi\ \sim\ g|\psi|^2\psi$.
 \item[{\bf (VA)}] $V$ is smooth and decays exponentially $|x|$ tends to  $\infty.$
\end{enumerate}
To ensure  the global well-posedness of the initial value problem for~(\ref{eq:NLS}) we impose:\begin{enumerate}
 \item[{\bf (fB)}]\ Subcritical nonlinearity for large amplitudes:\\
$|f(\xi)|\leq c(1+|\xi|^{\beta})$ for some $\beta\in[0,\frac{2}{d})$ and\\
  $|f^{'}(\xi)|\leq c(1+|\xi|^{\alpha-1})$ for some $\alpha\in
 [0,\frac{2}{(d-2)_{+}})$.\ Here, $s_{+}=\max\{s,0\}$.
\end{enumerate}
The following well-posedness theorem  can be found in ~\cite{Caz03,Cazenave,SulemSulem}.\\ 
\begin{theorem}\label{thm:wellposedness}
Assume that the nonlinearity $f$ satisfies condition (fB), and
the potential $V$ satisfies (VA). Then equation (~\ref{eq:NLS}) is
globally well-posed in ${H}^{1}$, {\it i.e.} the Cauchy problem
for Equation (~\ref{eq:NLS}) with a data $\psi(0)\in {H}^{1}$
has a unique solution $\psi(t)$ in the space ${H}^{1}$, which depends continuously on $\psi(0)$. Moreover, the solution $\psi(t)$ satisfies conservation of energy and conservation of particle number.
\end{theorem}

\section{ Bifurcation and Lyapunov Stability of Solitons in the Weakly Nonlinear Regime}\label{ExSt}
In this section we  discuss the existence of solitons in the weakly nonlinear regime. 
The following arguments are similar 
to those in \cite{RoWe,TsaiYau02} except that the
excited states are degenerate. We assume that the linear operator $-\Delta+V$ has
the following properties
\begin{enumerate}
\item[{\bf ($Eig_V$)}]
The linear operator $-\Delta+V$ has two eigenvalues $e_{0}<e_{1}<0$ with $2e_1>e_0$.
 $e_{0}$ is the lowest eigenvalue with
ground state $\phi_{lin}>0$, the eigenvalue $e_{1}$ is degenerate
with multiplicity $N$ and eigenfunctions
$\xi_{1}^{lin},\xi_{2}^{lin},\cdot\cdot\cdot,\xi_{N}^{lin}.$
\end{enumerate} 
\begin{remark}
In appendix ~\ref{subsec:example} we construct a class of  double-well examples, $V$,  in dimension $d=3$ and with multiplicity $N=2$.
\end{remark}

The following result shows that nonlinear bound state solutions $(\phi^\lambda,\lambda)$ of NLS/GP, (\ref{bs-eqn}), bifurcate from the zero state and the linear ground state energy $(0,\lambda=-e_0)$.

\begin{proposition}\label{LM:NearLinear1}
Suppose $-\Delta+V$ satisfies the
conditions in {\bf ($Eig_V$)} above. Then there exists a constant
$\delta_{0}>0$ and a nonempty interval 
$\mathcal{I}_{\delta_0}\subset [-e_{0}-\delta_{0},
-e_{0}+\delta_{0}]$ such that for any $\lambda \in \mathcal{I}_{\delta_0}$ (~\ref{eq:NLS}) has solutions of the form
$\psi(x,t)=e^{i\lambda t}\phi^{\lambda}\in L^{2}$ with
\begin{equation}
\phi^{\lambda}\ =\ \delta(\lambda)\ \left(\ \phi_{lin}\ +\ \cO\left(\delta(\lambda)\right)\ \right),\ \ \delta(\lambda)= {\cal O}(\left|\ \lambda-|e_0|\ \right|^{1\over2})
\label{phi-lambda}
\end{equation}
for $\left|\ \lambda-|e_0|\ \right|$ small.\\
Moreover,
 \begin{equation}\label{expondecay}
|\phi^{\lambda}(x)|\leq ce^{-\delta|x|}\ \text{and}\
\left|\D_\lambda\phi^{\lambda}(x)\right|\leq ce^{-\delta|x|},
\end{equation}
and similarly for the spatial derivatives of $\phi^{\lambda}$ and
$\D_\lambda\phi^{\lambda}$.
\end{proposition}
\begin{remark}\label{rmk:bif-g}
Suppose $f(|\psi|^2)\psi = -g|\psi|^2\ +\ o(|\psi|^2)$.\\
 Then, for $g>0$ (repulsive case) we have
for $\mathcal{I}_{\delta_0}=(-e_0,-e_0+\delta_0)$.\\ 
For $g<0$ (attractive case), we have $\mathcal{I}_{\delta_0}=(-e_0-\delta_0,-e_0)$.
\end{remark}
Finally, we conclude this section by noting that for $\delta'\le\delta_0$
sufficiently small that soliton manifold, $\cM_{\delta'}$, 
see (\ref{phi-lambda-mfld}), is $H^1$ orbitally stable; see  the discussions in the introduction and \cite{Wei86,RoWe,GSS87}.
\section{$L(\lambda)=JH(\lambda)$, the Linearized Operator about $\phi^\lambda$}\label{SEC:Operator} We now turn to a discussion of the operator obtained by linearization around the soliton and the existence of  neutral modes with non-zero frequencies.
 Rewrite Equation (~\ref{eq:NLS}) as 
\begin{equation}
\frac{\D\psi}{\D t}=G(\psi),
\nn\end{equation}
where the nonlinear map $G(\psi)$ is defined by
\begin{equation}\label{symmetryG}
G(\psi)=-i(-\Delta+\lambda+V)\psi+if(|\psi|^{2})\psi.
\end{equation}
Then the linearization of Equation (~\ref{eq:NLS}) can be written as
\begin{equation}
\frac{\partial\chi}{\partial t}=d G(\phi^{\lambda})\chi,
\label{linearized}
\end{equation}
where $d G(\phi^{\lambda})$ is the Fr\'echet derivative of
$G(\psi)$ at $\phi^{\lambda}$. It is computed to be
\begin{equation}\label{defineoperator}
d
G(\phi^{\lambda})\chi=-i(-\Delta+\lambda+V)\chi+if[(\phi^{\lambda})^{2}]\chi+if^{'}[(\phi^{\lambda})^{2}](\phi^{\lambda})^{2}(\chi+\bar\chi).
\end{equation}
This is a real linear but not complex linear operator. To convert it
to a linear operator we pass from complex functions to real
vector-functions
$$\chi\longleftrightarrow \vec{\chi}=\left(
\begin{array}{lll}
\chi_{1}\\
\chi_{2}
\end{array}
\right),$$ where $\chi_{1}=Re\chi$ and $\chi_{2}=Im\chi.$ Then
$d G(\phi^{\lambda})\chi\longleftrightarrow
L(\lambda)\vec{\chi}$ where the operator $L(\lambda)$ is given by
\begin{equation}\label{eq:LinOpera}
L(\lambda)=JH(\lambda)
\end{equation} where $J$ is a skew-symmetric matrix $$J:=\left(
\begin{array}{lll}
0&1\\
-1&0
\end{array}
\right)$$ and $H(\lambda)$ is a self-adjoint matrix
$$H(\lambda):=\left(
\begin{array}{lll}
L_{+}(\lambda)&0\\
0&L_{-}(\lambda)
\end{array}
\right)$$ with
$$L_{-}(\lambda):=-\Delta+\lambda+V-f[(\phi^{\lambda})^{2}]$$ and
$$L_{+}(\lambda):=-\Delta+\lambda+V-f[(\phi^{\lambda})^{2}]-2f^{'}[(\phi^{\lambda})^{2}](\phi^{\lambda})^{2}.$$
We extend the operator $L(\lambda)$ to the complex space
$ H^{2}(\mathbb{R}^{d},\mathbb{C})\oplus
 H^{2}(\mathbb{R}^{d},\mathbb{C}).$

\subsection{The Spectrum of $L(\lambda)$}\label{sec:weaklynonlin} The operator $L(\lambda)$ has the neutral modes:
\begin{proposition}\label{Prop:NeuMode}
Let $L(\lambda)$, or more explicitly,  $L(\lambda(\delta),\delta)$
denote the linearized operator about the the bifurcating state
$\phi^\lambda, \lambda=\lambda(\delta)$. Note that $\lambda(0)=
-e_0$. Corresponding to the degenerate e-value, $e_1$, of
$-\Delta+V$, the matrix operator $$L(\lambda=-e_0,\delta=0)$$ has
degenerate eigenvalues $\pm iE(-e_0)=\pm i(e_1-e_0)$, each of
multiplicity $N$. For $\delta>0$ and small these bifurcate to
(possibly degenerate) eigenvalues $\pm iE_1(\lambda),\dots,$ $\pm
iE_{N}(\lambda)$ with eigenfunctions
$$\left(
\begin{array}{lll}
\xi_{1}\\
\pm i\eta_{1}
\end{array}
\right),\ \left(
\begin{array}{lll}
\xi_{2}\\
\pm i\eta_{2}
\end{array}
\right),\ \cdot\cdot\cdot, \left(
\begin{array}{lll}
\xi_{N}\\
\pm i\eta_{N}
\end{array}
\right)$$ with $$\langle \xi_{m},\eta_{n}\rangle =\delta_{m,n}$$ and
\begin{equation}\label{eq:GoToNear}
0\not=\displaystyle\lim_{\lambda\rightarrow
e_{0}}\xi_{j}=\lim_{\lambda\rightarrow e_{0}}\eta_{j}\in
span\{\xi^{lin}_{j},\ j=1,2,\cdot\cdot\cdot, N\}\ \text{in}\
{H}^{k}\ \text{spaces for any}\ k>0.
\end{equation}
Moreover, for $\delta$ sufficiently small  $2E_j(\lambda)>\lambda,\
j=1,2,\cdot\cdot\cdot, N,$ (nonlinear coupling of discrete to continuous spectrum at second order).
\end{proposition}
For the case of a radial potential, $V=V(|x|)$, the neutral modes have the following structure:
\begin{proposition}\label{mainLem2}
If the potential is radial, $V=V(|x|)$,  then $\phi^{\lambda}$, hence $\partial_{\lambda}\phi^{\lambda}$, is spherically symmetric.  If the degenerate linear excited states $\xi^{lin}_{n}$ are of the form $\xi^{lin}_{j}=\frac{x_{j}}{|x|}\xi^{lin}(|x|)$ for some function $\xi^{lin}$, then  $E_{j}=E_{1}$ for any $j=1,2,\cdot\cdot\cdot,N=d$ and we can choose $\xi_{j},\eta_{j}$ such that $\xi_{j}=\frac{x_{j}}{|x|}\xi(|x|)$ and $\eta_{j}=\frac{x_{j}}{|x|}\eta(|x|)$ for some real functions $\xi$ and $\eta.$
\end{proposition}
\begin{remark}
For $d=3$, the hypothesis on the linear excited states states that these are of the proportional to $\xi^{lin}(|x|)\ Y_1^m(\theta,\phi),\ \ m=-1,0,1$, where $Y_1^m$ are the spherical harmonics of degree one.
\end{remark}
\begin{proof}
We sketch the proof. If $V$ is spherically symmetric then by the uniqueness of the ground states and the fact $-\Delta+V$ is invariant under unitary transformation we have $\phi^{\lambda}$, hence $\partial_{\lambda}\phi^{\lambda},$ is spherically symmetric.

We now outline a proof of the existence of $\xi_{j}$ and $\eta_{j}$ with desired structure. Define a linear space 
\begin{equation}
Y^k=\left\{\ J\in {H}^{k},\ J(x)=\frac{x_{1}}{|x|}g(|x|)\right\}.
\nn\end{equation}
 By the definition, $L(\lambda): Y^2\to Y^0$. Note  that, restricted to $Y^2$,  $\frac{x_1}{|x|}\xi^{lin}(|x|)$ is an eigenfunction of $-\Delta+V$
  of multiplicity one. Application of bifurcation theory to $Y^2$, we prove there exists an eigenfunction $(\xi_{1},i\eta_{1})^{T}\in Y$ with eigenvalue $E_{1}.$ The other eigenfunctions with the same eigenvalue are obtained by noting that this computation can be carried out for any $x_j,\ j=1,\dots,d$. \end{proof}
\bigskip

Based on the above discussion, we assume the following:\\ \\
{\bf (SA)\ Structure of the the discrete spectrum of $L(\lambda)=JH(\lambda)$}
\begin{enumerate}
\item $\sigma_d(L(\lambda))$ consists of an eigenvalue at $0$
 and complex conjugate eigenvalues at $\pm iE(\lambda)$.
 \item  The discrete subspace, corresponding to the eigenvalue $0$ is spanned by the associated eigenfunctions $\left(
\begin{array}{lll}
0\\
\phi^{\lambda}
\end{array}
\right)$ and $\left(
\begin{array}{lll}
\partial_{\lambda}\phi^{\lambda}\\
0
\end{array}
\right)$ 
\item The discrete subspace, corresponding to the eigenvalue $ iE(\lambda),$ $E(\lambda)>0$, is $N-$ dimensional and is spanned by the  (complex) eigenfunctions
 $v_{1},\ v_{2}\cdot\cdot\cdot, v_{N}$ 
 \item  Thus, $\overline{v_{1}},\ \overline{v_{2}},\cdot\cdot\cdot,
\overline{v_{N}}$ are eigenfunctions which span the discrete subspace corresponding to the eigenvalue $-iE(\lambda).$ 
\item Moreover we observe that
$Jv_{n}$ are eigenfunctions of the adjoint operator $L(\lambda)^*$ with
eigenvalue $-iE(\lambda)$
$$L(\lambda)^*Jv_{n}=-JL(\lambda)v_{n}=-iE(\lambda)Jv_{n}.$$
\end{enumerate}

Concerning the continuous spectrum of $L(\lambda)$, we apply Weyl's Theorem on the stability of the essential spectrum for localized perturbations of $J(-\Delta)$ 
\cite{HS96,RS3-4} to obtain
$$\sigma_{ess}(L(\lambda))=(-i\infty,-i\lambda]\cap
[i\lambda,i\infty)$$ if the potential $V$ in Equation
(~\ref{eq:NLS}) decays sufficiently rapidly as $|x|$ tends to infinity.
\bigskip

The end points of the essential spectrum are called threshold energies. 
\begin{definition} Let $d\ge3$.
A function $h$ is called a threshold resonance function of
$L(\lambda)$ at $\mu=\pm i\lambda$, the endpoint of the essential spectrum, if $h\not\in {L}^{2}$,
$|h(x)|\leq c\langle x\rangle^{-(d-2)_+}$ and $h$ is $C^{2}$ and
solves the equation
$$(L(\lambda)-\mu)h=0.$$
\end{definition}

\nit In this paper we make the following  spectral assumption on the thresholds $\pm i\lambda$:
\begin{enumerate}
\item[{{\bf Thresh}$_{\lambda}$}]:\  There exists $\delta'$, with $0<\delta'\le\delta_0$ (see Proposition \ref{LM:NearLinear1}), such that for $\lambda\in{\cal I}_{\delta'}$,\\  $L(\lambda)$ has no threshold resonances at $\pm i\lambda$
\end{enumerate}

In the weak amplitude limit, property ({\bf Thresh}$_\lambda$)
can be referred to the question of whether the scalar operator, $-\Delta+V(x)$ has a threshold (zero energy) resonance. In \cite{Jensen-Kato:79}  it was shown that $-\Delta+V$ has a zero energy resonance or eigenvector if and only if the operator $I+(-\Delta+i0)^{-1}V: \langle x\rangle^2  L^2\rightarrow \langle x\rangle^2  L^2$ is not invertible. Moreover, 
 this operator is generically invertible. That is, if we replace $V$ by $qV$, where $q$ is a real number, then we have non-invertibility for only a discrete set of $q$ values \cite{Rauch:78,Jensen-Kato:79}
 
 The reduction from the properties of $L(\lambda)$ to those of $-\Delta+V$ is seen as follows.  Let 
 \begin{equation}
 A:=\frac{1}{\sqrt{2}}\left(
\begin{array}{lll}
1&i\\
i&1
\end{array}
\right),\ \ \ A^*\ A\ =\ I
\label{Adef}
\end{equation}
Then
\begin{equation}\label{trans}
H := -iA^{*}L(\lambda)A,
\end{equation} 
It follows that $\pm i\lambda$ are threshold resonances of $L(\lambda)$ if and only if $\pm \lambda$ are threshold resonances of $H$. 

We next observe that $H$ is a small perturbation of $\sigma_3(-\Delta+V+\lambda)$, where $\sigma_3=\left(\begin{array}{lll}
1&0\\
0&-1
\end{array}
\right)$.
Indeed, a computation of  $H$ yields
\begin{equation}\label{defineH}
H=H_{0}+V_1+V_{small},
\end{equation}
where
\begin{equation}\label{h0w}
H_{0}:= (-\Delta+\lambda)\sigma_3
, \ V_1:=V\sigma_3
\end{equation} and for some $c>0$
\begin{equation}
|V_{small}|\leq e^{-c|x|}\ o(1),\nonumber
\end{equation}
where $o(1)\rightarrow 0$ as $|\lambda-|e_0||\rightarrow 0$.

Therefore, the generic validity of ({\bf Thresh}$_\lambda$) from the 
generic absence of zero energy threshold resonances for $-\Delta+V$ by the following result, proved for $d=3$ using results in ~\cite{CPV:05}. The proof for general dimensions is  similar.
\begin{proposition}\label{Prop:resonance}
Let $d=3$. If the operator $I+(-\Delta+i0)^{-1}V: \langle x\rangle^2 {L}^2\rightarrow \langle x\rangle^2 {L}^2$ is invertible, then ({\bf Thresh}$_{\lambda}$) holds when $|\lambda-|e_0||$ is sufficiently small. 
\end{proposition}
\begin{proof} We begin by proving that the operator $I+(H_0\pm \lambda+i0)^{-1}[V_1+V_{small}]: \langle x\rangle^2 {L}^2\rightarrow \langle x\rangle^2 {L}^2$ is invertible.
Observe that $-2\lambda\approx 2e_0$ is not an eigenvalue of the operator $-\Delta+V$, hence $I+(-\Delta+2\lambda)^{-1}V$ is invertible.  This, together with the hypothesis, implies that $I+(H_0\pm \lambda+i0)^{-1}V_1$ is invertible with a uniformly bounded inverse. On the other hand the norm of the operator $ (H_0\pm \lambda+i0)^{-1}V_{small}$ is small when $|e_0+\lambda|$ is small. Hence $I+(H_0\pm \lambda+i0)^{-1}[V_1+V_{small}]=[I+(H_0\pm \lambda+i0)^{-1}V_1][1+(1+(H_0\pm \lambda+i0)^{-1}V_1)^{-1}V_{small}]$ is invertible when $|\lambda-|e_{0}||$ is small.
Moreover in ~\cite{CPV:05} it is proved that  the operator $L(\lambda)$ has no threshold resonance functions if the operator $I+(H_0\pm \lambda+i0)^{-1}[V_1+V_{small}]: \langle x\rangle^2 {L}^2\rightarrow \langle x\rangle^2 {L}^2$ is invertible.  This completes the proof.
\end{proof}
\nit{\bf Choice of basis for  degenerate subspaces}\\
In our analysis, it is important that we choose an appropriate bases
of the degenerate eigenspaces corresponding to $\pm iE(\lambda)$.
We present this choice of basis and its construction here.

\begin{proposition}\label{prop:OrthonormalBasis}
There exist real functions $\xi_{n}$, $\eta_{n},\
n=1,2,\cdot\cdot\cdot, N,$ such that 
\begin{equation}
span\left\{\left(
\begin{array}{lll}
\xi_{n}\\
i\eta_{n}
\end{array}
\right)\right\}=span\{v_{1},v_{2},\cdot\cdot\cdot,v_{N}\},
\nn\end{equation}
 and for any $m,n\in \{1,2,\cdot\cdot\cdot, N\},$
\begin{equation}\label{eq:UnexpectedFact}
\int f^{'}[(\phi^{\lambda})^{2}](\phi^{\lambda})^{2}(\xi_{m}\eta_{n}-\xi_{n}\eta_{m})dx=0,
\end{equation}
\begin{equation}\label{eq:orthogonality}
\langle \phi^{\lambda},\xi_{n}\rangle=\langle \partial_{\lambda}\phi^{\lambda},\eta_{n}\rangle=0,\ \ \
\langle \xi_{n},\eta_{m}\rangle=\delta_{m,n}.
\end{equation}
\end{proposition}
The proof is given in the Appendix ~\ref{subsec:choice}. 

\begin{remark}
If $\phi^{\lambda}$ is spherically symmetric, then  $\xi_{n}=\frac{x_{n}}{|x|}\xi(|x|)$ and $\eta_{n}=\frac{x_{n}}{|x|}\eta(|x|)$, $n\in \{1,2,\cdot\cdot\cdot,N=d\}$; see Lemma ~\ref{mainLem2}). Therefore (~\ref{eq:UnexpectedFact}) trivially holds because $\xi_{m}\eta_{n}-\xi_{n}\eta_{m}=0.$
\end{remark}
We conclude this section with  the explicit form of the projection $P_{disc}$, whose proof for dimension one can be found in
~\cite{GS05}. The proof for general dimension is similar, and hence
omitted. Recall that $\langle \xi_{m},\eta_{n}\rangle=\delta_{m,n}$.
\begin{proposition}\label{Riesz-project}
For the non self-adjoint operator $L(\lambda)$ the (Riesz)
projection onto the discrete spectrum subspace of $L(\lambda)$, 
$P_{disc}=P_{disc}(L(\lambda))=P_{disc}^\lambda$, is given by
\begin{equation}\label{eq:PdProjection}
\begin{array}{lll}
P_{disc}&=&\frac{2}{\partial_{\lambda}\| \phi^{\lambda}\|^{2}}\left(\ \left|
\begin{array}{lll}
0\\
\phi^{\lambda}
\end{array}
\right\rangle \left\langle
\begin{array}{lll}
0\\
\partial_{\lambda}\phi^{\lambda}
\end{array}
\right|\ +\ \left|
\begin{array}{lll}
\partial_{\lambda}\phi^{\lambda}\\
0
\end{array}
\right\rangle \left\langle
\begin{array}{lll}
\phi^{\lambda}\\
0
\end{array}
\right|\ \right)\\
& &\\
& &-\frac{1}{2}i\displaystyle\sum_{n=1}^{N}\left(\ \left|
\begin{array}{lll}
\xi_{n}\\
i\eta_{n}
\end{array}
\right\rangle\left\langle
\begin{array}{lll}
-i\eta_{n}\\
\xi_{n}
\end{array}
\right| \ -\  \left|
\begin{array}{lll}
\xi_{n}\\
-i\eta_{n}
\end{array}
\right\rangle\left\langle
\begin{array}{lll}
i\eta_{n}\\
\xi_{n}
\end{array}
\right|\ \right).
\end{array}
\end{equation}
\end{proposition}
We define the projection onto the continuous spectral  subspace of $L(\lambda)$ by
\begin{equation} P_c\ =\ P_c(L(\lambda))\ =\ P_c^\lambda\ \equiv\ I\ -\ P_{disc}\label{Pcdef}.
\end{equation}

\subsection{Estimates on the Propagator}\label{Sec:Propagator} We will need estimates of the evolution operator
$U(t):=e^{tL(\lambda_{1})}$ for $\lambda_{1}\in \mathcal{I}$. Recall
that $L(\lambda_{1})$ has two branches of essential spectrum
$[i\lambda_{1},i\infty)$ and $(-i\infty,-i\lambda_{1}]$. We denote by
$P_{+}=P_{+}^{\lambda_{1}}$ and $P_{-}=P_{-}^{\lambda_{1}}$ the spectral projections associated with these two branches of the
essential spectrum. Hence,
$P_{c}^{\lambda_{1}}=P_{+}+P_{-}$. 

\begin{theorem}\label{prop:decayest}
Let  $d\geq
3$ and define $k:=[\frac{d}{2}]+1$ and
$\nu:=\frac{5+d}{2}$. Assume that $2E(\lambda_1)>\lambda$, so that $\pm 2iE(\lambda_1)\in\sigma_{ess}(L(\lambda_1))$.\\ Then, for any time $t\geq 0$ and $\lambda_1\in \mathcal{I}$ there exists a constant $c$ such that
\begin{align}\label{eq:SingularEst}
\|\langle x\rangle^{-\nu}(-\Delta+1)^{\frac{k}{2}}U(t)(L(\lambda_1)\pm
2iE(\lambda_1)-0)^{-n}P_{\pm}h\|_{2} \leq c(1+t)^{-\frac{d}{2}}    \|\langle x\rangle^{\nu}(-\Delta+1)^{\frac{k}{2}}h\|_{2}
\end{align} with $n=0,1,2$.\\
For any time $t\in (-\infty,\infty)$ and $\lambda_1\in \mathcal{I}$ there exists a constant $C_\mathcal{I}$ such that 
\begin{align}
&\|\langle x\rangle^{-\nu}(-\Delta+1)^{\frac{k}{2}} U(t)P_{\pm} h\|_{2}
\leq
C_\mathcal{I}\ (1+|t|)^{-\frac{d}{2}}\|\langle x\rangle^{\nu}(-\Delta+1)^{\frac{k}{2}}h\|_{2}\label{second}\\
&\|U(t)P_{\pm} h\|_{\infty}
\leq C_\mathcal{I}\ |t|^{-\frac{d}{2}}\|h\|_{1}\label{third}\\
&\|U(t)P_{\pm}h\|_{\infty}\leq
C_\mathcal{I}\ (1+|t|)^{-\frac{d}{2}}(\|h\|_{{H}^{k}}+\|h\|_{1})
\label{lastestimate}\\
&\|U(t)P_{\pm}h\|_{3} \leq
C_\mathcal{I}\ (1+|t|)^{-\frac{d}{6}}(\|h\|_{{H}^{k}}+\|h\|_{1})
\label{L3Est}\\
&\|\langle x\rangle^{-\nu}U(t)P_{\pm} h\|_{2}\leq
C_\mathcal{I}\ (1+|t|)^{-\frac{d}{2}}(\|h\|_{1}+\|h\|_{2}).
\label{finalestimate}
\end{align}
\end{theorem}
\nit We refer the proof of the estimates to
~\cite{SoWe99,GS3,TY02,GS04}. That the constant, $C_\mathcal{I}$ can be taken to hold uniformly for $\lambda_1\in \mathcal{I}$, see ~\cite{Cu2001, Cu03, Cu2005}.
%
%

\section{Matrix Fermi Golden Rule}\label{SEC:FGR}
As highlighted in the introduction,  the decay of neutral mode components, associated with the linearized NLS/GP equation, is necessary for asymptotic stability of the soliton manifold ${\cal M}_{\cal I}$. We shall prove that, after near-identity transformations, the system governing these neutral mode amplitudes is (\ref{new-nf}): 
\begin{equation}
\partial_{t}z =-iE(\lambda) z -\Gamma( z ,\bar{ z }) z +\Lambda( z ,\bar{ z })
z\ +\ \cO\left((1+t)^{-\frac{3}{2}-\delta}\right),\ \ \delta>0\label{new-nf1}
\end{equation}
where $\pm iE(\lambda)$ are complex conjugate $N-fold$ degenerate
neutral eigenfrequencies of $L(\lambda)=JH(\lambda)$, $\Gamma$ is symmetric and $\Lambda$ is skew symmetric. It follows that 
\begin{equation}
\partial_{t}\ |z(t)|^2\ =\ -2\ z^*\ \Gamma(z,\bar{z})\ z\ +\ \dots.
\label{energy-id}\end{equation}
Our strategy to show that $|z(t)|$ tends to zero, is based on proving that  $\Gamma$ is positive definite and that the corrections to (\ref{energy-id}) decay sufficiently rapidly as $t$ tends to infinity.
If  $L(\lambda)$ has a  complex conjugate pair of simple neutral eigenvalues,  then $\Gamma$ reduces to a non-negative scalar. If $L(\lambda)$ has multiple, well-separated pairs of neutral modes, then $\Gamma$ reduces to a diagonal matrix
 \cite{SoWe99,TY02,BuSu03,TsaiYau02,SoWe04,TsaiYau02-2,Tsai03}.
 The present case of problem of degenerate neutral modes is more
involved due to coupling among the various discrete modes and with the continuous spectrum. Our computation yields a non-diagonal FGR matrix, $\Gamma$. In this section, we  display the expression for  $\Gamma$, state a result on its general properties. The  detailed derivation of the expression for $\Gamma$  is carried out in section ~\ref{SEC:NormalForm}. 

\subsection{The FGR matrix, $\Gamma(z,\bar{z})$ }
To construct $\Gamma$ we must first introduce some notation.

Define vector functions $G_{k},\ k=1,2,\cdot\cdot\cdot, N$, as
\begin{equation}\label{eq:Fk2}
G_{k}(z,x):=\left(
\begin{array}{lll}
B(k)\\
D(k)
\end{array}
\right)
\end{equation} with the functions $B(k)$ and $D(k)$ defined as $$\begin{array}{lll}
B(k)&:=&-if^{'}\left[(\phi^{\lambda})^{2} \right]\phi^{\lambda}\ \ \left[\ \rhoo\ \eta_{k}+\omegaa\ \xi_{k}\ \right]\ , \\
D(k)&:=&-f^{'}[(\phi^{\lambda})^{2}]\phi^{\lambda}
\left[\ 3\rhoo\xi_{k}-\omegaa\eta_{k}\ \right]\
 -\ 2f^{''}\left[(\phi^{\lambda})^{2}\right](\phi^{\lambda})^{3}\ \rhoo\xi_{k}\ ,
\end{array}$$
where 
$$
z\cdot\xi\ :=\ \displaystyle\sum_{n=1}^{N}z_{n}\xi_{n},\ z\cdot\eta\  :=\ \displaystyle\sum_{n=1}^{N}z_{n}\eta_{n}.
  $$ 
In terms of the column 2-vector, $G_{k}$, we define
a $N \times N$ matrix $Z(z,\bar{z})$ as
\begin{equation}\label{eq:zMatrix}
Z(z,\bar{z})=(Z^{(k,l)}(z,\bar{z})),\ \ 1\le k,l\le N
\end{equation} and
\begin{equation}
Z^{(k,l)}(z,\bar{z})\ \equiv\ -\left\langle
(L(\lambda)+2iE(\lambda)-0)^{-1}P_{c}G_{l}, iJG_{k}\right\rangle
\label{Zkl}\end{equation}
Since 
  $P_c(L)^*J\ =\ J\ P_c(L)$, a consequence of $L=JH$ and $ H^*=H$ (see Proposition \ref{projection-prop} ), we have
the more symmetric expression for $Z^{(k,l)}$:
\begin{equation}
Z^{(k,l)}\ =\ -\left\langle
(L(\lambda)+2iE(\lambda)-0)^{-1}P_{c}G_{l}, iJP_cG_{k}\right\rangle
\label{Zkl-sym}
\end{equation}
Finally, we define $\Gamma(z,\bar{z})$ as follows:
\begin{equation}
\Gamma(z,\bar{z})\ :=\ \frac{1}{2}[Z(z,\bar{z})+Z^{*}(z,\bar{z})].
\label{Gammadef}
\end{equation}
Thus,
 \begin{equation}
[\ \Gamma(z,\bar{z})\ ]_{kl}\ =\ -\ Re\ \left\langle
(L(\lambda)+2iE(\lambda)-0)^{-1}P_{c}G_{l}, iJP_cG_{k}\right\rangle
\label{Gamma-kl}
\end{equation}
\\ \\
Concerning the properties of $\Gamma$, we have the following general result:
\begin{theorem}\label{Gamma-general} (Matrix nonlinear Fermi Golden Rule)
\begin{itemize}
\item[(1)]\ $\Gamma(z,\bar{z})=\Gamma(z,\bar{z};\lambda)$ is a non-negative symmetric $N \times N$ matrix, displayed in (\ref{Gamma-kl}).
\item[(2)]\ Define  $$K(\lambda,\vec{G}):=\displaystyle\min_{s,z\not=0}\frac{s^*\ \Gamma(z,\bar{z})\ s}{|s|^2|z|^2}.$$
Then, $K(\lambda,\vec{G})$ depends continuously on $\lambda$ and $\vec{G}$ (in the space $\langle x\rangle^3 L^{\infty}$), where $\vec{G}=(G_1,\dots,G_N)$, defined in (\ref{eq:Fk2}).
\end{itemize}
\end{theorem}

We shall require the following {\bf Fermi Golden Rule resonance condition}:\\ \\

\nit{\bf (FGR)}
There exists $\delta'$, with $0<\delta'\le\delta$ (see Proposition \ref{LM:NearLinear1}) and 
a constant $C>0$, such 
that for any $s=(s_{1},\cdot\cdot\cdot,s_{N})^{T},\
z=(z_{1},\cdot\cdot\cdot,z_{N})^{T}\in \mathbb{C}^{N}$,
\begin{equation}
s^*\ \Gamma(z,\bar{z};\lambda)\ s\geq\ C\ |s|^{2}|z|^{2},\label{eq:FGR}
\end{equation}
where $\lambda\in{\cal I}_{\delta'}$.

\begin{remark}
In the weakly nonlinear regime, see section \ref{sec:weaklynonlin},
 $E(\lambda)\sim e_1-e_0$, $\lambda\sim -e_0$ and therefore the condition for resonance with the continuous spectrum at second order is:
  $2E(\lambda)-\lambda\sim 2(e_1-e_0)+e_0=2e_1-e_0>0$.
  \end{remark}
  Our next result is a reduction of the condition  {\bf (FGR)}, for the  class of double-well potentials discussed in appendix \ref{subsec:example}, to an explicit condition on the operator $V$.
  
\begin{proposition}\label{prop:example-fgr}
Let $V$ denote the double-well potential satisfying condition {\bf ($Eig_V$)}, and constructed in Appendix ~\ref{subsec:example}.
Thus, $-\Delta+V$ has two negative eigenvalues, $e_0<e_1<0$, with 
 $2e_1-e_0>0$. The excited state eigenvalue, $e_2$, is degenerate of multiplicity $N=2$, with spanning eigenfunctions $\{\xi_1^{lin},\xi_2^{lin}\}$.\\
  Let $f(|\psi|^2)=-g|\psi|^2$. 
  Assume the matrix 
  \begin{equation}
\left(\ Im \left\langle (-\Delta+V-[2e_1-e_0]-i0)^{-1}P_{c}
\ \phi_{lin}\ \xi_m^{lin}\xi_n^{lin}, \phi_{lin}\ \xi_m^{lin}\xi_n^{lin}\right\rangle\ \right)_{1\le m,n\le2}\label{pdef}
 \end{equation}
 is positive definite.
Then, there exists $\delta'>0$, such that for the  $\phi^\lambda$ denotes the soliton of Proposition \ref{LM:NearLinear1},  if
 $|\ \lambda-|e_0|\ |<\delta'$ then $K(\lambda,\vec{G})>0$.
  and  the Fermi Golden Rule condition {\bf (FGR)}, holds by taking  
 $C=\inf_{\lambda\in\overline{\mathcal{I}_{\delta'}}}K(\lambda,\vec{G(\lambda)})>0$ in (\ref{eq:FGR}). Here, $\mathcal{I}_{\delta'}$ denotes a sufficiently small subinterval of the range of $\lambda-$ values for which the soliton exists; see Proposition \ref{LM:NearLinear1}.
 \end{proposition}
 \begin{remark}
Positive definiteness of  the matrix (\ref{pdef}) is equivalent to 
  \begin{equation}
 Im \left\langle (-\Delta+V-[2e_1-e_0]-i0)^{-1}P_{c}\ \phi_{lin}\ (z_1 \xi_{1}^{lin}+z_2\xi_2^{lin})^2, \phi_{lin}\ (z_1\xi_1^{lin}+ z_2\xi_2^{lin})^2\right\rangle> C|z|^2,\nn
 \end{equation}
 for all $z_1,z_2\in\C^1$.
 \end{remark}

\nit{\bf Proof of Proposition \ref{prop:example-fgr}:}
In what follows we sketch the proof, which is very similar to the case $N=1$, see ~\cite{SoWe99, TsaiYau02}). \\ 
Recall the transformation of $L(\lambda)$ in equation~(\ref{trans}):
\begin{equation}
\begin{array}{lll}
(L(\lambda)+2iE(\lambda)-0)^{-1}&=& [iA HA^* +2iE(\lambda)-0]^{-1}\\
&=&-iA [H+2E(\lambda)+i0]^{-1}A^*\\
&=&-iA [H_0+2E(\lambda)+i0]^{-1}A^*+iA [H+2E(\lambda)+i0]^{-1}V_{small} [H_0+2E(\lambda)+i0]^{-1}A^*
\end{array}
\end{equation} 
and 
\begin{equation}
[H_0+2E(\lambda)+i0]^{-1}=\left(
\begin{array}{lll}
[\ -\Delta+V-\ (-\lambda-2E(\lambda))\ ]^{-1}&0\\
0&- [-\Delta-\ (2E(\lambda)-\lambda)\ -i0]^{-1}
\end{array}
\right). \label{matrixx}
\end{equation}
 On the other hand
  by Propositions \ref{LM:NearLinear1} and ~\ref{Prop:NeuMode} we have that in $ H^{2}$ space $\frac{1}{\|\phi^{\lambda}\|_{{H}^2}}\phi^{\lambda}\rightarrow \frac{1}{\|\phi_{lin}\|_{ H^{2}}}\phi_{lin}$
and $(\frac{1}{\|\xi_{n}\|_{ H^{2}}}\xi_{n},\ \frac{1}{\|\eta_{n}\|_{ H^{2}}}\eta_{n})\rightarrow\frac{1}{\|\xi^{lin}_n\|_{ H^{2}}} (\xi^{lin}_n,\xi^{lin}_n)$ for some $\xi^{lin}_{n}$ as $|\lambda-|e_{0}||\rightarrow 0$. If the nonlinearity $f(\tau)=\tau^{\sigma},\ \sigma\geq 1$ we have $$A^{*}P_c \sum_{l}z_l G_{l}=C\|\phi_{lin}\|_{{H}^2}^{2\sigma-1}\left(
\begin{array}{lll}
*\\
P_{c} \phi_{lin}^{2\sigma-1}(z_1\xi_1^{lin}+z_2\xi_2^{lin})^2
\end{array}
\right)(1+o(1)) $$ for some constant $C\in \mathbb{C}.$ 

In considering (\ref{matrixx}), note that $-\lambda-2E(\lambda)\sim e_0-2(e_1-e_0)<0$ and 
 $2E(\lambda)-\lambda\sim 2e_1-e_0<0$.
Thus, $Im\langle (-\Delta+V+\lambda+2E(\lambda))^{-1}F, F\rangle=0$ for any $F$. Furthermore, $\|e^{-c|x|}V_{small}\|_{{L}^\infty}$ is small for some $c>0$, we have
\begin{equation}
K(\lambda,\vec{G})=|C|^2 \|\phi_{lin}\|_{{H}^2}^{4\sigma-2} K_{0}(1+o(1))
\label{Keqn}
\end{equation}
with $$K_{0}:=
Im \left\langle (-\Delta+V+e_0-2e_{1}-i0)^{-1}P_{c}(\phi_{lin})^{2\sigma-1}(z_1 \xi_{1}^{lin}+z_2\xi_2^{lin})^2, (\phi_{lin})^{2\sigma-1}(z_1\xi_1^{lin}+ z_2\xi_2^{lin})^2\right\rangle(1+o(1)).$$
The proof is complete.
In the appendix ~\ref{subsec:FGR} we have a simpler formula for (FGR) when the potential $V$ is spherical symmetric.

 The proof of Theorems \ref{Gamma-general}   is deferred to Appendix ~\ref{subsec:Gamma-general}.

\section{Main Theorem}\label{MainTHM}
In this section we state precisely the main theorem of this paper. 
Recall the notations $\xi=(\xi_1,\cdots,\xi_{N})$ and $\eta=(\eta_1,\dots,\eta_{N})$ for components of the neutrally stable modes of frequencies $\pm iE(\lambda)$ of the linearized operator. Recall the definition of the interval $\mathcal{I}$ in ~(\ref{eq:stab1}).

\begin{theorem}\label{THM:MainTheorem}
Assume conditions:\\
{\bf  (fA) (fB)} on the nonlinearity $f(|\psi|^2)$;\ section \ref{HaGWP}.\\
{\bf (VA)} on the potential $V(x)$;\ section \ref{HaGWP}.\\
{\bf  (SA)} on the structure of the discrete spectral subspace of the linearization about $\phi^\lambda$; section \ref{SEC:Operator}.\\
{\bf ({{Thresh}$_{\lambda}$})} on the absence of threshold resonances;\ section \ref{sec:weaklynonlin}.  \\
{\bf (FGR)}, the nonlinear Fermi Golden Rule resonance condition, (\ref{eq:FGR}); section \ref{SEC:FGR}.\\ \\
 Fix  $\nu>0$  sufficiently large and
let $k=[\frac{d}{2}]+1$, where $d\ge3$ denotes the spatial dimension.
\\ Then there exist constants
$c,\epsilon_{0}>0$ such that, if for some $\lambda_{0}\in \mathcal{I}$
\begin{equation}\label{InitCond}
\inf_{\gamma\in
\mathbb{R}}
\left\|\psi_0-e^{i\gamma}
\left(\phi^{\lambda_{0}}+\ (Re\ z^{(0)})\cdot \xi+i\ (Im\ z^{(0)})\cdot\eta
\right)\right\|_{{H}^{k,\nu}}\ \leq\ 
c\ | z^{(0)}| \le \epsilon_0,
\end{equation} 
  then there exist smooth functions 
  \begin{align}
  &\lambda(t):\mathbb{R}^{+}\rightarrow
\mathcal{I},\ \ \ \gamma(t): \mathbb{R}^{+}\rightarrow
\mathbb{R},\  z(t):\mathbb{R}^{+}\rightarrow \mathbb{C}^{N},\nn\\
 &\ \  R(x,t):\mathbb{R}^{d}\times\mathbb{R}^{+}\rightarrow
\mathbb{C}
\nn\end{align}
 such that the solution of NLS evolves in the form:
\begin{align}\label{Decom}
\psi(x,t)&\ =\ e^{i\int_{0}^{t}\lambda(s)ds}e^{i\gamma(t)}\nn\\
&\ \ \ \ \ \ \ \times[\ \ \phi^{\lambda}+a_{1}(z,\bar{z})
\D_\lambda\phi^{\lambda}+ia_{2}(z,\bar{z})\phi^{\lambda}\ +\ (Re\ \tilde{z})\cdot\xi\ +\ i(Im \tilde{z})\cdot\eta\ +\ R\ \ ],\end{align}
where  $\lim_{t\rightarrow \infty}\lambda(t)=\lambda_{\infty},$
for some $\lambda_{\infty}\in \mathcal{I}$.\\ 
Here, $a_{1}(z,\bar{z}),\ a_{2}(z,\bar{z}): \mathbb{C}^{N}\times\mathbb{C}^{N}\rightarrow \mathbb{R}$ and $\tilde{z}-z: \mathbb{C}^{N}\times\mathbb{C}^{N}\rightarrow \mathbb{C}^{N}$
 are some polynomials of $z$ and $\bar{z}$, beginning with terms of order $|z|^{2}$. \\
Moreover:
\begin{enumerate}
\item[(A)] $|z(t)|\leq c(1+t)^{  -\frac{1}{2}  }$ and
$z$ satisfies the initial value problem 
\begin{equation}\label{eq:detailedDescription1}
{\partial_t} z =-iE(\lambda) z -\Gamma( z ,\bar{ z }) z +\Lambda( z ,\bar{ z })
z\ +\ \cO((1+t)^{-\frac{19}{5}})
\end{equation} where $\Gamma( z ,\bar{ z })$ is a
symmetric and positive definite  matrix defined in (~\ref{eq:zMatrix}), $\Lambda( z ,\bar{ z })$ is a skew
symmetric matrix.
 \item[(B)] $\vec{R}(t)=( Re\ R(t)\ ,\ Im\ R(t))^T$ lies in the essential spectral part of $L(\lambda(t))$. Equivalently, $R(\cdot,t)$ satisfies the symplectic orthogonality conditions:
 \begin{align}\label{s-Rorthogonal}
&\omega\langle R,i\phi^{\lambda}\rangle\ =\ \omega\langle
R,\partial_{\lambda}\phi^{\lambda}\rangle\ =\ 0, \nn\\
&\omega\langle
R,i\eta_{n}\rangle=\omega\langle R,\xi_{n}\rangle=0,\
n=1,2,\cdot\cdot\cdot, N,
\end{align} 
 where $\omega\langle X,Y\rangle=Im\int X\overline{Y}$
 and satisfies the decay estimate:
 \begin{equation}
\|(1+x^{2})^{-\nu}\vec{R}(t)\|_{2}\leq c(1+t)^{-1}.
\label{Rdecay}
\end{equation}
\end{enumerate}
\end{theorem}

\begin{remark}
We conclude this section by stating that all the hypothesis except (FGR) in our main result applies to the double-well example of appendix ~\ref{subsec:example}; see Proposition \ref{prop:example-fgr} for a reduction of {\bf (FGR)} is to an explicit condition on the spectral condition on $-\Delta +V$.  We expect {\bf (FGR)} to hold generically in an appropriate sense.
\end{remark}

\section{Reformulation of The Main Theorem}\label{SEC:ReformMain}
In proving Theorem ~\ref{THM:MainTheorem} we  establish more detailed characterization of the perturbation about $\mathcal{M}_{\mathcal{I}}$.\\ \\
First, we introduce the following simplifying notation:
    \textbf{we always use $z$
to stand for a complex $N$-dimensional vector $z=(z_{1},z_{2},\cdot\cdot\cdot,z_{N})$ and an upper case letter or a
Greek letter with two subindices, for example $Q_{m,n}$, to stand
for
$$Q_{a,b}(\lambda)=\sum_{
 |a|=m,\ |b|=n
}q_{a,b}(\lambda)\ 
 \prod_{k=1}^{N}z_{k}^{a_{k}}\bar{z}_{k}^{b_{k}},$$ where $a,b\in \mathbb{N}^N$, $|a|:=\displaystyle\sum_{k=1}^{N}a_{k}$.
 We refer to this kind term as $(m,n)$ term.}
\begin{theorem}\label{GOLD:maintheorem}
The following 
more precise decomposition of the solution in Theorem ~\ref{THM:MainTheorem}  holds.  The perturbation $R$ in (~\ref{Decom})   can be decomposed as
\begin{equation}\label{eq:expanR}
\vec{R}=\displaystyle\sum_{ m+n=2}R_{m, n}(\lambda)+\tilde{R}
\end{equation} where $R_{m,n}$ are functions of the form
$$R_{m,n}=[L(\lambda)+iE(\lambda)(m-n)-0]^{-1}\phi_{m,n}$$
$\phi_{m,n}$ are polynomials of $z$ and $\bar{z}$ with coefficients
being smooth, exponentially decaying functions. The function
$\tilde{R}$ satisfies the equation
\begin{equation}\label{eq:tildeR}
\begin{array}{lll}
\D_t\tilde{R}&=&L(\lambda)\tilde{R}+M_{2}(z,\bar{z})\tilde{R}+P_{c}N_{2}(\vec{R},z)+P_{c}S_{2}(z,\bar{z}),
\end{array}
\end{equation} where
\begin{enumerate}
\item[(1)] $S_{2}(z,\bar{z})=O(|z|^{3})$ is a polynomial in $z$ and
 $\bar{z}$ with $\lambda$-dependent coefficients, and each coefficient can be written as the sum of
 functions of the form
 \begin{equation}\label{eq:unusual}
(L(\lambda)+2iE(\lambda)-0)^{-k}P_{c}\phi_{+k}(\lambda) \
\text{or}\ (L(\lambda)-2iE(\lambda)-0)^{-k}P_{c}\phi_{-k}(\lambda),
\end{equation}
where $k=0,1,2$ and the functions $\phi_{\pm k}(\lambda)$ are smooth
and decay exponentially fast at $\infty$;
 \item[(2)] $M_{2}(z,\bar{z})$ is an operator defined by
 \begin{equation}\label{eq:M2}
 M_{2}(z,\bar{z}):=\dot{\gamma}P_{c}J+\dot{\lambda}P_{c\lambda}+X,
 \end{equation}
 where $X$ is a $2\times 2$ matrix, satisfying the bound
   $$|X|\leq c|z|e^{-\epsilon_{0}|x|}.$$
 \item[(3)] $N_{2}(\vec{R},z)$ can be separated into
 localized term and nonlocal term
 \begin{equation}\label{eq:N2Decomposition}
 N_{2}=Loc+NonLoc
 \end{equation} where $Loc$ consists of terms spatially
 localized (exponentially) function of $x\in \mathbb{R}^{d}$ as a factor and satisfies the estimate
 \begin{equation}
 \|\langle x\rangle^{\nu}(-\Delta+1)Loc\|_{2}+\|Loc\|_{1}+
 \|Loc\|_{\frac{4}{3}}\leq c\ \left(\ |z(t)|^{3}+|z(t)|\ \|\langle x\rangle^{-\nu}(-\Delta+1)\vec{R}\|_{2}\ \right),
 \label{eq:Loc-est}\end{equation}
  and $NonLoc$ is given by
\begin{equation}\label{eq:NonlocDef}
NonLoc:=f(R_{1}^{2}+R_{2}^{2})J\vec{R},
\end{equation}
and consists of purely nonlinear terms in $\vec{R}$, with no spatially
localized factors. Here $\nu$ is the same as in Theorem
~\ref{THM:MainTheorem}.

\item[(4)] Denote by $Remainder(t)$ any quantity which satisfies the estimate:
\begin{equation}\label{remainder}
\begin{array}{lll}
|Remainder(t)|&\lesssim& |z(t)|^{4}+\|\langle
x\rangle^{-\nu}(-\Delta+1)\vec{R}(t)\|_{2}^{2}+\|\vec{R}(t)\|_{\infty}^{2}
+|z(t)|\ \ \|\langle
x\rangle^{-\nu}\tilde{R}(t)\|_{2}
\end{array}
\end{equation} 
The functions $\lambda$, $\gamma,$ $z$ have the following properties
\begin{enumerate}
\item[(A)]
\begin{equation}\label{ExpanLambda}
\dot\lambda=Remainder(t);
\end{equation}
\item[(B)]
\begin{equation}\label{ExpanGamma}
\dot\gamma=\Upsilon_{1,1}+Remainder(t)
\end{equation} with
\begin{equation}\label{eq:Gamma11}
\Upsilon_{1,1}:=
\frac{\left\langle
\phi^{\lambda}
\left[\frac{3}{2}f^{'}[(\phi^{\lambda})^{2}]+f^{''}[(\phi^\lambda)^{2}]
 (\phi^\lambda)^2 \right] |z\cdot\xi|^{2}
  +\frac{1}{2}f^{'}[(\phi^{\lambda})^2]
|z\cdot\eta|^{2} , \partial_{\lambda}\phi^{\lambda}\right\rangle}
{\langle\phi^{\lambda},\partial_{\lambda}\phi^{\lambda}\rangle}
,
\end{equation} 
\item[(C)] the
vector $z$ satisfies the equation
\begin{equation}\label{eq:detailedDescription}
{\partial_t}z=-iE(\lambda)z-\Gamma(z,\bar{z})z+\Lambda(z,\bar{z})z+Remainder(t)
\end{equation} where $\Gamma(z,\bar{z})$ is the $N \times N$ positive definite
matrix defined in (~\ref{Gammadef}), $\Lambda(z,\bar{z})$ is skew
symmetric.\end{enumerate}
\end{enumerate}
\end{theorem}
\section{The Effective
Equations for $\dot{z},\ \dot\lambda,\ \dot\gamma$ and $
R$}\label{SEC:effective}
In this section we derive equations for $\dot{z},\ \dot\lambda,\ \dot\gamma$ and $R$.

We decompose the solution as 
\begin{align}
\psi(x,t)&\ =\ e^{i\int_{0}^{t}\lambda(s)ds}e^{i\gamma(t)}\nn\\
&\ \ \ \ \ \times\left[\phi^{\lambda}+a_{1}
\phi_{\lambda}^{\lambda}+ia_{2}\phi^{\lambda}+\sum_{n=1}^{N}(\alpha_{n}+p_{n})\xi_{n}+
i\sum_{n=1}^{N}(\beta_{n}+q_{n})\eta_{n}+R\right]\nn\\
&\ =\ e^{i\int_{0}^{t}\lambda(s)ds}e^{i\gamma(t)}\left[\phi^{\lambda}+a_{1}
\phi_{\lambda}^{\lambda}+ia_{2}\phi^{\lambda}+(\alpha+p)\cdot\xi+
(\beta+q)\cdot\eta+R\right]
\label{Decom1}
\end{align} 
Here and going forward, we use the notations:
\begin{align}
\alpha=(\alpha_1,\dots,\alpha_N)^T,\ \ \beta=(\beta_1,\dots,\beta_N)^T,\nn\\
\xi=(\xi_1,\dots,\xi_{N})^T,\ \ \eta=(\eta_1,\dots,\eta_{N})^T.\nn
\end{align}
Introducing 
\begin{equation} z=\alpha+i\beta,\nn\end{equation}
we have 
\begin{equation}
\alpha=\frac{1}{2}(z+\overline{z}),\ \ \beta=\frac{1}{2i}(z-\overline{z}),
\nn\end{equation}
and we seek $a_j=a_j(z,\overline{z})=\cO(|z|)^2,\ j=1,2$ and $p_n=p_n(z,\overline{z})=\cO(|z|^2)$, $q_n=q_n(z,\overline{z})=\cO(|z|^2)$, polynomials in $z$ and $\overline{z}$, which are of degree higher than or equal to two.
Substitution of the Ansatz (~\ref{Decom1}) into NLS,
equation (~\ref{eq:NLS}), we have the following system of equations
for  $\vec{R}$ with 
\begin{equation}
\vec{R}\equiv \left(
\begin{array}{lll}
R_{1}\\
R_{2}
\end{array}
\right),\ R_{1}\equiv Re R,\ \ R_{2}\equiv Im R:
\nn\end{equation}
\begin{equation}\label{Eq:R}
\begin{array}{lll}
\D_t\vec{R}&=&L(\lambda)\vec{R}+\dot\gamma
J\vec{R}-J\vec{N}(\Vec{R},z)
-\left(
\begin{array}{lll}
\partial_{\lambda}\phi^{\lambda}[\dot\lambda+\partial_{t}a_{1}]\\
\phi^{\lambda} [ \dot{\gamma}+\partial_{t} a_{2}-a_{1} ]
\end{array}
\right)\\
& & +\ \left(
\begin{array}{lll}
\xi\cdot[E(\lambda)(\beta+q)-\partial_{t}(\alpha+p)]\\
-\eta\cdot[E(\lambda)(\alpha+p)+\partial_{t}(\beta+q)]
\end{array}
\right)\\
& &+\dot\gamma\ \left(
\begin{array}{lll}
(\beta+q)\cdot\eta\\
- (\alpha+p)\cdot\xi
\end{array}\right)
-\dot\lambda \left(
\begin{array}{lll}
a_{1}\partial_{\lambda}^{2}\phi^{\lambda}+
(\alpha+p)\cdot\partial_{\lambda}\xi\\
 a_{2}\partial_{\lambda}\phi^{\lambda}+
(\beta+q)\cdot\partial_{\lambda}\eta
\end{array}
\right),
\end{array}
\end{equation}
Here, 
\begin{align}
\vec{N} &=  \left(
\begin{array}{lll}
Re\ N(\Vec{R},z)\\
Im\ N(\Vec{R},z)
\end{array}
\right),\ \ \  J\vec{N}(\Vec{R},z)\ =\ \left(
\begin{array}{lll}
Im\ N(\Vec{R},z)\\
-Re\ N(\Vec{R},z)
\end{array}
\right)\label{JvecN}\\
ImN(\Vec{R},z)&:=
f(|\phi^{\lambda}+I_{1}+iI_{2}|^{2})I_{2}-f[(\phi^{\lambda})^{2}]I_{2}\nn\\
Re N(\Vec{R},z)&:=
[f(|\phi^{\lambda}+I_{1}+iI_{2}|^{2})-f[(\phi^{\lambda})^{2}]](\phi^{\lambda}+I_{1})
-2f^{'}[(\phi^{\lambda})^{2}](\phi^{\lambda})^{2}I_{1}\nn\\
I_{1}\ &=\ A_{1}+A_{2}+R_{1},\ \ \ \ 
I_{2}\ =\ B_{1}\ +\ B_{2}\ +\ R_{2}\nn\\
A_{1}\ &=\ \alpha\cdot\xi, \ \ \ \ \ A_{2}\ =\ a_{1}\partial_{\lambda}\phi^{\lambda}+
p\cdot\xi,\nn\\
 B_{1}\ &=\ \beta\cdot \eta,\ \ \ \ \ 
B_{2}\ =\ a_{2}\phi^{\lambda}\ +\ q\cdot\eta.
\nn\end{align}

From the system of equations (~\ref{Eq:R}) and the orthogonality conditions
(~\ref{eq:orthogonality}) we obtain equations for $
\dot\lambda,\ \dot\gamma$ and $z_{n}=\alpha_n+i\beta_n,\ n=1,\dots, N: $
\begin{equation}\label{eq:z1}
\partial_{t}(\alpha_{n}+p_{n})-E(\lambda)(\beta_{n}+q_{n})+\langle ImN(\Vec{R},z),\eta_{n}\rangle=F_{1n};
\end{equation}
\begin{equation}\label{eq:z2}
\partial_{t}(\beta_{n}+q_{n})+E(\lambda)(\alpha_{n}+p_{n})-\langle ReN(\Vec{R},z),\xi_{n}\rangle=F_{2n};
\end{equation}
\begin{equation}\label{eq:gamma}
\dot\gamma+\partial_{t}a_{2}-a_{1}-\frac{1}{\langle
\phi^{\lambda},\D_{\lambda}\phi^{\lambda}\rangle}\langle
ReN(\Vec{R},z),\partial_{\lambda}\phi^{\lambda}\rangle=F_{3};
\end{equation}
\begin{equation}\label{eq:lambda}
\dot\lambda+\partial_{t}a_{1}+\frac{1}{\langle
\phi^{\lambda},\D_{\lambda}\phi^{\lambda}\rangle}\langle
ImN(\Vec{R},z),\phi^{\lambda}\rangle=F_{4}.
\end{equation}
Finally, the scalar functions $F_{j,n},\ j=1,2, F_3,
 F_4,$ are defined as
\begin{align}
F_{1n}\ &=\ \dot\gamma\langle
(\beta+q)\cdot\eta,\eta_{n}\rangle-\dot{\lambda}a_{1}\langle
\partial_{\lambda}^{2}\phi^{\lambda},\eta_{n}\rangle
-\dot\lambda\langle
(\alpha+p)\cdot\partial_{\lambda}\xi,\eta_{n}\rangle
-\dot\gamma\langle
R_{2},\eta_{n}\rangle+\dot\lambda\langle
R_{1},\partial_{\lambda}\eta_{n}\rangle,\nn\\
F_{2n}\ &=\ -\dot\gamma\langle
(\alpha+p)\cdot\xi,\xi_{n}\rangle
-\dot{\lambda}a_{2}\langle\phi^{\lambda}_{\lambda},
\xi_{n}\rangle
-\dot\lambda\langle(\beta+q)\cdot\partial_{\lambda}\eta,\xi_{n}\rangle+\dot\gamma\langle
R_{1},\xi_{n}\rangle+\dot\lambda\langle
R_{2},\partial_{\lambda}\xi_{n}\rangle, \nn\\
F_{3}\ &=\ \frac{1}{\langle
\phi^{\lambda},\phi^{\lambda}_{\lambda}\rangle}
\left[\ \dot\lambda \langle
R_{2},\phi_{\lambda\lambda}^{\lambda}\rangle -\dot\gamma\langle
R_{1},\phi_{\lambda}^{\lambda}\rangle
-\langle\dot\gamma(\alpha+p)\cdot\xi+
\dot{\lambda}a_{2}\phi^{\lambda}_{\lambda}
+\dot\lambda (\beta+q)\cdot\partial_{\lambda}\eta,\phi^{\lambda}_{\lambda}\rangle\ \right], \nn\\
F_{4}\ &=\ \frac{1}{\langle
\phi^{\lambda},\phi^{\lambda}_{\lambda}\rangle}
\left[\ \dot\lambda\langle
R_{1},\phi_{\lambda}^{\lambda}\rangle +\dot\gamma\langle
R_{2},\phi^{\lambda}\rangle
+\langle
\dot\gamma (\beta+q)\cdot \eta-\dot{\lambda}a_{1}
\partial_{\lambda}^{2}\phi^{\lambda}-\dot\lambda
(\alpha+p)\cdot\partial_{\lambda}\xi,\phi^{\lambda}\ \rangle\ \right].
\end{align}
\begin{remark}
\begin{itemize}
\item[(a)] Recall the estimate of $Remainder$ in (~\ref{remainder}). By Equations (~\ref{eq:z1})-(~\ref{eq:lambda}) we have
\begin{equation}\label{eq:EstRough}
\dot\lambda,\ \dot\gamma,\
\partial_{t}z_{n}+iE(\lambda)z_{n}=O(|z|^{2})+Remainder.
\end{equation}
\item[(b)] The functions $a_j(z,\overline{z}), j=1,2,\ p_n(z,\overline{z}),\ 
 q_n(z,\overline{z}), n=1,\dots, N$ will be chosen to eliminate ``non-resonant'' terms: $z^m\overline{z}^n, \ \ 2\le |m|+|n|\le3$.
 \end{itemize}
\end{remark}
Finally, we derive an equation for 
\begin{equation}
\vec{R}=P_c^{\lambda(t)}\vec{R}=P_c\vec{R},
\nn\end{equation}
the continuous spectral part of the solution, relative to the operator, $L(\lambda(t))$. Applying $P_c=P_c^{\lambda(t)}$ to (~\ref{Eq:R}) and using  the commutator identity:
\begin{equation}
P_{c}\D_t\vec{R}=\D_t\vec{R}-\dot\lambda
\D_\lambda P_{c}\vec{R}
\label{Pccomdt}\end{equation}
we obtain
\begin{equation}\label{RAfProj}
\D_t\vec{R}\ =\ 
 L( \lambda(t) )\vec{R}\ -\ P_{c}^{\lambda(t)}J\vec{N}(\Vec{R},z)\ +\ L_{(\dot\lambda,\dot\gamma)}\vec{R}\ +\ \mathcal{G}.
\end{equation}
The operator $L_{(\dot\lambda,\dot\gamma)}$ and the 
 vector function $\mathcal{G}$ are defined 
as 
\begin{align}
& L_{(\dot\lambda,\dot\gamma)}\ =\ \dot\lambda
(\D_\lambda P_{c}^{\lambda(t)})+\dot\gamma P_{c}^{\lambda(t)}J,\label{Lgdld}\\
& \mathcal{G}\ =\ P_{c}^{\lambda(t)}\left(
\begin{array}{lll}
\dot\gamma (\beta+q)\cdot\eta-\dot{\lambda}a_{1}
\partial_{\lambda}^{2}\phi^{\lambda}
-\dot\lambda (\alpha+p)\cdot\partial_{\lambda}\xi\\
-\dot\gamma (\alpha+p)\cdot\xi-\dot{\lambda}a_{2}\phi^{\lambda}_{\lambda}-\dot\lambda
(\beta+q)\cdot\partial_{\lambda}\eta
\end{array}
\right).\label{Gdef}
\end{align}
We now summarize the preceding calculation in the following:
\begin{proposition} {\bf (Reformulation of NLS)}\\
Using the Ansatz (\ref{Decom1})
\begin{equation} 
\psi(x,t)\ =\ e^{i\int_{0}^{t}\lambda(s)ds}e^{i\gamma(t)}\left[\phi^{\lambda}+a_{1}
\phi_{\lambda}^{\lambda}+ia_{2}\phi^{\lambda}+(\alpha+p)\cdot\xi+
(\beta+q)\cdot\eta+R\right]
\label{Decom2}
\end{equation} 
 NLS can be equivalently expressed as a coupled system of equations (\ref{eq:z1},\ref{eq:z2},\ref{eq:gamma},\ref{eq:lambda}) for modulating solitary wave parameters:
 $\lambda(t),\ \gamma(t)$, neutral mode amplitudes $z_n(t)=\alpha_n(t)+i\beta_n(t),\ n=1,\dots, N$ and an  equation (\ref{RAfProj})
,  governing ``dispersive part'', $\vec{R}(t)$, which evolves in the continuous spectral subspace of 
  $L(\lambda(t))$,
  {\it i.e.} $P_c^{\lambda(t)}\vec{R}(t)=\vec{R}(t)$; see (\ref{Pcdef}).\\
  Finally, the functions $a_j=a_j(z,\overline{z}), j=1,2,\  p(z,\overline{z}),q(z,\overline{z})\ =\ (p_n,q_n)_{n=1,\dots, N}$ are $\cO(|z|^2)$ and are
  polynomials chosen (in what follows) to eliminate ``non-resonant'' 
  terms of the form $z^a\overline{z}^b,\ \ 2\le |a|+|b|\le3$.
\end{proposition}
\subsection{Extracting the $\cO(|z|^2)$ part of $\vec{R}(t)$; Proof of Equation (~\ref{eq:tildeR})}
For fixed $z(t)\in\C^{N}$, the equation for $\vec{R}(t)$ is forced by terms
of order $\cO(|z(t)|^2)$; linear terms are removed due to the equations satisfied by $z(t)=\alpha(t)+i\beta(t)$.
  In our analysis, we need to explicitly extract the quadratic in $z,\overline{z}$ part of $\vec{R}(t)$.

Thus, we consider the quadratic terms generated by the nonlinearity:
\begin{equation}\label{eq:SecondOrderTerm}
\begin{array}{lll}
& & \vec{N}_{m,n}\ =\ \left(\ \begin{array}{c} N^{Re}_{m,n}\\ N^{Re}_{m,n}\end{array}\ \right)\nn\\
& &\displaystyle\sum_{m+n=2}\ J\vec{N}_{m,n}\ =\ J\vec{N}_{2,0}+J\vec{N}_{1,1}+J\vec{N}_{0,2}\\
&=&\left(
\begin{array}{ccc}
2f^{'}[(\phi^{\lambda})^{2}]\phi^{\lambda}A_{1}B_{1}\\
-\left(\ 3f^{'}[(\phi^{\lambda})^{2}]\phi^{\lambda}+2f^{''}[(\phi^{\lambda})^{2}]
(\phi^{\lambda})^{3}\ \right)A_{1}^{2}-f^{'}[(\phi^{\lambda})^{2}] \phi^{\lambda}B_{1}^{2}
\end{array}\right),
\end{array} 
\end{equation} 
where $A_{1}=\alpha\cdot\xi,\ B_{1}=\beta\cdot\eta$.\\
Now we have the main theorem
\begin{theorem}
Define 
\begin{equation}\label{eq:Rform}
\vec{R}_{m,n}:=[L(\lambda)+iE(\lambda)(m-n)-0]^{-1}P_{c}J\vec{N}_{m,n},
\end{equation}
and decompose $\vec{R}(t)$ as
\begin{equation}
\vec{R}\ =\ \sum_{m+n=2}\ \vec{R}_{m,n}\ +\ \tilde{R}
\label{RtR}
\end{equation}
The function $\tilde{R}(x,t)$
 satisfies (~\ref{eq:tildeR}).
\end{theorem}
\begin{proof}
We observe that  to get
\begin{equation}
\begin{array}{lll}
\displaystyle\D_t\tilde{R} &=&
L(\lambda)\tilde{R}+L_{(\dot\lambda,\dot\gamma)}\tilde{R}+\displaystyle\sum_{m+n=2}L_{(\dot\lambda,\dot\gamma)}R_{m,n}+\mathcal{G}\\
&
&-\displaystyle\sum_{m+n=2}[\partial_{t}\vec{R}_{m,n}+iE(\lambda)(m-n)\vec{R}_{m,n}]-P_{c}J\vec{N}_{>2}
\end{array}
\end{equation} where we used the definitions of $\vec R_{m,n}$ in (~\ref{eq:Rform}), the operator $L_{(\dot\lambda,\dot\gamma)}$ and
the term $\mathcal{G}$ are defined in (~\ref{RAfProj}),
\begin{equation}
J\vec{N}_{>2}:=J\vec{N}(\vec{R},z)-\displaystyle\sum_{m+n=2}J\vec N_{m,n}.
\nn\end{equation}

Next we further decompose $J\vec{N}_{>2}$ and find $M_{2}$, $S_{2}$
and $N_{2}$ in (~\ref{eq:tildeR}). We consider the functions $JN_{m,n},\
m+n=3$,   the third order terms of $J\vec{N}_{>2}:$
\begin{equation}\label{eq:JNm+n=3}
\sum_{m+n=3}J\vec N_{m,n}\ =\ X\left(\ \sum_{m+n=2} \vec R_{m,n}+\left(
\begin{array}{lll}
A_{2}\\
B_{2}
\end{array}
\right)\ \right)+\left(
\begin{array}{lll}
G_{1}(A_{1}^{2},B_{1}^{2})B_{1}\\
-G_{2}(A_{1}^{2},B_{1}^{2})A_{1}
\end{array}
\right)
\end{equation} where, recall the definitions of $A_1,\ B_1, \ A_2, \ B_2$ from ~\eqref{JvecN},
$$G_{1}(A_{1}^{2},B_{1}^{2}):=f^{'}[(\phi^{\lambda})^{2}](A_{1}^{2}+B_{1}^{2})+
2f^{''}[(\phi^{\lambda})^{2}](\phi^{\lambda})^{2}A_{1}^{2},$$ and
$$
\begin{array}{lll}
G_{2}(A_{1}^{2},B_{1}^{2})&:=&[f^{'}[(\phi^{\lambda})^{2}]+2f^{''}[(\phi^{\lambda})^{2}]
(\phi^{\lambda})^{2}](A_{1}^{2}+B_{1}^{2})\\
&
&+[2f^{''}[(\phi^{\lambda})^{2}](\phi^{\lambda})^{2}+\frac{4}{3}f^{'''}[(\phi^{\lambda})^{2}](\phi^{\lambda})^{4}]
A_{1}^{2}
\end{array}
$$
and $X$ is a $2\times 2$ matrix of order $|z|$, defined as
\begin{equation}\label{eq:XDef}
X\ =\ X_{0,1}\ +\ X_{1,0}\ =\ \left(
\begin{array}{lll}
2f^{'}[(\phi^{\lambda})^{2}]\phi^{\lambda}B_{1}& 2f^{'}[(\phi^{\lambda})^{2}]\phi^{\lambda}A_{1}\\
-[6f^{'}[(\phi^{\lambda})^{2}]\phi^{\lambda}+4f^{''}[(\phi^{\lambda})^{2}](\phi^{\lambda})^{3}]A_{1}&
-2f^{'}[(\phi^{\lambda})^{2}]\phi^{\lambda}B_{1}
\end{array}
\right).
\end{equation}
We define the linear operator $M_{2}(z,\bar{z})$ as
$M_{2}(z,\bar{z}):=X+L_{(\dot\lambda,\dot\gamma)}$ which satisfies
the equation (~\ref{eq:M2}). 

The function $S_{2}$, in the statement of Theorem 
\ref{THM:MainTheorem}, is defined as
\begin{equation}\label{eq:s2}
\begin{array}{lll}
S_{2}(z,\bar{z})&:=&\displaystyle\sum_{m+n=2}L_{(\dot\lambda,\dot\gamma)}R_{m,n}+\mathcal{G}-\sum_{m+n=2}[\partial_{t}R_{m,n}+iE(\lambda)(m-n)R_{m,n}]\\
& &+\displaystyle\sum_{m+n=3}JN_{m,n}.
\end{array}
\end{equation} By Equation (~\ref{eq:EstRough}) and
$$
\begin{array}{lll}
& &[\partial_{t},(L(\lambda)\pm iE(\lambda)-0)^{-1}]\\
&\equiv&\partial_{t}(L(\lambda)\pm iE(\lambda)-0)^{-1}-(L(\lambda)\pm
iE(\lambda)-0)^{-1}\partial_{t}\\
&=&\dot\lambda(L(\lambda)\pm
iE(\lambda)-0)^{-1}\partial_{\lambda}[L(\lambda)\pm
iE(\lambda)](L(\lambda)\pm iE(\lambda)-0)^{-1}
\end{array}
$$ we have that
$S_{2}(z,\bar{z})$ satisfies the estimate in part 1 of Theorem 
\ref{GOLD:maintheorem}. For the details we refer to ~\cite{GS3}. 

Lastly, we define the nonlinear
term 
\begin{equation}\label{eq:NonlinearRemainder}
\vec N_{2}(\vec{R},z):=-\left(\ J\vec{N}(\Vec{R},z)-\sum_{m+n=2,3}JN_{m,n}\right).
\end{equation}
 Using the smoothness of the nonlinearity $f[\cdot]$ and the removal of  
  $\cO(|z|^2)$ and $\cO(|z|^3)$ terms, we have that $N_{2}(\vec{R},z)=Loc\ +\ NonLoc$, see (\ref{eq:N2Decomposition}),  satisfying 
  (\ref{eq:Loc-est}) and (\ref{eq:NonlocDef}). The computation is straightforward, but tedious, and is therefore omitted.

Collecting the various definitions and estimates above we have
(~\ref{eq:tildeR}).
\end{proof}
\subsection{$z(t)$ dependence of equations for  $\lambda(t)$ and $\gamma(t)$}
In this section we present the proofs of equations (~\ref{ExpanLambda}) and (~\ref{ExpanGamma}), crucial to controlling the large time behavior. 

Here's the idea. Central to our claim about the large time dynamics of NLS, is that the solution settles into
an asymptotic solitary wave, $\phi^{\lambda_\infty}$, where $\lambda(t)\to\lambda_\infty$. We show this by establishing the integrability and uniform smallness of  $\dot\lambda$. Since we expect the neutral mode amplitudes, $z(t)$, to decay with a rate $t^{-\frac{1}{2}}$, we require that there be no $\cO(|z(t)|^2)$ in the equation (\ref{eq:lambda}):
  $\dot\lambda(t) + \D_t a_1(z,\overline{z})\ =\ \dots$. The strategy is
  to choose the quadratic part of the polynomial $a_1(z,\overline{z})$
  so as to eliminate all quadratic terms non-resonant terms. The latter are terms whose $z$-behavior is: $(z_k)^2$ or $(\overline{z_k})^2$
  and are oscillatory with frequencies $\sim \pm 2iE(\lambda)$.
  But what of the terms of the form: $z_k\overline{z_m}$, which are resonant (non-oscillatory)? This is where we use the choice of basis
  for the degenerate subspace; section \ref{subsec:choice}. A consequence of this choice is that
  there are {\bf no} resonant quadratic terms appearing in the equation for $\lambda$! The calculation is carried out below; see
   Lemma~\ref{coordinate-lemma}.
  
In what follows we use the notations $N^{Im}_{m,n},\ N^{Re}_{m,n}$
to stand for functions satisfying 
$$\left(
\begin{array}{lll}
N^{Im}_{m,n}\\   
-N^{Re}_{m,n}
\end{array}
\right)=JN_{m,n}.$$

We define the polynomials $a_{1}$, $a_{2}$, $p_{k}$ and $q_{k},\
k=1,2,\cdot\cdot\cdot, N$ in \eqref{Decom1}, see also (~\ref{Decom}), as
\begin{equation}\label{eq:pkmn}
\begin{array}{lll}
a_{k}(z,\bar{z})&:=&\displaystyle\sum_{m+n=2,3,m\not=n}A^{(k)}_{m,n}(\lambda),
\
k=1,2,\\
p_{k}(z,\bar{z})&:=&\displaystyle\sum_{m+n=
2,3}P^{(k)}_{m,n}(\lambda),
\ k=1,2,\cdot\cdot\cdot, N,\\
q_{k}(z,\bar{z})&:=&\displaystyle\sum_{m+n=
2,3}Q^{(k)}_{m,n}(\lambda),
\ k=1,2,\cdot\cdot\cdot, N\\
\end{array}
\end{equation}
with the explicit forms
\begin{equation}\label{eq:A1}
\begin{array}{lll}
2iE(\lambda)A_{2,0}^{(1)}&:=&\frac{1}{\langle
\phi^{\lambda},\partial_{\lambda}\phi^{\lambda}\rangle}\langle
N^{Im}_{2,0},\phi^{\lambda}\rangle;\\
3iE(\lambda)A_{3,0}^{(1)}&:=&\frac{1}{\langle
\phi^{\lambda},\partial_{\lambda}\phi^{\lambda}\rangle}\langle
N^{Im}_{3,0},\phi^{\lambda}\rangle;\\
i E(\lambda)A_{2,1}^{(1)}&:=&\frac{1}{\langle
\phi^{\lambda},\partial_{\lambda}\phi^{\lambda}\rangle}[\langle
N^{Im}_{2,1},\phi^{\lambda}\rangle-\frac{i}{2}\Upsilon_{1,1}\langle
z\cdot\eta,\phi^{\lambda}\rangle]
\end{array}
\end{equation} where, recall $\Upsilon_{1,1}$ from ~\eqref{eq:Gamma11},
\begin{equation}
\begin{array}{lll}
-2iE(\lambda)A_{2,0}^{(2)}+A_{2,0}^{(1)}&:=&\frac{1}{\langle
\phi^{\lambda},\partial_{\lambda}\phi^{\lambda}\rangle}\langle
N^{Re}_{2,0},\partial_{\lambda}\phi^{\lambda}\rangle;\\
-3iE(\lambda)A_{3,0}^{(2)}+A_{3,0}^{(1)}&:=&\frac{1}{\langle
\phi^{\lambda},\partial_{\lambda}\phi^{\lambda}\rangle}\langle
N^{Re}_{3,0},\partial_{\lambda}\phi^{\lambda}\rangle;\\
-iE(\lambda)A_{2,1}^{(2)}+A_{2,1}^{(1)}&:=&\frac{1}{\langle
\phi^{\lambda},\partial_{\lambda}\phi^{\lambda}\rangle}[\langle
N^{Re}_{2,1},\partial_{\lambda}\phi^{\lambda}\rangle
-\frac{1}{2}\Upsilon_{1,1}\langle
z\cdot\xi,\partial_{\lambda}\phi^{\lambda}\rangle];
\end{array}
\end{equation}
$$A_{k,l}^{(n)}:=\overline{A_{l,k}^{(n)}}\ \text{for}\ n=1,2,\ k+l=2,3,
\ k\not=l;$$ and
\begin{equation}\label{eq:pq20and02}
\begin{array}{lll}
-2iE(\lambda)P_{2,0}^{(n)}-E(\lambda)Q_{2,0}^{(n)}&:=&-\langle
N_{2,0}^{Im},\eta_{n}\rangle,\\
-2iE(\lambda)Q_{2,0}^{(n)}+E(\lambda)P_{2,0}^{(n)}&:=&\langle
N_{2,0}^{Re},\xi_{n}\rangle,\\
-3iE(\lambda)P_{3,0}^{(n)}-E(\lambda)Q_{3,0}^{(n)}&:=&-\langle
N_{3,0}^{Im},\eta_{n}\rangle,\\
-3iE(\lambda)Q_{3,0}^{(n)}+E(\lambda)P_{3,0}^{(n)}&:=&\langle
N_{3,0}^{Re},\xi_{n}\rangle,\\
\end{array}
\end{equation}
\begin{equation}
\begin{array}{lll}
2iE(\lambda)P_{1,2}^{(n)}-2E(\lambda)Q_{1,2}^{(n)}&:=&-\langle
N_{1,2}^{Im},\eta_{n}\rangle+i\langle N_{1,2}^{Re},\xi_{n}\rangle\\
& &+i\Upsilon_{1,1}\displaystyle\sum_{k=1}^{N}\bar{z}_{k}[\langle\eta_{k},\eta_{n}\rangle-\langle\xi_{k},\xi_{n}\rangle],\\
E(\lambda)Q_{1,1}^{(n)}&:=&\langle N_{1,1}^{Im},\eta_{n}\rangle,\\
E(\lambda)P_{1,1}^{(n)}&:=&\langle N_{1,1}^{Re},\xi_{n}\rangle.
\end{array}
\end{equation}
$$P_{k,l}^{(n)}:=\overline{P_{l,k}^{(n)}},\ Q_{l,k}^{(n)}:=\overline{Q_{k,l}^{(n)}}.$$
The following is the main result.
\begin{proposition}\label{Prop:ExplicitPolyno}
Define the polynomials $a_{1}(z,\bar{z}),\ a_{2}(z,\bar{z}),\ p_{n}(z,\bar{z}),\ q_{n}(z,\bar{z})$ as
above. Then, 
 (~\ref{ExpanLambda})-(~\ref{ExpanGamma}) hold and
moreover
\begin{align}
\label{ExpanLambdaGamma}
&\D_t\lambda=Remainder(t),\nn\\
& \D_t\gamma=\Upsilon_{1,1}+Remainder(t)\\
\label{eq:ZNequation}
\partial_{t}z_{n}+iE(\lambda)z_{n}&=-\left< JN_{2,1},
\left(\begin{array}{lll}\eta_{n}\\-i\xi_{n}\end{array}\right)\right>\nn\\
&+
\frac{1}{2}\Upsilon_{1,1}\displaystyle\sum_{m=1}^{N}z_{m}\left<\left(
\begin{array}{lll}
-i\eta_{m}\\ \xi_{m} \end{array}\right), 
\left(
\begin{array}{lll}
\eta_{n}\\
i\xi_{n}
\end{array}
\right)\right>+Remainder(t),
\end{align}
where $\Upsilon_{1,1}$ is defined in (\ref{eq:Gamma11}) 
and there are no $z\overline{z}$ terms in $Remainder(t)$. Moreover, 

\begin{equation}\label{remainder1}
\begin{array}{lll}
|Remainder(t)|&\lesssim& |z(t)|^{4}+\|\langle
x\rangle^{-\nu}(-\Delta+1)\vec{R}(t)\|_{2}^{2}+\|\vec{R}(t)\|_{\infty}^{2}
+|z(t)|\ \ \|\langle
x\rangle^{-\nu}\tilde{R}(t)\|_{2}
\end{array}
\end{equation} 
\end{proposition}

\nit Before proving the proposition we state the following key
observation.
\begin{lemma}\label{coordinate-lemma}
\begin{equation}\label{eq:unknownresult}
\langle N_{1,1}^{Im},\phi^{\lambda}\rangle=0.
\end{equation}
\end{lemma}
\begin{proof}Recall that
\begin{equation}
A_1=\alpha\cdot\xi=\frac{1}{2}(z\cdot\xi+\overline{z}\cdot\xi),\ \ 
B_1=\beta\cdot\eta=\frac{1}{2i}(z\cdot\eta-\overline{z}\cdot\eta)
\nn\end{equation}
The explicit form of $JN_{2,0}+JN_{1,1}+JN_{0,2}$ in
(~\ref{eq:SecondOrderTerm}) implies that $$
\begin{array}{lll}
N^{Im}_{2,0}+N^{Im}_{1,1}+N^{Im}_{0,2}&=&2f^{'}[(\phi^{\lambda})^{2}]\phi^{\lambda}A_{1}B_{1}\\
&=&\frac{1}{2i}f^{'}[(\phi^{\lambda})^{2}]\phi^{\lambda}(\displaystyle\sum_{n=1}^{N}z_{n}\xi_{n}+\sum_{n=1}^{N}\bar{z}_{n}\xi_{n})
(\sum_{m=1}^{N}z_{m}\eta_{m}-\sum_{m=1}^{N}\bar{z}_{m}\eta_{m}).
\end{array}
$$ By taking the relevant terms we have
$$
\begin{array}{lll}
N^{Im}_{1,1}&=&\frac{1}{2i}f^{'}[(\phi^{\lambda})^{2}]\phi^{\lambda}
(\displaystyle\sum_{n=1}^{N}\bar{z}_{n}\xi_{n}\sum_{m=1}^{N}z_{m}\eta_{m}-
\sum_{n=1}^{N}z_{n}\xi_{n}\sum_{m=1}^{N}\bar{z}_{m}\eta_{m})\\
&=&\frac{1}{2i}\displaystyle\sum_{n=1}^{N}\sum_{m=1}^{N}\bar{z}_{n}z_{m}
f^{'}[(\phi^{\lambda})^{2}]\phi^{\lambda}(\xi_{n}\eta_{m}-\xi_{m}\eta_{n}).
\end{array}
$$ which together with (~\ref{eq:UnexpectedFact}) yields
$$\langle
N^{Im}_{1,1},\phi^{\lambda}\rangle=\frac{1}{2i}\displaystyle\sum_{n=1}^{N}\sum_{m=1}^{N}\bar{z}_{n}z_{m}\int
f^{'}[(\phi^{\lambda})^{2}](\phi^{\lambda})^{2}(\xi_{n}\eta_{m}-\xi_{m}\eta_{n})=0.$$
This completes the  proof of Lemma \ref{coordinate-lemma}.
\end{proof}
\textbf{Proof of Proposition ~\ref{Prop:ExplicitPolyno}}

Recall the estimate of $Remainder$ in (~\ref{remainder}). We put
(~\ref{eq:gamma}) (~\ref{eq:lambda}) in the matrix form
\begin{equation}\label{eq:lambdagamma}
[Id+M(z,\vec{R},p,q)]\left(
\begin{array}{lll}
\dot\lambda\\
\dot\gamma-\Upsilon_{1,1}
\end{array}
\right)=\Omega+Remainder
\end{equation} with the matrix $\Omega$ defined as
\begin{equation}
\Omega:=\left(
\begin{array}{lll}
-\frac{1}{\langle
\phi^{\lambda},\partial_{\lambda}\phi^{\lambda}\rangle}[\ \langle
ImN,\phi^{\lambda}\rangle+\frac{i}{2}\Upsilon_{1,1}\langle\ (z-\bar{z})\cdot\eta,\phi^{\lambda}\ \rangle\ ]-\partial_{t}a_{1}\\
\frac{1}{\langle\phi^{\lambda},\partial_{\lambda}\phi^{\lambda}\rangle}[\langle
ReN,\partial_{\lambda}\phi^{\lambda}\rangle-\frac{1}{2}\Upsilon_{1,1}\langle\
(z+\bar{z})\cdot\xi,\partial_{\lambda}\phi^{\lambda}\ \rangle]-\Upsilon_{1,1}-\partial_{t}a_{2}+a_{1}
\end{array}
\right)
\end{equation}
the term $Remainder$ is produced by
$\frac{\Upsilon_{1,1}}{\langle\phi^{\lambda},\partial_{\lambda}\phi^{\lambda}\rangle}\left(
\begin{array}{lll}
-\langle R_{1},\partial_{\lambda}\phi^{\lambda}\rangle+p\langle \xi,\partial_{\lambda}\phi^{\lambda}\rangle\\
\langle R_{2},\phi^{\lambda}\rangle+q\langle \eta,\phi^{\lambda}
\rangle 
\end{array}
\right),$ $Id$ is the $2\times 2$ identity matrix,
$M(z,\vec{R},p,q)$ is a vector depending on $z,\vec{R},p$ and $q$
and satisfying the estimate
\begin{equation}\label{eq:Mterm}
\|M(z,\vec{R},p,q)\|= \cO(|z|)+Remainder.
\end{equation}

Now by the definitions of $a_{1}$ and $a_{2}$ in (~\ref{eq:pkmn}),
we remove the lower order terms in $z,\bar{z}$ from $\langle
ImN,\phi^{\lambda}\rangle-\frac{i}{2}\Upsilon_{1,1}\langle\ 
(z-\bar{z})\cdot\eta,\phi^{\lambda}\rangle$ and $\langle
ReN,\partial_{\lambda}\phi^{\lambda}\rangle+\frac{1}{2}\Upsilon_{1,1}\langle\ 
(z+\bar{z})\cdot\xi,\partial_{\lambda}\phi^{\lambda}\rangle$ to get $$
\Omega=D_{1}+D_{2}$$ with
$$D_{1}:=\frac{1}{\langle\phi^{\lambda},\partial_{\lambda}\phi^{\lambda}\rangle}\left(
\begin{array}{lll}
-\langle ImN-\displaystyle\sum_{m+n=2,3}N_{m,n}^{Im},\phi^{\lambda}\rangle\\
\langle
ReN-\displaystyle\sum_{m+n=2,3}N_{m,n}^{Re},\partial_{\lambda}\phi^{\lambda}\rangle
\end{array}
\right),$$ where we used (~\ref{eq:unknownresult});
$$D_{2}:=-\displaystyle\sum_{m+n=2,3}\left(
\begin{array}{lll}
\partial_{t}A_{m,n}^{(1)}+iE(\lambda)(m-n)A_{m,n}^{(1)}\\
\partial_{t}A_{m,n}^{(2)}+iE(\lambda)(m-n)A_{m,n}^{(2)}
\end{array}
\right).$$ 

Now we claim that 
\begin{equation}\label{eq:D1D2}
D_{1}, D_2=Remainder.
\end{equation} 
If the claim holds then estimates (~\ref{ExpanLambda}) and (~\ref{ExpanGamma}) follow from
the estimates (~\ref{eq:lambdagamma}), (~\ref{eq:Mterm}) and the
facts $\Omega=D_{1}+D_{2}$.

In the next we prove claim ~\eqref{eq:D1D2} together with ~\eqref{eq:ZNequation}.

By the fact we removed all the second and third order terms  of $J\vec{N}$ we obtain
$
D_{1}=Remainder.
$ Recall the estimate of $Remainder$ in \eqref{remainder}. 

To estimate $D_{2}$ we have to start with studying the equation for $z$. By the fact that
$$\partial_{t}z_{n}+iE(\lambda)z_{n}=\cO(|z|^{2})+Remainder$$ in (~\ref{eq:EstRough}) we obtain $D_{2}=\cO(|z|^{3})+Remainder$, hence
\begin{equation}\label{eq:LambdaGammaRough}
\begin{array}{lll}
\dot\lambda&=&\cO(|z|^{3})+Remainder\\
\dot\gamma-\Upsilon_{1,1}&=&\cO(|z|^{3})+Remainder
\end{array}
\end{equation}
which together with the expansion of $J\vec{N}$ in
(~\ref{eq:NonlinearRemainder}) yields
\begin{equation}
\partial_{t}(\alpha_{n}+p_{n})-E(\lambda)(\beta_{n}+q_{n})+\sum_{k+l=2,3}\langle N^{Im}_{k,l},\eta_{n}\rangle
=-\frac{i}{2}\Upsilon_{1,1}\langle\ (z-\bar{z})\cdot\eta,\eta_{n}\ \rangle+Remainder;
\end{equation}
\begin{equation}
\partial_{t}(\beta_{n}+q_{n})+E(\lambda)(\alpha_{n}+p_{n})-\sum_{k+l=2,3}\langle N^{Re}_{k,l},\xi_{n}\rangle
=-\frac{1}{2}\Upsilon_{1,1}\langle\ (z+\bar{z})\cdot\xi,\xi_{n}\ \rangle+Remainder
\end{equation} where, recall the real function $\Upsilon_{1,1}$ from
(~\ref{ExpanGamma}). Choose $p_{n}$ and $q_{n}$ as in
(~\ref{eq:pq20and02}) to remove the lower order terms as in the
equations of $\dot\lambda$ and $\dot\gamma,$ which together with the
definition $z_{n}=\alpha_{n}+i\beta_{n}$ enables us to obtain
\begin{align}\label{eq:ZnRough}
\partial_{t}z_{n}+iE(\lambda)z_{n}&=-\left\langle JN_{2,1}
 +\frac{1}{2}\Upsilon_{1,1}\left(
\begin{array}{lll}
iz\cdot\eta\\
z\cdot\xi
\end{array}
\right), \left(
\begin{array}{lll}
\eta_{n}\\
-i\xi_{n}
\end{array}
\right)\right\rangle\nn\\
&+ D_{3}(n)+Remainder
\end{align}
with $D_{3}(n)$ defined as
$$D_{3}(n):=-\sum_{k+l=2,3}[\partial_{t}P_{k,l}^{(n)}+i(k-l)E(\lambda)P_{k,l}^{(n)}]-i\sum_{k+l=2,3}[\partial_{t}Q_{k,l}^{(n)}+i(k-l)E(\lambda)Q_{k,l}^{(n)}].$$

We claim that this together with the equations for $\dot\lambda$ in
(~\ref{eq:lambdagamma}) implies that
\begin{equation}\label{eq:D234Remainder}
|D_{2}|,\ |D_{3}(n)|=Remainder.
\end{equation} Indeed, by ~\eqref{eq:EstRough} we have $\partial_{t}z_{n}+iE(\lambda)z_{n}=\cO(|z|^{2})+Remainder$ which together with the equation for $\dot\lambda$ in ~\eqref{eq:LambdaGammaRough} implies $D_{3}=O(|z|^{3})+Remainder$, in turn we have an improved equation
for $z_{n}$ as
$$\partial_{t}z_{n}=-iE(\lambda)z_{n}+\cO(|z|^{3})+Remainder.$$ Using this and repeating the
analysis we find there is no $\cO(|z|^{3})$ term in $D_{2}$ and
$D_{3}$, hence (~\ref{eq:D234Remainder}) holds which leads to
(~\ref{eq:ZNequation}) and ~\eqref{eq:D1D2}.
The proof is complete.
\begin{flushright}
$\square$
\end{flushright}
\section{Proof of the Normal Form Equation (~\ref{eq:detailedDescription})}\label{SEC:NormalForm}
Recall the definitions of
%
%
 the functions $B(n)$ and $D(n)$ after
(~\ref{eq:Fk2}). Then the function $JN_{2,0}$ in
(~\ref{eq:SecondOrderTerm}) admits the form
\begin{equation}\label{eq:JN20}
JN_{2,0}= \displaystyle\sum_{n=1}^{N}z_{n}\left(
\begin{array}{lll}
B(n)\\
D(n)
\end{array}
\right).
\end{equation}

The following is the result establishing the desired normal form of the 
differential equation for the neutral mode amplitudes, $z(t)$. 
\begin{theorem}\label{THM:FGR}
By defining the polynomials $a_{1}, a_{2},\ p_{n},\ q_{n},\
n=1,2,\cdot\cdot\cdot, N,$ as in
(~\ref{eq:pkmn})-(~\ref{eq:pq20and02}), Equation
(~\ref{eq:detailedDescription}) holds.
\end{theorem}
\begin{proof}
Recall the definitions of $JN_{m,n},\ m+n=3,$ from
(~\ref{eq:JNm+n=3}) and the equations for $z_{n}$ in
(~\ref{eq:ZNequation}) whose first two terms on the right hand side
admit the expansion
$$\sum_{k=1}^{5}K_{k}(n)
$$ with
$$
K_{1}(n) :=-\left< X_{0,1}R_{2,0},\left(
\begin{array}{lll}
\eta_{n}\\
-i\xi_{n}
\end{array}
\right)\right>=-\left< R_{2,0},X^{*}_{0,1}\left(
\begin{array}{lll}
\eta_{n}\\
-i\xi_{n}
\end{array}
\right)\right>;$$ $$K_{2}(n):=-\left< X_{1,0}\left(
\begin{array}{lll}
\displaystyle\sum_{k=1}^{N}P_{1,1}^{(k)}\xi_{k}\\
\displaystyle\sum_{k=1}^{N}Q_{1,1}^{(k)}\eta_{k}
\end{array}
\right)+X_{0,1}\left(
\begin{array}{lll}
\displaystyle\sum_{k=1}^{N}P_{2,0}^{(k)}\xi_{k}+A_{2,0}^{(1)}\partial_{\lambda}\phi^{\lambda}\\
\displaystyle\sum_{k=1}^{N}Q_{2,0}^{(k)}\eta_{k}+A_{2,0}^{(2)}\phi^{\lambda}
\end{array}
\right), \left(
\begin{array}{lll}
\eta_{n}\\
-i\xi_{n}
\end{array}
\right)\right>,
$$ where, recall the definition of $X$ in (~\ref{eq:XDef}) and we divide it into two terms
$X=X_{1,0}+X_{0,1}$
$$X_{1,0}:=\left(
\begin{array}{lll}
-if^{'}[(\phi^{\lambda})^{2}]\phi^{\lambda}\ z\cdot\eta& f^{'}[(\phi^{\lambda})^{2}]\phi^{\lambda}\ z\cdot\xi\\
-[3f^{'}[(\phi^{\lambda})^{2}]\phi^{\lambda}+2f^{''}[(\phi^{\lambda})^{2}](\phi^{\lambda})^{3}]\ z\cdot\xi&
if^{'}[(\phi^{\lambda})^{2}]\phi^{\lambda}\ z\cdot\eta
\end{array}
\right)$$ and $\overline{X_{0,1}}:=X_{1,0}$;
$$
\begin{array}{lll} 
K_{3}(n)&=&-\frac{1}{8}\left\langle\ [f^{'}[(\phi^{\lambda})^{2}]+
2f^{''}[(\phi^{\lambda})^{2}](\phi^{\lambda})^{2}]
 ((z\cdot\xi)^{2}-(z\cdot\eta)^{2})\left(
\begin{array}{lll}
i\bar{z}\cdot\eta\\ -\bar{z}\cdot\xi
\end{array}
\right), \left(
\begin{array}{lll}
\eta_{n}\\-i\xi_{n}
\end{array} \right)\ \right\rangle\\
&&+\left\langle\ 
\frac{1}{4}[f^{'}[(\phi^{\lambda})^{2}]+2f^{''}[(\phi^{\lambda})^{2}]
(\phi^{\lambda})^{2}](|z\cdot\xi|^2+|z\cdot\eta|^2)\left(
\begin{array}{lll}
i z\cdot\eta\\ z\cdot\xi
\end{array} \right), \left(
\begin{array}{lll}
\eta_{n}\\ -i\xi_{n}
\end{array} \right)\right\rangle\\
& &-\left\langle
\frac{3}{4}if^{''}[(\phi^{\lambda})^{2}](\phi^{\lambda})^{2}\ (z\cdot\eta)^{2}\left( \begin{array}{lll}
\bar{z}\cdot\eta\\ 0 \end{array} \right), \left(
\begin{array}{lll}
\eta_{n}\\ -i\xi_{n}
\end{array} \right)\right\rangle\\
& &-\left\langle [\frac{3}{4} f^{''}[(\phi^{\lambda})^{2}](\phi^{\lambda})^{2}+\frac{1}{2}
f^{'''}[(\phi^{\lambda})^{2}](\phi^{\lambda})^{4}]\ (z\cdot\xi)^{2}\left(
\begin{array}{lll}
0\\ -\bar{z}\cdot\xi
\end{array}
\right), \left(
\begin{array}{lll}
\eta_{n}\\
-i\xi_{n}
\end{array}
\right)\right\rangle
\end{array}
$$
\begin{equation}
K_{4}(n):=\frac{1}{2}\ \Upsilon_{1,1}\left\langle\left(
\begin{array}{lll}
-i z\cdot\eta\\
z\cdot\xi
\end{array}
\right), \left(
\begin{array}{lll}
\eta_{n}\\
i\xi_{n}
\end{array}
\right)\right\rangle,
\nn\end{equation}
$$K_{5}(n):=-\left\langle R_{1,1},X^{*}_{1,0}\left(
\begin{array}{lll}
\eta_{n}\\
-i\xi_{n}
\end{array}
\right)\right\rangle$$ 
where, recall the real function $\Upsilon_{1,1}$ from
(~\ref{ExpanGamma}).

Next we study $K_{j}(n),$ $j=1,2,3,4,5.$

We start with the most important term $K_{1}(n)$. Recall the
definition of $G_{n}$ in (~\ref{eq:Fk2}). By direct computation we
obtain
\begin{equation}\label{eq:MN20}
\begin{array}{lll}
X^{*}_{0,1}\left(
\begin{array}{lll}
\eta_{n}\\
-i\xi_{n}
\end{array}
\right)=-iJ\left(
\begin{array}{lll}
B(n)\\
D(n)
\end{array}
\right)
\end{array}=-iJG_{n}
\end{equation} which together with (~\ref{eq:Rform}) and (~\ref{eq:JN20}) implies that
$$
K_{1}(n)=\sum_{k=1}^{N}z_{k}\langle
(L(\lambda)+2iE(\lambda)-0)^{-1}P_{c}G_{k},iJG_{n}\rangle.
$$ Define $Z(k,n):=-\langle
(L(\lambda)+2iE(\lambda)-0)^{-1}P_{c}G_{n},iJG_{k}\rangle$ and a $N \times N$ matrix 
\begin{equation}\label{eq:K121}
\Gamma(z,\bar{z}):=[A(k,l)]\ \text{with}\ A(k,l):=\frac{1}{2}[Z(k,l)+\bar{Z}(l,k)],\ 1\leq k,\ l\leq d.
\end{equation}

For $\displaystyle\sum_{j=2}^{5}K_{j}(n)$ we claim that it can
decomposed into the matrix form
\begin{equation}\label{eq:SkewSym}
\displaystyle\sum_{j=2}^{5}K_{j}(n)=(S(n,1),S(n,2),\cdot\cdot\cdot,S(n,d))z\
\text{with}\ S(k,l)+\overline{S(l,k)}=0.
\end{equation} Define a $N \times N$ skew symmetric matrix $$\Lambda(z,\bar{z}):=[\Lambda(j,k)]\ \text{with} \ \Lambda(j,k):=S_{k,l}+\frac{1}{2}[Z(k,l)-\bar{Z}(l,k)].$$  This together with (~\ref{eq:ZNequation}) and
(~\ref{eq:K121}) yields the equation for $z$ in
(~\ref{eq:detailedDescription})

What is left is to prove (~\ref{eq:SkewSym}). To avoid tedious, but
simple, computations, we only analyze part of $K_{2}(n)$ and
$K_{3}(n).$

\begin{enumerate}
\item[(A)]
The part of $K_{2}(n)$ we study is
$$\Psi_{2,1}(n):=
-\left\langle X_{0,1}\left(
\begin{array}{lll}
A_{2,0}^{(1)}\partial_{\lambda}\phi^{\lambda}\\
A_{2,0}^{(2)}\phi^{\lambda}
\end{array}
\right), \left(
\begin{array}{lll}
\eta_{n}\\
-i\xi_{n}
\end{array}
\right)\right\rangle,
$$ the analysis of the other terms are
similar. By (~\ref{eq:MN20}) we rewrite $\Psi_{2,1}(n)$ as
$$
\begin{array}{lll}
\Psi_{2,1}(n)&=&\left\langle\left(
\begin{array}{lll}
A_{2,0}^{(1)}\partial_{\lambda}\phi^{\lambda}\\
A_{2,0}^{(2)}\phi^{\lambda}
\end{array}
\right), 4i\left(
\begin{array}{lll}
D(n)\\
-B(n)
\end{array}
\right)\right\rangle\\
&=&-4iA_{2,0}^{(1)}\langle
\partial_{\lambda}\phi^{\lambda},D(n)\rangle+4iA_{2,0}^{(2)}\langle \phi^{\lambda},B(n)\rangle.
\end{array}
$$

Equation (~\ref{eq:A1}) relates $\langle
N^{Re}_{2,0},\phi^{\lambda}_{\lambda}\rangle$ and $\langle
N^{Im}_{2,0},\phi^{\lambda}\rangle$ to $A^{(1)}_{2,0}$ and
$A^{(2)}_{2,0}$ which together with the expression of $JN_{2,0}$ in
Equation (~\ref{eq:JN20}) yields
\begin{equation}\label{eq:Psi21}
\Psi_{2,1}(n)=\sum_{k=1}^{N}\Psi(n,k)z_{k}
\end{equation} with
$$
\begin{array}{lll}
\Psi(n,k)&:=&\frac{2}{E(\lambda)\langle
\phi^{\lambda},\partial_{\lambda}\phi^{\lambda}\rangle}
\displaystyle[\langle
B(k),\phi^{\lambda}\rangle\langle\partial_{\lambda}\phi^{\lambda},D(n)\rangle-\langle
D(k),\partial_{\lambda}\phi^{\lambda}\rangle\langle\phi^{\lambda},B(n)\rangle]\\
& &+\frac{i}{E^{2}(\lambda)\langle
\phi^{\lambda},\partial_{\lambda}\phi^{\lambda}\rangle}\displaystyle\langle
B(k),\phi^{\lambda}\rangle\langle\phi^{\lambda},B(n)\rangle.
\end{array}
$$ By straightforward computation we have
\begin{equation}\label{eq:zeroPart1}
\Psi(n,k)+\overline{\Psi(k,n)}=0.
\end{equation}

By (~\ref{eq:Psi21}) and (~\ref{eq:zeroPart1}) we complete the proof
for $\Psi_{2,1}(n).$
\item[(B)] 
To simplify the notation we introduce $\rho$ and $\omega$
 \begin{equation}\label{eq:RhoSigma}
 \begin{array}{lll}
 \rho&:=\ \frac{1}{2}\rhoo\ =\ \frac{1}{2}\displaystyle\sum_{n=1}^{N}z_{n}\xi_{n},\\
\omega &:=\ \frac{1}{2}\omegaa\ =\ \frac{1}{2}\displaystyle\sum_{n=1}^{N}z_{n}\eta_{n}.
\end{array}
\end{equation} 
This implies that
$$\rho^{2}-\omega^{2}=\frac{1}{2}\sum_{n=1}^{N}z_{n}[\rho\xi_{n}-\omega\eta_{n}], \
\rho\bar\rho+\omega\bar\omega=\frac{1}{2}\sum_{n=1}^{N}z_{n}[\xi_{n}\bar\rho+\eta_{n}\bar\omega],$$
$$\rho^{2}=\frac{1}{2}\sum_{n=1}^{N}z_{n}\xi_{n}\rho,\
\omega^{2}=\frac{1}{2}\sum_{n=1}^{N}z_{n}\eta_{n}\omega.$$ By the
definition of $K_{3}(n)$ it is not hard to get
\begin{equation}\label{eq:K321}
K_{3}(n)=\displaystyle\sum_{k=1}^{N}z_{k}\Phi(n,k)
\end{equation} with
$$
\begin{array}{lll}
\Phi(n,k)&:=&\frac{i}{2}\langle[f^{'}[(\phi^{\lambda})^{2}]+2f^{''}[(\phi^{\lambda})^{2}](\phi^{\lambda})^{2}]
(\rho\xi_{k}-\omega\eta_{k}),(\rho\xi_{n}-\omega\eta_{n})\rangle\\
& &+i\langle
[f^{'}[(\phi^{\lambda})^{2}]+2f^{''}[(\phi^{\lambda})^{2}](\phi^{\lambda})^{2}]
[\bar\omega \eta_{k}+\bar\rho\xi_{k}\rangle],[\bar\omega
\eta_{n}+\bar\rho\xi_{n}\rangle]\\
& &+i\langle
[3f^{''}[(\phi^{\lambda})^{2}](\phi^{\lambda})^{2}+2f^{'''}[(\phi^{\lambda})^{2}](\phi^{\lambda})^{4}]
\rho\xi_{k},\rho\xi_{n}\rangle\\
& &-i\langle
3f^{''}[(\phi^{\lambda})^{2}](\phi^{\lambda})^{2}\omega\eta_{k},\omega
\eta_{n}\rangle.
\end{array}
$$
Immediately we have
\begin{equation}
\Phi(n,k)+\overline{\Phi(k,n)}=0.
\end{equation} This together with (~\ref{eq:K321}) completes the
proof for $K_{3}(n).$
\end{enumerate}
\end{proof}
\section{Proof of the Main Theorem ~\ref{THM:MainTheorem}}\label{ProveMain}
For simplicity, we present  the proof 
 of Theorem ~\ref{THM:MainTheorem} for the case $d=3$;
 the proof can be easily modified to cover $d\geq 3$. The main difference is that, in controlling  $\|\vec{R}(t)\|_{L^\infty(\R^d)}$ by $\|\vec{R}(t)\|_{H^k(\R^d)}$, for $d=3$ we take $k=2$, while in general we need $k=[\frac{d}{2}]+1$;  see section
~\ref{Sec:Propagator}. 
\subsection{Estimation strategy}
In this subsection, discuss our strategy for studying the large time behavior of solutions.\\ \\
We begin by introducing a family of space-time norms, $ Z(T),\ {\cal R}_j(T)$,  for measuring the decay of the $z(t)$ and  $\vec{R}(t)$  for $0\le t\le T$, with $T$ arbitrary.
We then prove that this family of norms satisfy a set of coupled inequalities, 
from which we can infer
  the desired large time asymptotic behavior. 
Define
\begin{equation}
T_{0}:=|z^{(0)}|^{-1},\label{Todef}
\end{equation}
where
$z^{(0)}$ and the constant $\nu$
are defined in  Theorem ~\ref{THM:MainTheorem}.\\ \\

\nit{\bf Family of Norms:}
\begin{equation}\label{majorant}
\begin{array}{lll}
Z(T):=\displaystyle\max_{t\leq T}(T_{0}+t)^{\frac{1}{2}}|z(t)|,& &
\mathcal{R}_{1}(T):=\displaystyle\max_{t\leq
T}(T_{0}+t)\|\langle x\rangle^{-\nu} \vec{R}(t)\|_{ H^{2}},\\
\mathcal{R}_{2}(T):=\displaystyle\max_{t\leq
T}(T_{0}+t)\|\vec{R}(t)\|_{\infty},& &
\mathcal{R}_{3}(T):=\displaystyle\max_{t\leq
T}(T_{0}+t)^{\frac{7}{5}}\|\langle x\rangle^{-\nu}\tilde{R}(t)\|_{2}\\
\mathcal{R}_{4}(T):=\displaystyle\max_{t\leq
T}\|\vec{R}(t)\|_{ H^{2}},& &
\mathcal{R}_{5}(T):=\displaystyle\max_{t\leq
T}\frac{(T_{0}+t)^{\frac{1}{2}}}{\log(T_{0}+t)}\|\vec{R}(t)\|_{3}\ .\\
\end{array}
\end{equation} 
\\
\nit{\bf Remark on choice of norms:}\ 
It is clear that a combination of $H^2$, spatially weighted $H^2$ and $L^\infty$ norms of $\vec{R}(t)$, as well as a bound on $|z(t)|$, are  plausible choices of norms to control the large time behavior. This accounts for the  definitions: $Z(T),  \cR_1, \cR_2(T)$ and $ \cR_4(T)$. Our list of norms also includes estimation of the time decay of 
$\| \vec{R}(t)\|_3$  ($\cR_5$) and 
 the local $L^2$ norm of an auxiliary function, $\tilde{R}(t)$ ($\cR_3$).
Why these two additional norms? As will be seen, the $\zeta(t)=|z(t)|$ satisfies an equation of the form $\dot\zeta\sim-\kappa^2\zeta^3\ +\ c(t)$, where $c(t)$ consists of various coupling terms (products)  involving neutral mode amplitudes ($z(t)$),  the ground state ($\phi^{\lambda(t)}$) and dispersive terms ($\vec{R}(t)$). First, neglecting $c(t)$, we observe that $\zeta(t)\sim t^{-\frac{1}{2}}$. To treat $c(t)$ as a small perturbation for large $t$, it is necesssary that it decay more rapidly than the term $\zeta^3(t)\sim  t^{-\frac{3}{2}}$. Without any further decomposition of $\vec{R}(t)$, we find among the coupling terms one of order  $|z(t)|\ \|\langle x\rangle^{-\nu}\ \vec{R}(t)\|_2$. The expected decay rates of each factor, imply this term is of order $t^{-\frac{3}{2}}$ for large $t$, which is of the {\it same} order as $\zeta^3(t)$. The resolution is to expand $\vec{R}(t)$ as  a leading order part consisting of terms $R_{m,n}=z^m\bar{z}^n,\ \ m+n=2$ plus a more rapidly decaying correction $\tilde{R}(t)$,
 with $\|\langle x\rangle^{-\nu}\ \tilde{R}(t)\|_2=\cO(t^{-1-\delta}),\ \delta>0$; see  equation (\ref{eq:expanR}). This modification yields an equation with an improved correction term of order $|z(t)|\ \|\langle x\rangle^{-\nu}\ \tilde{R}(t)\|_2=t^{-\frac{3}{2}-\delta},\ \delta>0$, which can be treated as a small peturbation in the large time dynamics.
\\ \\
\nit{\bf Remark on the estimation strategy; see also, 
\cite{BuSu03,SoWe04}.}
 Estimation of ${\cal R}_j(T)$ proceeds by using the DuHamel, equivalent integral equation, formulation relative to a fixed operator linearized operator $L(\lambda_1)$, where $\lambda_1=\lambda(T)$ 
and $T>0$ is fixed and arbitrary.  Namely,
\begin{align}
&\D_t\vec{R}=L(\lambda_{1})\vec{R}+\dots\nn\\
&\implies\ \ \D_tP^{\lambda_{1}}_{c}\vec{R}=L(\lambda_{1})P^{\lambda_{1}}_{c}\vec{R}+\ P_c^{\lambda_1}\ \left(\ L(\lambda(t))-L(\lambda_1)\ \right)\vec{R}+\dots\nn\\
&\implies\ \ P^{\lambda_{1}}_{c}\vec{R}(t)\ =\ e^{L(\lambda_{1})t}\vec{R}(0)\ +\ \int_0^t\ e^{L(\lambda_1)(t-s)}\ \left(\cdot\cdot\cdot\right)\ ds
\end{align}
We can therefore apply the time-decay estimates of Proposition \ref{prop:decayest} to obtain bounds on local decay and $L^\infty$ norms of $P_c^{\lambda_1}\vec{R}(t)$.  However, we need bounds on $\vec{R}(t)=P_c^{\lambda(t)}\vec{R}(t)$. Since $\vec{R}(t)=P_c^{\lambda_1}\vec{R}(t)+P_{disc}^{\lambda_1}\vec{R}(t)$, it suffices to bound 
$P_{disc}^{\lambda_1}\vec{R}(t)$. This is done as follows.
\begin{align}
P_{disc}^{\lambda_1}R&=(P_{disc}^{\lambda_1}-P_{disc}^{\lambda(t)})R(t)
 \ +\ P_{disc}^{\lambda(t)}R(t)\nn\\
  &=\ (P_{disc}^{\lambda_1}-P_{disc}^{\lambda(t)})R(t),\ \ \ \ ({\rm because}\ P_{disc}^{\lambda(t)}R(t)=0)\nn\\
  &=\ (P_{disc}^{\lambda_1}-P_{disc}^{\lambda(t)})\ P_{disc}^{\lambda_1}R(t) 
  \ +\ (P_{disc}^{\lambda_1}-P_{disc}^{\lambda(t)})\ P_c^{\lambda_1}R(t)\nn\\
  &\implies\ \left(\ I-(P_{disc}^{\lambda_1}-P_{disc}^{\lambda(t)})\ \right)P_{disc}^{\lambda_1}R\ =\ (P_{disc}^{\lambda_1}-P_{disc}^{\lambda(t)})\ P_c^{\lambda_1}R(t)
  \end{align}
  Therefore,
    \begin{align}
 & P_{disc}^{\lambda_1}R(t)\ =\
 \left(I-\delta(\lambda,\lambda_1\right) )^{-1}\  \delta(\lambda,\lambda_1) 
\ P_c^{\lambda_1}R(t)
  \end{align}
 and we estimate $R(t)$ in either a local energy, 
 $H^2(\R^d;\langle x\rangle^{-\sigma}dx)$ or $L^\infty(\R^d)$ via 
  \begin{align}
  \|R(t)\|_X &\le
   \| P_c^{\lambda_1}R(t) \|_X\ +\ \|P_{disc}^{\lambda_1}R(t)\|_X\nn\\
  &\le  \| P_c^{\lambda_1}R(t)  \|_X + 
  \|P_c^{\lambda_1}R(t)\|_X.
\end{align}
 Here, $\delta(\lambda,\lambda_1)=P_{disc}^{\lambda_1}-P_{disc}^{\lambda(t)}$ is finite rank and of  small norm proportional to $\int_t^T |\dot\lambda(s)|\ ds$.\\ \\
 We now derive the integral equation for $P_c^{\lambda_1}$, which is the basis for our time-decay estimates. 
 If we write
$L(\lambda(t))=L(\lambda_{1})+L(\lambda(t))-L(\lambda_{1})$, then
Equation (~\ref{RAfProj}) for $\vec{R}$ takes the form
$$\D_tP^{\lambda_{1}}_{c}\vec{R}=L(\lambda_{1})P^{\lambda_{1}}_{c}\vec{R}+(\lambda-\lambda_{1}+\dot\gamma)P^{\lambda_{1}}_{c}J\vec{R}+\cdot\cdot\cdot$$
 Recall that $L(\lambda)$ has two branches of essential
spectrum $[i\lambda,i\infty)$ and $(-i\infty,-i\lambda]$, we use
$P_{+}$ and $P_{-}$ to denote the projection operators onto these
two branches of the essential spectrum of $L(\lambda_{1})$. Then we
have
\begin{lemma}\label{LM:ApproOp} For any function $h$ and any large constant $\nu>0$ we have
\begin{equation}\label{eq:PosiNegBran}
\left\|\ \langle x\rangle^{\nu}(-\Delta+1)\left(\ P_{c}^{\lambda_{1}}J-i(P_{+}-P_{-})\ \right)h\ \right\|_{2}\leq
c\ \left\|\ \langle x\rangle^{-\nu}(-\Delta+1)h\ \right\|_{2}.
\end{equation}
\end{lemma}
\nit For $d=1$ the proof of this lemma can be found in ~\cite{BuSu03}, the proof of $d\geq 3$ is similar, hence omitted here.
\\ \\
Equation (~\ref{RAfProj}) can be rewritten as
\begin{equation}\label{estimater21r22}
\begin{array}{lll}
\D_tP_{c}^{\lambda_{1}}\vec{R}&=&L(\lambda_{1})P_{c}^{\lambda_{1}}
\vec{R}+[\dot{\gamma}+\lambda-\lambda_{1}]i(P_{+}-P_{-})\vec{R}\\
&
&+P_{c}^{\lambda_{1}}O_{1}\vec{R}+P_{c}^{\lambda_{1}}P_{c}^{\lambda(t)}\mathcal{G}-P_{c}^{\lambda_{1}}P_{c}^{\lambda(t)}JN(\vec{R},z),
\end{array}
\end{equation}
where $O_{1}$ is the operator defined by
\begin{equation}\label{EstO1}
O_{1}:=\dot{\lambda}P_{c\lambda}+L(\lambda)-L(\lambda_{1})+
\dot{\gamma}P_{c}^{\lambda}J-[\dot{\gamma}+\lambda-\lambda_{1}]i(P_{+}-P_{-}).
\end{equation} 
By Equation (~\ref{estimater21r22}) and the observation that the
operators $P_{+},$ $P_{-}$ and $L(\lambda_{1})$ commute with each
other, we have
\begin{equation}\label{rewriter21r22}
\begin{array}{lll}
P_{c}^{\lambda_{1}}\vec{R}
&=&e^{tL(\lambda_{1})+a(t,0)(P_{+}-P_{-})}P_{c}^{\lambda_{1}}\vec{R}(0)\\
&
&+\int_{0}^{t}e^{(t-s)L(\lambda_{1})+a(t,s)(P_{+}-P_{-})}P_{c}^{\lambda_{1}}[O_{1}\vec{R}+P_{c}^{\lambda}\mathcal{G}-P_{c}^{\lambda}JN(\vec{R},z)]ds,
\end{array}
\end{equation} with $a(t,s)\ =\ i\ \int_{s}^{t}
[\dot{\gamma}(\tau)+\lambda(\tau)-\lambda_{1}]
\ d\tau.$ We observe that
$P_{+}P_{-}=P_{-}P_{+}=0$ and for any times $t_{1}\leq t_{2}$ the
operator
$$
e^{ a(t_{2},t_{1})(P_{+}-P_{-})}
=e^{a(t_{2},t_{1})}P_{+}+e^{-a(t_{2},t_{1})}P_{-}:
 H^{2}\rightarrow  H^{2}
$$ is uniformly bounded. \\ \\

We conclude this subsection by recording the following result which is used repeatedly in our estimates:
\begin{proposition}\label{prop:conv-est}
 Let $T_0\ge 2$. There exists a constant $c>0$, such that 
\begin{align}
 \int_0^t\ \frac{1}{(1+t-s)^{3\over2}}\ \frac{1}{(T_0+s)^{\sigma}}\ ds\ &\le\  \frac{c}{(T_0+t)^{\sigma}},\ \sigma\in [0,\frac{3}{2}],
\label{conv-est}\\
\int_{0}^{t}(t-s)^{-\frac{1}{2}}(T_{0}+s)^{-1}ds\ &\leq\ c\ (T_{0}+t)^{-\frac{1}{2}}\log(T_{0}+t).
\label{log-conv-est}
\end{align}
\end{proposition}
Similar versions can be found in many literature, for example ~\cite{SoWe99,BuSu03}.
The proof is given in the Appendix ~\ref{subSEC:Time-Con}. 

\subsection{Estimate for
$\mathcal{R}_{1}(T)\ :=\ 
\displaystyle\max_{t\leq T}(T_{0}+t)\|\langle x\rangle^{-\nu} \vec{R}(t)\|_{{H}^{k}}$} 
%
\begin{proposition}\label{prop:controlM1}
\begin{equation}\label{EstR1}
\mathcal{R}_{1}\leq c\left(\ T_{0}\|\langle x\rangle^{\nu}\vec{R}(0)\|_{ H^{2}}+\mathcal{R}^{2}_{4}\mathcal{R}_{2}+Z^{2}+T_{0}^{-\frac{1}{2}}[Z^{3}+Z\mathcal{R}_{1}+\mathcal{R}_{4}\mathcal{R}_{2}^{2}]\ \right).
\end{equation}
\end{proposition}

 With a view toward proving the time decay estimate of Proposition \ref{prop:controlM1}, we now first give appropriate norm-estimates of the latter terms in equation (\ref{estimater21r22}).

First,  the norm definitions (~\ref{majorant}) and Lemma \ref{LM:ApproOp}, we estimate the $O_1\vec{R}$ and ${\cal G}$ terms 
\begin{equation}\label{estimateonO1}
\|\langle x\rangle^{\nu}(-\Delta+1)O_{1}\vec{R}\|_{2}\leq
c(T_{0}+t)^{-\frac{3}{2}}Z\mathcal{R}_{1},
\end{equation}
\begin{equation}\label{estimateonO2}
\|\langle x\rangle^{\nu}(-\Delta+1)\mathcal{G}\|_{2}\leq
c(T_{0}+t)^{-\frac{3}{2}}Z^{3}.
\end{equation} 

Next, we estimate the nonlinear term,
$JN$, via
\begin{lemma}\label{LM:JNNonlinear}
\begin{equation}\label{estimateonO3}
\begin{array}{lll}
& &\|(-\Delta+1)JN(\vec{R},z)\|_{1}+\|(-\Delta+1)JN(\vec{R},z)\|_{2}\\
&\leq&
c(T_{0}+t)^{-1}[\mathcal{R}_{4}^{2}\mathcal{R}_{2}+Z^{2}]+c(T_{0}+t)^{-\frac{3}{2}}[Z\mathcal{R}_{1}+\mathcal{R}_{4}\mathcal{R}_{2}^{2}].
\end{array}
\end{equation}
\end{lemma}
\begin{proof}
Recall the definition $N_{2}(\vec{R},z):=-J\vec{N}(\Vec{R},z)+\sum_{m+n=2,3}JN_{m,n},$ in (~\ref{eq:NonlinearRemainder}) and the decomposition $N_{2}$ as the sum of $Loc$ and $NonLoc$ in (~\ref{eq:N2Decomposition}). By the fact $JN_{m,n}$, $m+n=2,3,$ are localized functions we have the estimate
 $$
 \begin{array}{lll}
 & &\|(-\Delta+1)[JN(\vec{R},z)-NonLoc]\|_{1}+\|(-\Delta+1)[JN(\vec{R},z)-NonLoc]\|_{2}\\
 &\leq &c\ |z|\left(\ |z|+\|\langle x\rangle^{-\nu}\vec{R}\|_{2}\ \right)\\
 &\leq &c\left[\ (T_{0}+t)^{-1}Z^{2}+(T_{0}+t)^{-\frac{3}{2}}Z\mathcal{R}_{1}\ \right].
 \end{array}
 $$
More challenging is the term $NonLoc$, defined in
(~\ref{eq:NonlocDef}), which is purely nonlinear, having no spatially localized factors. We use the estimate
$$
\begin{array}{lll}
& &\|(-\Delta+1)NonLoc\|_{1}+\|(-\Delta+1)NonLoc\|_{2}\nn\\
  & &\leq
c\left(\ \|\vec{R}\|_{ H^{2}}^{2}\|\vec{R}\|_{\infty}
   \ +\ \|\vec{R}\|_{ H^{2}}\|\vec{R}\|_{\infty}^{2}\ \right)\ 
\nn\\
& &\leq c(T_{0}+t)^{-1}\mathcal{R}_{4}^{2}\mathcal{R}_{2}+c(T_{0}+t)^{-\frac{3}{2}}\mathcal{R}_{4}\mathcal{R}_{2}^{2}\nn
\end{array}
$$
by the fact $f(x^{2})x$ is of the order $x^{3}$ around $x=0$ for
$d=3$.\\ Collecting the estimates above we have (~\ref{estimateonO3}).
\end{proof}

\nit {\bf{Proof of
Proposition ~\ref{prop:controlM1}}}. By Equation (~\ref{rewriter21r22}),
Estimates (~\ref{second}) and (~\ref{finalestimate}) for $d=3$ we
have
\begin{equation}\label{estimaterinnorm2}
\begin{array}{lll}
& &\|\langle x\rangle^{-\nu}(-\Delta+1)P_{c}^{\lambda_{1}}\vec{R}(t)\|_{2}\\
&\leq&
\|\langle x\rangle^{-\nu}(-\Delta+1)e^{tL(\lambda_{1})}P_{c}^{\lambda_{1}}\vec{R}(0)\|_{2}\\
&
&+\|\int_{0}^{t}\langle x\rangle^{-\nu}(-\Delta+1)e^{(t-s)L(\lambda_{1})}P_{c}^{\lambda_{1}}[O_{1}(s)\vec{R}+P_{c}^{\lambda}\mathcal{G}-P_{c}^{\lambda}JN(\vec{R},z)]ds\|_{2}\\
&\leq&c(1+t)^{-\frac{3}{2}}\|\langle x\rangle^{\nu}(-\Delta+1)\vec{R}(0)\|_{2}\\
&
&+\int_{0}^{t}(1+t-s)^{-\frac{3}{2}}\|\langle x\rangle^{\nu}(-\Delta+1)[O_{1}\vec{R}
+P_{c}^{\lambda}\mathcal{G}]ds\|_{2}\\
&
&+\int_{0}^{t}(1+t-s)^{-\frac{3}{2}}(\|(-\Delta+1)P_{c}^{\lambda}JN(\vec{R}(s),z)\|_{1}
+\|(-\Delta+1)P_{c}^{\lambda}JN(\vec{R}(s),z)\|_{2})ds.
\end{array}
\end{equation}
Therefore by estimates (~\ref{estimateonO1})-(~\ref{estimateonO2})
and (~\ref{estimateonO3}) we have
$$
\begin{array}{lll}
& &\|\langle x\rangle^{-\nu}(-\Delta+1)P_{c}^{\lambda_{1}}\vec{R}\|_{2}\\
&\leq&c[(1+t)^{-\frac{3}{2}}\|\langle x\rangle^{\nu}(-\Delta+1)\vec{R}(0)\|_{2}\\
& &
+\int_{0}^{t}(1+t-s)^{-\frac{3}{2}}(T_{0}+s)^{-1}ds\ \left(\ \mathcal{R}_{4}^{2}\mathcal{R}_{2}+Z^{2}+T_{0}^{-\frac{1}{2}}[Z^{3}+Z\mathcal{R}_{1}+\mathcal{R}_{4}\mathcal{R}_{2}^{2}]\ \right).
\end{array}
$$
Using the time convolution estimate ~\eqref{conv-est} we obtain
$$
\begin{array}{lll}
& &\|\langle x\rangle^{-\nu}(-\Delta+1)P_{c}^{\lambda_{1}}\vec{R}\|_{2}\\
&\leq&
c(T_{0}+t)^{-1}\left[\ T_{0}\|\langle x\rangle^{\nu}\vec{R}(0)\|_{2}+\mathcal{R}_{4}^{2}\mathcal{R}_{2}+Z^{2}+T_{0}^{-\frac{1}{2}}[Z^{3}+Z\mathcal{R}_{1}+\mathcal{R}_{4}\mathcal{R}_{2}^{2}]\ \right].
\end{array}
$$
This implies Proposition
~\ref{prop:controlM1}.
\begin{flushright}
$\square$
\end{flushright}
\subsection{ Estimate for 
$\mathcal{R}_{2}(T):=\displaystyle\max_{t\leq
T}(T_{0}+t)\|\vec{R}(t)\|_{\infty}$} 

\begin{proposition}\label{controlM2}
\begin{equation}\label{EstR2}
\mathcal{R}_{2}\leq c\ \left[\ T_{0}\left(\|\vec{R}(0)\|_{1}+\|\vec{R}(0)\|_{ H^{2}}\right)\ +\ Z^{2}+\mathcal{R}_{4}^{2}\mathcal{R}_{2}+T_{0}^{-\frac{1}{2}}[Z^{3}+Z\mathcal{R}_{1}+\mathcal{R}_{1}^{2}]\ \right].
\end{equation}
\end{proposition}
\nit To prove this, we use the following result, whose proof very similar to that of Lemma ~\ref{LM:JNNonlinear}, and hence omitted.
\begin{lemma}
\begin{equation}\label{eq:JNForR2}
\begin{array}{lll}
&
&\|P_{c}^{\lambda}JN(\vec{R},z)\|_{1}+\|P_{c}^{\lambda}JN(\vec{R},z)\|_{ H^{2}}\\
&\leq &c(T_{0}+t)^{-1}[Z^{2}+\mathcal{R}_{4}^{2}\mathcal{R}_{2}]+c
(T_{0}+t)^{-\frac{3}{2}}[Z^{3}+Z\mathcal{R}_{1}+\mathcal{R}_{1}^{2}].
\end{array}
\end{equation}
\end{lemma}

\nit \textbf{Proof of Proposition ~\ref{controlM2}} By Estimate
(~\ref{lastestimate}) for $d=3$ and Equation
(~\ref{estimater21r22}) we have that
\begin{equation}
\begin{array}{lll}
& &\|P_{c}^{\lambda_{1}}\vec{R}(t)\|_{\infty}\\
&\leq& \|e^{tL(\lambda_{1})}P_{c}^{\lambda_{1}}\vec{R}(0)\|_{\infty}
+\int_{0}^{t}\|e^{(t-s)L(\lambda_{1})}P_{c}^{\lambda_{1}}[O_{1}(s)\vec{R}+P_{c}^{\lambda}\mathcal{G}-P_{c}^{\lambda}JN(\vec{R},z)]\|_{\infty}ds\\
&\leq&c(1+t)^{-\frac{3}{2}}(\|\vec{R}(0)\|_{1}+\|\vec{R}(0)\|_{ H^{2}})\\
& &+c\int_{0}^{t}(1+t-s)^{-\frac{3}{2}}[\|O_{1}(s)\vec{R}+P_{c}^{\lambda}\mathcal{G}\|_{1}+\|O_{1}(s)\vec{R}+P_{c}^{\lambda}\mathcal{G}\|_{ H^{2}}]ds\\
&
&+c\int_{0}^{t}(1+t-s)^{-\frac{3}{2}}(\|P_{c}^{\lambda}JN(\vec{R},z)\|_{1}+\|P_{c}^{\lambda}JN(\vec{R},z)\|_{ H^{2}})ds.
\end{array}
\end{equation}
By the properties of $O_{1}$ (Equation (~\ref{EstO1})) and $\mathcal{G}$
(Equation (~\ref{RAfProj})) we have
$$\|O_{1}(s)\vec{R}+P_{c}^{\lambda}\mathcal{G}\|_{1}+\|O_{1}(s)\vec{R}
+P_{c}^{\lambda}\mathcal{G}\|_{ H^{2}}\leq
c(T_{0}+t)^{-\frac{3}{2}}[Z\mathcal{R}_{1}+Z^{3}].$$ This together
with (~\ref{eq:JNForR2}) yields
$$
\begin{array}{lll}
\|P_{c}^{\lambda_{1}}\vec{R}(t)\|_{\infty} &\leq&
c(T_{0}+t)^{-1}[T_{0}\|\vec{R}(0)\|_{1}+T_{0}\|\vec{R}(0)\|_{ H^{2}}+Z^{2}+\mathcal{R}_{4}^{2}\mathcal{R}_{2}\\
& &
+T_{0}^{-\frac{1}{2}}[Z^{3}+Z\mathcal{R}_{1}+\mathcal{R}_{1}^{2}]].
\end{array}
$$
This implies the proposition.
\begin{flushright}
$\square$
\end{flushright}

\subsection{Estimate for $\mathcal{R}_{5}(T):=\displaystyle\max_{t\leq
T}\frac{(T_{0}+t)^{\frac{1}{2}}}{\log(T_{0}+t)}\|\vec{R}(t)\|_{3}$}

\begin{proposition}\label{prop:estR5}
\begin{equation}\label{EstR5}
\mathcal{R}_{5}\leq c\ \left[\ T_{0}\left(\ \|\vec{R}(0)\|_{1}+
\|\vec{R}(0)\|_{ H^{2}}\ \right)+Z^{2}+T_{0}^{-\frac{1}{2}}[\mathcal{R}_{5}^{2}\mathcal{R}_{2}+Z^{3}+Z\mathcal{R}_{1}+\mathcal{R}_{1}^{2}+\mathcal{R}_{2}^{2}]\ \right].
\end{equation}
\end{proposition}
\nit We use the following lemma to prepare for the proof.
\begin{lemma}
\begin{equation}\label{eq:JNforR5}
\|JN(\vec{R},z)\|_{\frac{3}{2}}\leq
c(T_{0}+t)^{-1}Z^{2}+c(T_{0}+t)^{-\frac{3}{2}}\left[\ \mathcal{R}_{5}^{2}\mathcal{R}_{2}+Z^{3}+Z\mathcal{R}_{1}+\mathcal{R}_{1}^{2}+\mathcal{R}_{2}^{2}\ \right].
\end{equation}
\end{lemma}
\begin{proof} As in the proof of Lemma ~\ref{LM:JNNonlinear} we decompose $JN$ into the localized term $JN(\vec{R},z)-NonLoc$ and non-localized term $NonLoc.$ The estimate of the first term is similar to that of Lemma ~\ref{LM:JNNonlinear}, hence omitted.
The nonlocal term $NonLoc$ defined in (~\ref{eq:NonlocDef}) admits
the estimate
$$\|NonLoc\|_{\frac{3}{2}}\leq c\ 
\left(\ \int |\vec{R}|^{\frac{5}{4}} \right)^{\frac{2}{3}}\ \leq\ c\ \|\vec{R}\|_{3}^{2}\|\vec{R}\|_{\infty}.$$

By using the definitions of estimating functions on all the terms above we have (~\ref{eq:JNforR5}).
\end{proof}
\textbf{Proof of Proposition ~\ref{prop:estR5}} By Estimate
(~\ref{L3Est}) for $d=3$ and Equation
(~\ref{estimateonO3}) we have that
\begin{equation}
\begin{array}{lll}
& &\|P_{c}^{\lambda_{1}}\vec{R}(t)\|_{3}\\
&\leq& \|e^{tL(\lambda_{1})}P_{c}^{\lambda_{1}}\vec{R}(0)\|_{3}
+\int_{0}^{t}\|e^{(t-s)L(\lambda_{1})}P_{c}^{\lambda_{1}}[O_{1}(s)\vec{R}+P_{c}^{\lambda}\mathcal{G}-
P_{c}^{\lambda}JN(\vec{R},z)]\|_{3}ds\\
&\leq&c(1+t)^{-\frac{1}{2}}(\|\vec{R}(0)\|_{1}+\|\vec{R}(0)\|_{ H^{2}})\\
&
&+c\int_{0}^{t}(t-s)^{-\frac{1}{2}}\|O_{1}(s)\vec{R}+P_{c}^{\lambda}\mathcal{G}\|_{\frac{3}{2}}ds+
\int_{0}^{t}(t-s)^{-\frac{1}{2}}\|P_{c}^{\lambda}JN(\vec{R},z)\|_{\frac{3}{2}}ds.
\end{array}
\end{equation}
By the properties of $O_{1}$ (Equation (~\ref{EstO1})) and
$\mathcal{G}$ (Equation (~\ref{RAfProj})) we have
$$\|O_{1}(s)\vec{R}+P_{c}^{\lambda}\mathcal{G}\|_{\frac{3}{2}}\leq c(T_{0}+t)^{-\frac{3}{2}}[Z\mathcal{R}_{1}+Z^{3}].$$ This
together with (~\ref{eq:JNForR2}) and (~\ref{log-conv-est})
implies
\begin{align}
\|P_{c}^{\lambda_{1}}\vec{R}(t)\|_{3} &\leq
c(T_{0}+t)^{-\frac{1}{2}}\log(T_{0}+t)[T_{0}^{\frac{1}{2}}\|\vec{R}(0)\|_{1}
+T_{0}\|\vec{R}(0)\|_{ H^{2}}\nn\\
& +Z^{2}
+T_{0}^{-\frac{1}{2}}[\mathcal{R}_{5}^{2}\mathcal{R}_{2}+Z^{3}+Z\mathcal{R}_{1}+\mathcal{R}_{1}^{2}+\mathcal{R}_{2}^{2}]].
\end{align}
This estimate and the definition of $\mathcal{R}_{5}$ yield the
proposition. 
\begin{flushright}
$\square$
\end{flushright}
\subsection{Estimate for $\mathcal{R}_{3}(T):=\displaystyle\max_{t\leq
T}(T_{0}+t)^{\frac{7}{5}}\|\langle x\rangle^{-\nu}\tilde{R}(t)\|_{2}$}

\begin{proposition}\label{controlM3} Let the constant $\nu$ the same as in
(~\ref{eq:SingularEst})-(~\ref{second}) with $d=3$. Then
\begin{equation}\label{EstR3}
\mathcal{R}_{3}\leq
c\ \left[\ T_{0}^{\frac{3}{2}}\ \left(\ \|\langle x\rangle^{\nu}\vec{R}(0)\|_{2}+\ |z(0)|^{2}\ \right) \right]+cT_{0}^{-\frac{1}{20}}
(Z^{3}
+Z\mathcal{R}_{3}+Z\mathcal{R}_{1}+\mathcal{R}_{5}^{3}+R_{2}^{2}\mathcal{R}_{4}).
\end{equation}
\end{proposition} As usual we estimate the nonlinear
term $N_{2}(\vec{R},z).$
\begin{lemma}\label{LM:NonlinearityEst}
\begin{equation}\label{eq:NonlinearEst}
\begin{array}{lll}
&
&\int_{0}^{t}\|\langle x\rangle^{-\nu}e^{(t-s)L(\lambda)}P_{c}^{\lambda_{1}}P_{c}^{\lambda}N_{2}(\vec{R},z)\|_{2}ds\\
&\leq &
c\ (T_{0}+t)^{-\frac{7}{5}}\ T_{0}^{-\frac{1}{20}}\ \left[\ Z^{3}+Z\mathcal{R}_{1}+\mathcal{R}_{5}^{3}+\mathcal{R}_{2}^{2}\mathcal{R}_{4}\ \right].
\end{array}
\end{equation}
\end{lemma}
\begin{proof}
We start with the function $N_{2}$. Recall that $N_{2} =Loc+NonLoc$ in (~\ref{eq:N2Decomposition}) and the estimate of $Loc$ after that.

The nonlocal term $NonLoc$ defined in (~\ref{eq:NonlocDef}) admits
the estimate
$$\|NonLoc\|_{1}+\|NonLoc\|_{2}\leq c\ \left[\ \|\vec{R}\|_{3}^{3}+\|\vec{R}\|_{6}^{3}\ \right]\leq c\ \left[\ \|\vec{R}\|_{3}^{3}+\|\vec{R}\|_{\infty}^{2}\|\vec{R}\|_{2}\ \right].$$ 
By the definition of estimating function we have $$
\begin{array}{lll}
\|NonLoc\|_{1}+\|NonLoc\|_{2}&\leq& c(T_{0}+t)^{-\frac{3}{2}}\ \left[
\log(T_{0}+t)\ \right]^{3/2}\mathcal{R}_{5}^{3}+(T_{0}+t)^{-2}\mathcal{R}_{2}^{2}\mathcal{R}_{4}\ \\
&\leq &c\ (T_{0}+t)^{-\frac{7}{5}}T_{0}^{-\frac{1}{20}}\ \left[ \mathcal{R}_{5}^{3}+\mathcal{R}_{2}^{2}\mathcal{R}_{4}\ \right].
\end{array}
$$

Finally we prove Equation (~\ref{eq:NonlinearEst}). By the propagator estimates (~\ref{lastestimate}) and (~\ref{finalestimate}) we have $$
\begin{array}{lll}
& &\int_{0}^{t}\|\langle x\rangle^{-\nu}e^{(t-s)L(\lambda)}P_{c}^{\lambda_{1}}P_{c}^{\lambda}N_{2}(\vec{R},z)\|_{2}ds\\
&\leq& c\int_{0}^{t}(1+t-s)^{-\frac{3}{2}}\ \left[\ \|NonLoc(\vec{R},z)\|_{1}+\|NonLoc(\vec{R},z)\|_{2}+\|\langle x\rangle^{\nu}Loc\|_{2}\ \right]\ ds.
\end{array}
$$ This together with the estimates of $Loc$ and $NonLoc$ above yields (~\ref{eq:NonlinearEst}).
\end{proof}
\textbf{Proof of Proposition ~\ref{controlM3}} By the same
techniques as in deriving Equation (~\ref{estimater21r22})
we have the following equation
\begin{equation}\label{estimateg1g2}
\begin{array}{lll}
\D_t P_{c}^{\lambda_{1}}\tilde{R}&=&L(\lambda_{1})P_{c}^{\lambda_{1}}\tilde{R}+(\dot\gamma+\lambda-\lambda_{1})i(P_{+}-P_{-})\tilde{R}\\
&
&+P(z,\bar{z})\tilde{R}+P_{c}^{\lambda_{1}}S_{2}(z,\bar{z})+P_{c}^{\lambda_{1}}P_{c}^{\lambda}N_{2}(\vec{R},z),
\end{array}
\end{equation} where the operator
$P(z,\bar{z})$ is defined as
$$P(z,\bar{z}):=P_{c}^{\lambda_{1}}M_{2}(z,\bar{z})-(\dot{\gamma}+\lambda-\lambda_{1})i(P_{+}-P_{-})+P^{\lambda_{1}}_{c}(L(\lambda)-L(\lambda_{1})),$$
and the terms $S_{2}(z,\bar{z})$, $M_{2}(z,\bar{z})$ and
$P_{c}^{\lambda}N_{2}(\vec{R},z)$ are defined in Theorem
~\ref{GOLD:maintheorem}.

Rewrite Equation (~\ref{estimateg1g2}) in the integral form by
the Duhamel principle to obtain
\begin{equation}\label{formulag}
\begin{array}{lll}
\|\langle x\rangle^{-\nu}P_{c}^{\lambda_{1}}\tilde{R}(t)\|_{2}
&\leq&\|\langle x\rangle^{-\nu}e^{tL(\lambda_{1})}P_{c}^{\lambda_{1}}\tilde{R}(0)\|_{2}
+\int_{0}^{t}\|\langle x\rangle^{-\nu}e^{(t-s)L(\lambda_{1})}\\
&
&\times[P(z,\bar{z})\tilde{R}+P_{c}^{\lambda_{1}}S_{2}(z,\bar{z})+P_{c}^{\lambda_{1}}P_{c}^{\lambda}N_{2}(\vec{R},z)]\|_{2}ds.
\end{array}
\end{equation}
For the left hand side we claim that
\begin{equation}\label{estimateg1g20}
\|\langle x\rangle^{-\nu}e^{tL(\lambda_{1})}P_{c}^{\lambda_{1}}\tilde{R}(0)\|_{2}\leq
c\ (1+t)^{-\frac{3}{2}}(\|\langle x\rangle^{\nu}\vec{R}(0)\|_{2}+|z(0)|^{2}).
\end{equation} Indeed, recall that
$$\tilde{R}=\vec{R}-\sum_{
 m+n=2 }R_{m,n}
$$ with $R_{m,n}$ defined in (~\ref{eq:Rform}).
Therefore, displaying the time-dependent of $\tilde{R}$, $\lambda$
and $z,$
$$
\|\langle x\rangle^{-\nu}e^{tL(\lambda_{1})}P_{c}^{\lambda_{1}}\tilde{R}(0)\|_{2}\leq
\|\langle x\rangle^{-\nu}e^{tL(\lambda_{1})}P_{c}^{\lambda_{1}}\vec{R}(0)\|_{2}+\displaystyle\sum_{
m+n=
2}\|\langle x\rangle^{-\nu}e^{tL(\lambda_{1})}P_{c}^{\lambda_{1}}R_{m,n}(0)\|_{2}.
$$
By (~\ref{eq:SingularEst}) and the fact that the notation
$R_{m,n}$ is the summation of terms of order $|z|^{2}$ we have
$$\|\langle x\rangle^{-\nu}e^{tL(\lambda_{1})}P_{c}^{\lambda_{1}}R_{m,n}(0)\|_{2}\leq
c|z(0)|^{2}(1+t)^{-\frac{3}{2}}.$$ This together with the estimate
$$\|\langle x\rangle^{-\nu}e^{tL(\lambda_{1})}P_{c}^{\lambda_{1}}\vec{R}(0)\|_{2}\leq c(1+t)^{-\frac{3}{2}}
\|\langle x\rangle^{\nu}\vec{R}(0)\|_{2}$$ implies (~\ref{estimateg1g20}).

Use (~\ref{eq:SingularEst}) on the right hand side of Equation
(~\ref{formulag}) to obtain
\begin{equation}\label{secondterm}
\begin{array}{lll}
&
&\int_{0}^{t}\|\langle x\rangle^{-\nu}e^{(t-s)L(\lambda_{1})}[P(z,\bar{z})\tilde{R}+P_{c}^{\lambda_{1}}S_{2}(z,\bar{z})
+P_{c}^{\lambda_{1}}P_{c}^{\lambda}N_{2}(\vec{R},z)]\|_{2}ds\\
&\leq&\int_{0}^{t}(1+t-s)^{-\frac{3}{2}}(\|\langle x\rangle^{\nu}P(z,\bar{z})\tilde{R}\|_{2}+\|N_{2}(\vec{R},z)\|_{1}
+\|N_{2}(\vec{R},z)\|_{2})ds\\
&
&+\int_{0}^{t}\|\langle x\rangle^{-\nu}e^{(t-s)L(\lambda_{1})}P_{c}^{\lambda_{1}}S_{2}(z,\bar{z})\|_{2}ds.
\end{array}
\end{equation}
We estimate these terms in detail:
\begin{enumerate}
\item[(A)] By the definition of $S_{2}(z,\bar{z})$ in Equation
(~\ref{eq:unusual}) and Estimate (~\ref{eq:SingularEst}) with $d=3$
we have that
$$
\begin{array}{lll}
|\int_{0}^{t}\|\langle x\rangle^{-\nu}e^{(t-s)L(\lambda_{1})}P_{c}^{\lambda_{1}}S_{2}(z,\bar{z})\|_{2}ds|&\leq&
c\int_{0}^{t}(1+t-s)^{-\frac{3}{2}}(T_{0}+s)^{-\frac{3}{2}}ds Z^{3}\\
&\leq&c(T_{0}+t)^{-\frac{3}{2}}Z^{3}.
\end{array}
$$
\item[(B)] By the definition of $P(z,\bar{z})$ and the estimate of $M_{2}(z,\bar{z})$ in Equation (~\ref{eq:M2})
$$
\begin{array}{lll}
\|\langle x\rangle^{\nu}P(z(s),\bar{z}(s))\tilde{R}(s)\|_{2}&\leq&
c|z|\|\langle x\rangle^{-\nu}\tilde{R}(s)\|_{2}\\
&\leq& c(T_{0}+s)^{-\frac{19}{20}}Z\mathcal{R}_{3}.
\end{array}
$$ Hence by ~\eqref{conv-est} $$
\begin{array}{lll}
& &\int_{0}^{t}(1+t-s)^{-\frac{3}{2}}\|\langle x\rangle^{\nu}P(z(s),\bar{z}(s))\tilde{R}\|_{2} ds\\
&\leq& c\int_{0}^{t}(1+t-s)^{-\frac{3}{2}}(T_{0}+s)^{-\frac{7}{5}}dsT_{0}^{-\frac{1}{20}}Z\mathcal{R}_{3}\\
&\leq &c(T_{0}+t)^{-\frac{7}{5}}T_{0}^{-\frac{1}{20}}Z\mathcal{R}_{3}\end{array}
$$
\end{enumerate} These together with (~\ref{eq:NonlinearEst}) imply $$
\begin{array}{lll}
\|\langle x\rangle^{-\nu}P_{c}^{\lambda_{1}}\tilde{R}\|_{2}
&\leq&c(1+t)^{-\frac{3}{2}}[\|\langle x\rangle^{\nu}\vec{R}(0)\|_{2}+|z|^{2}(0)]\\
& &+
c(T_{0}+t)^{-\frac{7}{5}} T_{0}^{-\frac{1}{20}}(Z\mathcal{R}_{1}+Z\mathcal{R}_{3}+Z^{3}+\mathcal{R}_{5}^{3}+\mathcal{R}_{2}^{2}\mathcal{R}_{4})\\
&\leq
&c(T_{0}+t)^{-\frac{7}{5}}[T_{0}^{\frac{7}{5}}\|\langle x\rangle^{\nu}\vec{R}(0)\|_{2}
+T_{0}^{\frac{7}{5}}|z|^{2}(0)\\
& &+T_{0}^{-\frac{1}{20}}(Z\mathcal{R}_{1}+Z\mathcal{R}_{3}+Z^{3}+\mathcal{R}_{5}^{3}+\mathcal{R}_{2}^{2}\mathcal{R}_{4})].
\end{array}
$$
This implies (~\ref{EstR3}). \begin{flushright}
$\square$
\end{flushright}

\begin{proposition}\label{controlR4}
\begin{equation}\label{EstR4}
\mathcal{R}_{4}^{2}\leq
\|\vec{R}(0)\|^{2}_{ H^{2}}+cT_{0}^{-1}[\mathcal{R}_{1}^{2}+Z^{2}\mathcal{R}_{1}+Z^{2}\mathcal{R}_{1}^{2}+\mathcal{R}_{4}^{2}\mathcal{R}_{2}^{2}].
\end{equation}
\end{proposition}
Before the proof we estimate the nonlinear terms.
\begin{lemma}
\begin{equation}\label{eq:K3}
|\langle
(-\Delta+1)P_{c}^{\lambda}JN(\vec{R},z),(-\Delta+1)\vec{R}\rangle|\leq
c(T_{0}+t)^{-2}[Z^{2}\mathcal{R}_{1}+\mathcal{R}_{4}^{2}\mathcal{R}_{2}^{2}].
\end{equation}
\end{lemma}
\begin{proof}
As in Lemma ~\ref{LM:JNNonlinear} we decompose $J\vec{N}$ into the localized term $Loc$
and the non-localized $NonLoc$ defined in (~\ref{eq:NonlocDef}). The Localized part satisfies the estimate $$|\langle
(-\Delta+1)Loc,(-\Delta+1)\vec{R}\rangle|\leq c[|z|^{2}+\|\langle x\rangle^{-\nu}\vec{R}\|_{2}^{2}].$$

By the definition of $NonLoc$ in (~\ref{eq:NonlocDef}) we obtain
$$ |\langle (-\Delta+1)NonLoc, (-\Delta+1)\vec{R}\rangle|\leq
c\|(-\Delta+1)\vec{R}\|_{2}^{2}\|\vec{R}\|_{\infty}^{2}.$$

This together with the definitions of estimating functions implies (~\ref{eq:K3}).
\end{proof}
\begin{proof}
By Equation (~\ref{RAfProj}), we have
$$
\begin{array}{lll}
& &\partial_{t}\langle
(-\Delta+1)\vec{R},(-\Delta+1)\vec{R}\rangle\\
&=&\langle
(-\Delta+1)\frac{d}{dt}\vec{R},(-\Delta+1)\vec{R}\rangle+\langle(-\Delta+1)\vec{R},(-\Delta+1)\frac{d}{dt}\vec{R}\rangle\\
&=&\displaystyle\sum_{n=1}^{4}K_{n}
\end{array}
$$ with $K_{n}$, $n=1,2,3,4,$ defined as $$K_{1}:=\langle
(-\Delta+1)(L(\lambda)+\dot\gamma
J)\vec{R},(-\Delta+1)\vec{R}\rangle +\langle (-\Delta+1)\vec{R},
(-\Delta+1)(L(\lambda)+\dot\gamma J)\vec{R}\rangle;$$
$$K_{2}:=\dot\lambda\langle
(-\Delta+1)P_{c\lambda}\vec{R},(-\Delta+1)\vec{R}\rangle+\dot\lambda\langle(-\Delta+1)\vec{R},(-\Delta+1)P_{c\lambda}\vec{R}\rangle;$$
$$K_{3}:=-\langle
(-\Delta+1)P_{c}^{\lambda}JN(\vec{R},z),(-\Delta+1)\vec{R}\rangle-\langle
(-\Delta+1)\vec{R},(-\Delta+1)P_{c}^{\lambda}JN(\vec{R},z)\rangle;$$
$$K_{4}:=\langle (-\Delta+1)P_{c}^{\lambda}\mathcal{G},
(-\Delta+1)\vec{R}\rangle+\langle
(-\Delta+1)\vec{R},(-\Delta+1)P_{c}^{\lambda}\mathcal{G}\rangle.$$

Recall the definition of the operator $L(\lambda)$ in
(~\ref{eq:LinOpera}). By the observation $J^{*}=-J$ and the fact
that $JL(\lambda)$ is self-adjoint we cancel all the nonlocal terms
in $K_{1}$:
$$
|K_{1}| \leq  c\|\langle x\rangle^{-\nu}\vec{R}\|^{2}_{ H^{2}}\leq
c(T_{0}+t)^{-2}\mathcal{R}_{1}^{2}.
$$
By observing that $|\dot\lambda|=O(|z|^{2})$ and $P_{c\lambda}\vec{R}$ is localized we have that
$$|K_{2}|\leq c|z(t)|^{2}\ \|\langle x\rangle^{-\nu}\vec{R}(t)\|^{2}_{ H^{2}}\leq
c(T_{0}+t)^{-2}Z^{2}(t)\mathcal{R}_{1}^{2}(t).$$ By (~\ref{eq:K3})
we have
$$|K_{3}|\leq
c(T_{0}+t)^{-2}[Z^{2}\mathcal{R}_{1}^{2}+\mathcal{R}_{4}^{2}\mathcal{R}_{2}^{2}].$$
 By the property of
$P_{c}^{\lambda}\mathcal{G}$ in (~\ref{RAfProj}) we have
$$|K_{4}|\leq c|z|^{2}\|\langle x\rangle^{-\nu}\vec{R}\|_{ H^{2}}\leq c(T_{0}+t)^{-2}Z^{2}\mathcal{R}_{1}.$$
\nit Collecting all the estimates above to obtain
$$
|\frac{d}{dt}\langle
(-\Delta+1)\vec{R},(-\Delta+1)\vec{R}\rangle|\leq
c(T_{0}+t)^{-2}[\mathcal{R}_{1}^{2}+Z^{2}\mathcal{R}_{1}+Z^{2}\mathcal{R}_{1}^{2}+\mathcal{R}_{4}^{2}\mathcal{R}_{2}^{2}].
$$
After integrating the equation above from $0$ to $t$ we have
(~\ref{EstR4}).
\end{proof}
\subsection{Estimate for $Z(T)=\displaystyle\max_{t\leq T}(T_{0}+t)^{\frac{1}{2}}|z(t)|$}\label{se:zdecay}
Recall that by (FGR), eqn (\ref{eq:FGR}), 
 $z^*[Z(z,\bar{z})+Z^*(z,\bar{z})]z\ \ge\ 
C|z|^{4}$.
\begin{proposition}\label{Ydecay}
There exists an order one constant $m>0$, such that if  
\begin{equation}
m<T_0<|z(0)|^{-2},\ \ \ {\rm then}\label{mT0z0}
\end{equation}
\begin{equation} \label{controlX}
Z(T)\ \le\ 1\ +\ \frac{K}{T_0^{2\over5}}\ Z(T)\ \left(\ Z(T)+ \cR_1^2(T)+\cR_2^2(T)+Z(T)\cR_3(T)\ \right)
\end{equation}
\end{proposition}
\begin{proof}
By the equation (~\ref{eq:detailedDescription}) we have that
\begin{equation}\label{prericatti}
\frac{d}{dt}|z|^{2}=-z^*[Z(z,\bar{z})+Z^*(z,\bar{z})]z+Re(\bar{z}\ Remainder(t))
\end{equation}
which can be transformed into a Riccati inequality
\begin{equation}\label{eq:Reccati}
\begin{array}{lll}
\partial_{t}|z(t)|^{2} \leq -C|z(t)|^{4}\ +\ 2\ |z(t)|\ |Remainder(t)|.
\end{array}
\end{equation}
By  (\ref{remainder}), 
\begin{align}
|z(t)|\  |Remainder(t)| \le \ \frac{c}{(T_0+t)^{2+\delta}}\ Z(T) \left(\ Z(T)+\cR_1^2(T)+\cR_2^2(T)+Z(T)\cR_3(T)\ \right),
\end{align}
where $\delta=2/5$. We now use the following Lemma, proved in the Appendix ~\ref{subSEC:Riccati}.
\begin{lemma}\label{LM:RiccatiEst}
Suppose that $z(t)$ is any function satisfying the equation
\begin{equation}\label{eq:Riccati}
\begin{array}{lll}
\partial_{t}|z(t)|^{2}&\leq &-|z(t)|^{4}+g(t),\ \ 
z(0)\ =\ z_0
\end{array}
\end{equation}  where $g(t)$ is a function satisfying the estimate 
\begin{equation}
|g(t)|\ \leq\ c_\#\ (T_{0}+t)^{-2-\delta}
\label{gbound}
\end{equation}
 with constants $c_\#, \  \delta>0.$
 Then, there exists $K>0$, independent of $T_0$ and $c_\#$, such that if $c_\# T_0^{-\delta}$ is sufficiently small, then
the function $z(t)$ in (~\ref{eq:Riccati}) admits the bound 
\begin{equation}
|z(t)|\ \leq\ \frac{1+\ K\ c_\#\ T_0^{-\delta}}{  (\kappa+t)^{\frac{1}{2}} }
\label{wbound}
\end{equation}
where $\kappa=\min\{T_{0},|z_{0}|^{-2}\}$.
\end{lemma}
We now apply this Lemma, choosing 
\begin{equation}
m<T_0<|z(0)|^{-2}
\end{equation} as in (~\ref{mT0z0})
 where $m$ is an order one positive constant. Then,
\begin{equation}
Z(T)\ \le\ 1\ +\ \frac{K}{T_0^{2\over5}}\ Z(T)\ \left(\ Z(T)+ \cR_1^2(T)+\cR_2^2(T)+Z(T)\cR_3(T)\ \right)
\nn\end{equation}

\end{proof}

\subsection{Closing the Estimates; Completing the Proof of the Main Theorem ~\ref{THM:MainTheorem}}\label{subsec:proofsubmaintheorem}

We seek to obtain $T$-independent bounds on ${\cal R}_j(T)$ and $Z(T)$, 
 defined in (\ref{majorant}). 
  This will be achieved by choosing  the parameter $T_0$ sufficiently large, appearing in the norm definitions, and the data $R(0)$ sufficiently small, with $T_0$ and $R(0)$ related in a manner to be specified.\\
 Define
\begin{equation}\label{defineS}
M(T):=\displaystyle\sum_{n\not=4}\mathcal{R}_{n}(T),\ S:=T_{0}^{\frac{3}{2}}(\|\vec{R}(0)\|_{ H^{2}}+\|\langle x\rangle^{\nu}\vec{R}(0)\|_{2}),
\end{equation} where, recall the definition of $T_{0}=|z^{(0)}|^{-1}$ in
(~\ref{Todef}). By the conditions on the datum (~\ref{InitCond})
we have that $\mathcal{R}_{4}(0)$ is small and $M(0)$ and $Z(0)$ are
bounded. 

Recall the estimates of $\mathcal{R}_{n},\ n=1,2,3,4,5$ and $Z$ in Equations (~\ref{EstR1}), (~\ref{EstR2}),
(~\ref{EstR3}), (~\ref{EstR4}), (~\ref{EstR5})  and (~\ref{controlX}). By plugging (~\ref{controlX}) ~\eqref{EstR4}, the estimate of $Z$ and $\mathcal{R}_{4}$, into (~\ref{EstR1}), (~\ref{EstR2}) and (~\ref{EstR5}) we
obtain 
\begin{equation}\label{eq:IMF}
\begin{array}{lll}
M(T)&\leq &c(S+1)+(R_{4}(T)+T_{0}^{-\frac{1}{20}})P(M(T),Z(T)),\\
Z(T)\ &\le&\ 1\ +\ T_0^{-\frac{1}{20}}P(M(T),Z(T)),\\
\mathcal{R}_{4}^{2}(T)&\leq &\|\vec{R}(0)\|^{2}_{ H^{2}}+T_{0}^{-1}P(M(T),Z(T))
\end{array}
\end{equation} where $P(x,y)>0$ is a polynomial of variables $x$ and $y.$
Using an implicit-function-theorem type argument (see below) we have that if
$S$ and $M(0)$ are bounded, then 
\begin{equation}\label{eq:conseIMF}
M(T)+Z(T)\leq \mu(S)\ \text{and}\ \mathcal{R}_{4}\ll 1,
\end{equation} where
$\mu$ is a bounded function for $S$ bounded.
By the definitions of  $\cR_j(T)$  and $Z(T)$ there exists some constant $c$ such that
\begin{equation}\label{equa619a}
\|\langle x\rangle^{-\nu}\vec{R}(t)\|_{2},\ \|\vec{R}(t)\|_{\infty}\leq
c(T_{0}+t)^{-1},\ |z(t)|\leq c(T_{0}+t)^{-\frac{1}{2}}
\end{equation} which is Statement (B) in Theorem
~\ref{THM:MainTheorem}. 

By the bound of $Remainder$ in (~\ref{remainder}) and the
estimates (~\ref{equa619a}) we have $$|Remainder|\leq c(T_{0}+t)^{-\frac{19}{5}}$$ which
together with (~\ref{eq:detailedDescription}) implies Statement (A). 

The convergence of $\lambda$ comes from (~\ref{ExpanLambda}) and the fact $Remainder$ is integrable at $\infty.$
\begin{flushright}
$\square$
\end{flushright}
In the following we prove \eqref{eq:IMF} implies ~\eqref{eq:conseIMF} by using implicit function theorem. For the other methods we refer to ~\cite{SoWe99,TsaiYau02,BuSu03,SoWe04,TsaiYau02-2,Tsai03,CuMi}. First we transform the inequalities by taking square root of the third equation of  ~\eqref{eq:IMF} and plugging it into the first one, then
there exists a fixed polynomial $P(x,y)$ with positive coefficients such that \begin{equation}\label{eq:IMF2}
\begin{array}{lll}
M(T)&\leq &c(S+1)+(\|\vec{R}(0)\|_{ H^{2}}+T_{0}^{-\frac{1}{20}})P(M(T),Z(T)),\\
Z(T)\ &\le&\ 1\ +\ T_0^{-\frac{1}{20}}P(M(T),Z(T)),\\
\mathcal{R}_{4}(T)&\leq &\|\vec{R}(0)\|_{ H^{2}}+T_{0}^{-\frac{1}{20}}P(M(T),Z(T)).
\end{array}
\end{equation} In what follows we use this equation instead of ~\eqref{eq:IMF}.
 Define a vector function $F_{\epsilon,\delta}(\tilde{M},\tilde{Z})$ as $$F_{\epsilon,\delta}(\tilde{M},\tilde{Z}):=(F^{(1)}_{\epsilon}(\tilde{M},\tilde{Z}),F^{(2)}_{\epsilon}(\tilde{M},\tilde{Z}),F^{(3)}_{\epsilon,\delta}(\tilde{M},\tilde{Z}))$$ with $$F^{(1)}_{\epsilon,\delta}(\tilde{M},\tilde{Z}):=c(S+1)+(\delta+\epsilon)P(\tilde{M},\tilde{Z})$$ $$F^{(2)}_{\epsilon}(\tilde{M},\tilde{Z}):=1+\epsilon P(\tilde{M},\tilde{Z})\ \text{and}\ F^{(3)}_{\epsilon,
\delta}:=\delta+\epsilon P(\tilde{M},\tilde{Z}).$$ Immediately we can see that $M_{0}=c(1+S),\ Z_{0}=1$ and $\tilde{R}_{0}=0$ is a solution to the equation $$(M_{0},Z_{0},R_{0})=F_{0,0}(M_{0},Z_{0},R_{0}).$$ Define a closed set $$\Sigma:=[0,2c(S+1)]\times [0,2]\times [0,1].$$ Now we have the result.
\begin{lemma}
There exists a $\delta_{0}\geq 0$ such that if $\epsilon,\delta\in [0,\delta_{0}]$ then 
\begin{equation}\label{FixPoint}
(\tilde{M},\tilde{Z},\tilde{R})=F_{\epsilon,\delta}(\tilde{M},\tilde{Z},\tilde{R})\end{equation} has a unique solution in $\Sigma$, moreover for any continuous functions $M,Z,\mathcal{R}:\mathbb{R}^{+}\rightarrow \mathbb{R}^{+}$ satisfying $$(M(0),Z(0),\mathcal{R}(0))\leq (\tilde{M},\tilde{Z},\tilde{R})\ \text{and} \ (M(t),Z(t),R(t)) \leq F_{\epsilon,\delta}(M(t),Z(t),\mathcal{R}(t))$$ we have for any time $t$, 
\begin{equation}\label{eq:Maximum}
(M(t),Z(t),\mathcal{R}(t))\leq (\tilde{M},\tilde{Z},\tilde{R}).
\end{equation}
\end{lemma}
\begin{proof}
The proof of existence and uniqueness of the solution is not difficult by observing $$\|(\partial_{M} F_{\epsilon,\delta}(M,Z),\partial_{Z}F_{\epsilon,\delta}(M,Z),\partial_{\mathcal{R}}F_{\epsilon,\delta}(M,Z))\|\leq c(\delta+\epsilon)$$ if $(M,Z,\mathcal{R})\in \Sigma.$ Hence by implicit function theorem we have that if $c(\epsilon+\delta)\leq \frac{1}{2}$ there exists a unique solution to ~\eqref{FixPoint}.

We next prove ~\eqref{eq:Maximum} by contradiction. Suppose that ~\eqref{eq:Maximum} fails at time $t$. Since $(M(t),Z(t),\mathcal{R}(t))$ is continuous, there exists a time $t_{1}\leq t$ such that $(M(t_{1}),Z(t_{1}),\mathcal{R}(t_{1}))\in \Sigma$ and ~\eqref{eq:Maximum} does not hold. Without loss of generality we assume $t=t_{1}.$ Then by subtracting the inequality for $(M(t),Z(t),\mathcal{R}(t))$ by ~\eqref{FixPoint} we get 
$$
\begin{array}{lll}
M(t)-\tilde{M}&\leq & (\delta+\epsilon)[K_{1}(M(t)-\tilde{M})+K_{2}(Z(t)-\tilde{Z})]\\
Z(t)-\tilde{Z},\ \mathcal{R}(t)-\tilde{R}&\leq& \epsilon K_{3}(M(t)-\tilde{M})+\epsilon K_{4}(Z(t)-\tilde{Z})
\end{array}
$$ for some $K_{n},\ n=1,2,3,4,$ depending on $(M(t),Z(t),\mathcal{R}(t))$ and $(\tilde{M},\tilde{Z},\tilde{R})$. By the fact that 
\begin{equation}
(\tilde{M},\tilde{Z},\tilde{R}),\ (M(t),Z(t),\mathcal{R}(t))\in \Sigma
\nn\end{equation}
 and $P(x,y)$ is a polynomial with positive coefficient we have that $K_{n},\ n=1,2,3,4$ are {\bf{positive}} and bounded. By these inequalities and the fact $0\leq \epsilon,\delta\ll 1$ we derive ~\eqref{eq:Maximum}. This contradicts our assumption. Thus ~\eqref{eq:Maximum} holds for any time $t\geq 0.$
\end{proof}
\section{Summary and Discussion}
We have extended the asymptotic stability / scattering theory of solitary waves of the nonlinear Schr\"odinger / Gross-Pitaevskii (NLS / GP) equation to the important case, where the linearized dynamics about the Lyapunov stable has degenerate neutral modes. This is the prevalent case in situation, where the equation is invariant under a nontrivial symmetry. We construct  a class of double-well potentials to which the theory applies. The current theory, as in all previous work on soliton scattering in systems with non-trivial neutral modes, requires a Fermi Golden Rule {\bf (FGR)} non-degeneracy hypothesis. The analytical verification of this hypothesis for specific or generic NLS/GP systems is an open question. Numerical experiments for the time-dependent NLS/GP equations, in which decay rates of neutral modes is measured,  are consistent with the typical validity  {\bf (FGR)} non-degeneracy hypothesis.

We conclude by mentioning an interesting direction for further exploration.

\nit {\it Semiclassical limits and higher order nonlinear Fermi Golden Rule:}\  A problem of great interest is NLS / GP on $\R^d$ in the semi-classical limit:
\begin{equation}\label{eq:FNNLS}
\begin{array}{lll}
i\partial_{t}\psi&=&-\Delta\psi+V(hx)\psi-f(|\psi|^{2})\psi,\ \ 0<h<<1\\
\psi(x,0)&=&\psi_{0}(x)
\end{array}
\end{equation}
The nonlinearity is taken to be focusing (attractive) but subcritical. Using the Lyapunov-Schmidt method it has been shown in~\cite{FW86,Oh88,ABC96} that for $h$ sufficiently small a soliton concentrated at a nondegenerate critical point  of $V$ can be constructed. The soliton, constructed in this manner, is soliton of the translation invariant nonlinear Schr\"odinger equation, scaled to be highly concentrated about the critical point of $V$. 
Therefore, the linearized operator $JH^h(\lambda)$ is expected to have spectrum, quite closely related to the linearization about the translation invariant NLS soliton. If the soliton is concentrated near a minimum of $V$, then it is Lyapunov stable \cite{Oh88}, and therefore the spectrum of $JH^h(\lambda)$ is a subset of the imaginary axis. As we have seen for NLS/GP, there is a two-dimensional generalized eigenspace corresponding to an eigenvalue zero. $h$ being small, implies that the $2\times d$ {\it zero modes}, associated with the translation symmetry ($\psi(x,t)\mapsto \psi(x+x_0,t)$) and Galilean symmetry ($\psi(x,t)\mapsto e^{iv\cdot(x-vt)}\ \psi(x-2vt,t)$) perturb to two complex conjugate eigenvalues, each degenerate, of multiplicity $N=d$. Although we expect semiclassical, highly localized solitons to be asymptotically stable and for the degenerate neutral modes to damp by resonant radiation damping, as elucidated in this article,  we note that for $h$ very small, the  complex conjugate neutral modes of $JH^h(\lambda)$ are very close to zero and the condition $2E(\lambda)-\lambda>0$, which is necessary (although not sufficient) for the Fermi Golden Rule resonance condition {\bf (FGR)} to hold, fails. It remains an open question to derive the normal form, when resonance of discrete modes with the continuum occurs at some arbitrary order in the coupling parameter, $g$; recall $f(|\psi|^2)\psi=-g|\psi|^2\psi$; see also the discussion in the introduction. For results in this direction, see \cite{G1,CuMi}.
%
%

\section{Appendix}\label{sec:appendix}
\subsection{A class of double-well potentials for which $-\Delta+V$ satisfies condition {\bf ($Eig_V$)} and $L(\lambda)$ satisfies (SA) and (Thresh$_{\lambda}$)}\label{subsec:example} In this section we find an example $-\Delta+V$ in a subspace of $L^{2}(\mathbb{R}^3)$ satisfying condition {\bf ($Eig_V$)}, motivated by the study of double well potentials. 
Define 
$$\mathcal{A}:=\{f: \mathbb{R}^3\rightarrow \mathbb{C}|\ f(-x)=f(x)\ \text{for any }\ x\}.$$ Observe that $\mathcal{A}$ is a self-closed subspace, i.e. if $f_1,\ f_2\in \mathcal{A}$ then $f_1+f_2,\ f_1f_2, \Delta f_1\in \mathcal{A}$. Hence we can study ~\eqref{eq:NLS} in the space $\mathcal{A}\cap {L}^{2}(\R^3)$ and obtain all the results. The following is the main result
\begin{proposition}\label{Prop:MainProp} There exists a potential $V$ such that the linear operator $-\Delta+V$ acting on the subspace $\mathcal{A}\cap {L}^{2}(\R^3)$ has two eigenvalues $e_{0}<e_{1}<0$ with $2e_1>e_0$.
 $e_{0}$ is the lowest eigenvalue, the eigenvalue $e_{1}$ is degenerate
with multiplicity $2.$ Moreover the operator $1+(-\Delta+i0)^{-1}V:\langle x\rangle^2  L^2\rightarrow \langle x\rangle^2 L^2$ is invertible.
\end{proposition} 
If the nonlinearity $f(x)=x$, 
and if $|\lambda-|e_0||$ is sufficiently small and $\phi^{\lambda}$ is the ground state satisfying $$-\Delta\phi^{\lambda}+V\phi^{\lambda}+\lambda\phi^{\lambda}-(\phi^{\lambda})^3=0$$ then we have the following results for the linearized operator $L(\lambda)$ defined in ~\eqref{eq:LinOpera}.
\begin{proposition}\label{Prop:SAandThresh}
The operator $L(\lambda)$ satisfies the spectral conditions (SA) and (Thresh$_{\lambda}$).
\end{proposition}
The Proposition ~\ref{Prop:MainProp} is implied by Proposition ~\ref{Prop: LastStep} below. Proposition ~\ref{Prop:SAandThresh} will be proved at the end of this section.

As proved in ~\cite{Albeverio-etal} (Theorem 1.1.4, Page 116) the operator $-\Delta-q\delta(x)$ has only one eigenfunction, i.e. the ground states, for any $q>0$. 
 %
 By this observation we have 
\begin{lemma} For any $q>0$,
there exists a constant $\lambda\in (0,\infty)$ such that the operators $-\Delta-q \lambda^{-\frac{3}{2}}e^{-\frac{|x|^2}{\lambda}},$ $-\Delta-\frac{1}{3}q \lambda^{-\frac{3}{2}}e^{-\frac{|x|^2}{\lambda}}$ both have only one eigenfunction in  $\phi_1,\ \phi_2\in \mathcal{A}$ respectively.
\end{lemma}  To facilitate later discussions we define $$W:=q\lambda^{-\frac{3}{2}}e^{-\frac{|x|^2}{\lambda}}.$$

We start with constructing a family of operators: 
 Define $M_1:=(m,0,0)$, $M_2:=(0,m,0)$ and $M_3:=(0,0,m).$ And define $W_{M_k}$ as  $$W_{M_k}(x):=\frac{1}{2}[ W(x+M_k)+ W(x-M_k)],\  k=1,2,3.$$

Now we have the following result.
\begin{lemma}\label{LM:GR}
If $m$ is sufficiently large, then in the subspace $\mathcal{A}\cap {L}^2(\mathbb{R}^3)$ the operators $-\Delta-W_{M_k},\ -\Delta-\frac{1}{3}W_{M_k},\ k=1,2,3,$ each has only one eigenfunction.
\end{lemma}
\begin{proof}
In that following we only prove the argument for $-\Delta-W_{M_1}.$ The proof of the other cases are similar, hence omitted.

First we have that if $m$ is sufficiently large then $$\langle (-\Delta-W_{M_1})[\phi(\cdot+M_1)+\phi(\cdot-M_1)],[\phi(\cdot+M_1)+\phi(\cdot-M_1)]\rangle<0.$$ This implies that the operator $-\Delta-W_{M_1}$ has at least one ground state.

Secondly the min-max principle implies that any function $f\perp \phi(\cdot+M_1),\ \phi(\cdot-M_1)$, $$\langle (-\Delta-W_{M_1})f,f\rangle=\frac{1}{2}[\langle (-\Delta-W(\cdot+M))f,f\rangle+\langle (-\Delta-W(\cdot-M))f,f\rangle]\geq 0.$$ 
This together with the facts  $\phi(\cdot+M_1)-\phi(\cdot-M_1)\perp {L}^{2}(\mathbb{R}^3)\cap \mathcal{A}$ and $\{\phi(\cdot-M_1),\ \phi(\cdot+M_1)\}=\{\phi(\cdot-M_1)\pm \phi(\cdot+M_1)\}$ yields that $\langle \langle (-\Delta-W_{M_1})f,f\rangle\geq 0$ for any $f\in \mathcal{A}\cap L^2$ and $f\perp \phi(\cdot+M_1)+\phi(\cdot-M_1).$

Collecting what was proved we have that the operator $-\Delta-W_{M_1}$ has only one eigenfunction, i.e. the ground state. 

The proof is complete.
\end{proof}
To prove the main result we have to define $V$. Define $V_{m}$ as $$V_{m}:=\frac{1}{3}[ W_{M_1}+W_{M_2}+ W_{M_3}].$$

Now we have
\begin{proposition}\label{Prop: LastStep}
There exists at least one $m\in [0,\infty)$ such that $-\Delta-V_{m}$ has all the properties in  Proposition ~\ref{Prop:MainProp}.
\end{proposition}
\begin{proof}
To prepare for the proof we list the following facts:
\begin{itemize}
\item[(A)]  For any $m\in [0,\infty)$ the operator $-\Delta-V_{m}$ has at most three eigenfunctions in $\mathcal{A}\cap {L}^2$. Recall in Lemma ~\ref{LM:GR} we proved that if $f\perp \phi(\cdot+M_k)+\phi(\cdot-M_k), \ k=1,2,3$ and $f\in \mathcal{A}\cap {L}^2$ then $\langle (-\Delta-W_{M_k})f,f\rangle\geq 0.$  Consequently if $f\perp \phi(\cdot+M_k)+\phi(\cdot-M_{k}),\ k=1,2,3$ then $$\langle (-\Delta-V_{m})f,f\rangle=\frac{1}{3}[\langle (-\Delta-W_{M_1})f,f\rangle+\langle (-\Delta-W_{M_2})f,f\rangle+ \langle (-\Delta-W_{M_3})f,f\rangle ]\geq 0.$$ The min-max principle ~\cite{RS3-4} implies that there are at most three eigenfunctions.
\item[(B)] If $m$ is sufficiently large, then in the space ${L}^2\cap \mathcal{A}$ the operator $-\Delta-V_{m}$ has three eigenfunctions and two eigenvalues: one ground state and two (degenerate) neutral modes. The fact $-\Delta-V_{m}$ has three eigenfunctions comes from the min-max principle, the proof is similar to that of double well potential (see ~\cite{Har, JaWe}), hence is omitted here. The hard part is to prove the neutral mode is degenerate.  Indeed, as $m\rightarrow \infty$ the three eigenfunctions converges to some linear combination of $\phi(\cdot+M_k)+\phi(\cdot-M_k),\ k=1,2,3$ and the ground states converges to $\sum_{k=1}^{3}\phi(\cdot+M_k)+\phi(\cdot-M_k)$. Moreover the ground state must be simple and orthogonal to the neutral modes, i.e. the neutral mode can not be invariant under the permutation $(x_1, x_2,x_3)\rightarrow (x_{n(1)},x_{n(2)},x_{n(3)})$ while the operator $-\Delta-V_{m}$ is. This enable us to obtain one  of the neutral modes by permuting any one of two, hence the eigenvalue of neutral modes must be the same.
\item[(C)] When $m=0$, $-\Delta-V_{m}$ has only one eigenfunction, the ground state. This is obvious by the fact $V_{m}=W$ when $m=0.$
\item[(D)] For any $m\geq 0,$ $-\Delta-V_{m}$ has at least one eigenfunction with eigenvalue less than some $-c_{0}<0$. Let $\phi_2$ be the normalized ground state of $-\Delta-\frac{1}{3}W_{M_2}$ with eigenvalue $-c_{0}<0$. Then we have $\langle (-\Delta-V_{m})\phi_2,\phi_2\rangle <-c_{0}$ by the facts $\phi_2>0$ and $W>0.$ By the min-max principle $-\Delta-V_{m}$ has one ground state.
\end{itemize}

The definition of $W$ implies that $(-\Delta+k)^{-1}W(\cdot+z)$ is analytic in $z$ if $k\in \mathbb{C}\backslash\mathbb{R}^{+}$. By ~\cite{RS3-4} we have that the eigenvalues are analytic functions of $z$ in a suitable subset of $\mathbb{C}$. Since the eigenvalue of the neutral modes is degenerate for sufficiently large $m$ (see (B)), it is degenerate for any $m$ before the neutral modes disappear into the essential spectrum.  Hence there exists at least one $m$ such that $-\Delta-V_{m}$ has one eigenvalue less than $-c_{0}$ (defined in (D)) and two degenerate neutral modes with one eigenvalue sufficiently close to the essential spectrum (see (A), (C)).

In the final step we find $m$ and $q$ such that the operator $1+(-\Delta+i0)^{-1}V_{m}:\langle x\rangle^2  L^2\rightarrow \langle x\rangle^2 L^2$ is invertible. Recall that $V_{m}=qV_2(m)$ with $V_2(m)$ independent of $q$ by its definition. For a fixed $q_0$ we proved that there exists at least one $m=m_0$ such that the eigenvalues of $-\Delta-q_0V_{2}(m_0)$ have the desired properties. Now we consider a family of operators $X(g):=1-q(-\Delta+i0)^{-1}V_{2}(m_0)$ which is analytic in $q$ and the operator $q(-\Delta+i0)^{-1}V_{2}(m_0):\langle x\rangle^2 L^2\rightarrow \langle x\rangle^{2} L^2$ is compact. By \cite{RS3-4} the operators $X(q): \langle x\rangle^{2} {L}^{2}\rightarrow \langle x\rangle^2 {L}^2$ are either invertible everywhere (i.e. no threshold resonance) except discrete points, or are not invertible anywhere. The first case holds because the operator is invertible when $q=0$.

Now we consider $-\Delta-qV_2(m_{0})$ with $q\in [q_{0}-\epsilon, q_0+\epsilon]$. Choose $\epsilon$ sufficiently small such that for every $q$ the operator $-\Delta-qV_2(m_0)$ has at least three eigenvectors. On the other hand by what we proved above it has at most three eigenvectors and the second eigenvalue must be degenerate. Now by applying the fact that $1-q(-\Delta+i0)^{-1} V_{2}(m_0)$ is not invertible only at discrete points we obtain the desired result. 

The proof is complete.
\end{proof}
Now we prove  Proposition ~\ref{Prop:SAandThresh}\\
{\bf{Proof of Proposition ~\ref{Prop:SAandThresh}}}
The fact $L(\lambda)$ has no resonances at $\pm i\lambda$ is due to the facts $1+(-\Delta+i0)^{-1}V$ is invertible and $|\lambda-|e_0||$ is small, see Proposition ~\ref{Prop:resonance}.

In the next we prove the neutral mode is degenerate. Recall that the potential we constructed is of the form $V=V_{m_0}$ for some $m_0.$ For each $m>0$ there are $\lambda=\lambda_m$ and $\phi^{\lambda}=\phi^{\lambda,m}$ satisfying the equation $$-\Delta\phi^{\lambda,m}+\lambda_{m}\phi^{\lambda,m}+V_m\phi^{\lambda,m}-(\phi^{\lambda,m})^3=0$$ with $\lambda_m$ and $\phi_m$ analytic in $m$ in some proper neighborhood of positive real axis. 

Recall that when $m$ is sufficiently large the neutral modes of $-\Delta+V_{m}$ can be generated by permuting one of them, hence the neutral modes of $L(\lambda)=L(\lambda,m)$ are degenerate when $m$ is large. Moreover the eigenvalues of $L(\lambda,m)$ are analytic in $m$, thus the neutral modes must be degenerate.

The proof is complete.
\begin{flushright}
$\square$
\end{flushright}
\subsection{Fermi Golden Rule\ -\ Proof of Theorem ~\ref{Gamma-general}}\label{subsec:Gamma-general}

The proof of Theorem \ref{Gamma-general} uses the following
\begin{proposition}\label{simplify-ip}
Given smooth functions $\mathcal{F},\ \mathcal{G}:\ \mathbb{R}^{d}\rightarrow \mathbb{C}^2$, there exists $\tilde{\mathcal{F}}=(\tilde{\mathcal{F}}_1,\tilde{\mathcal{F}}_2)$ and $\tilde{\mathcal{G}}=(\tilde{\mathcal{G}}_1,\tilde{\mathcal{G}}_2)$ (see definitions below), such that 
\begin{equation}
-Re\ \left\langle
(L(\lambda)+2iE(\lambda)-0)^{-1}P_{c}\ \mathcal{F}, iJP_c\ \mathcal{G}\right\rangle
\ =\ \pi\ \left\langle\ \delta(-\Delta-(2E(\lambda)-\lambda)\ ) 
  \tilde{\mathcal{F}}_2,\tilde{\mathcal{G}}_2\ \right\rangle
\label{fgr-ip}
\end{equation}
\end{proposition}
The proposition will be proved later.
\nit{\bf Proof of Theorem \ref{Gamma-general}:}\  We use Proposition \ref{simplify-ip} with $\mathcal{F}=G_k$ and $\mathcal{G}=G_l$, $\tilde{\mathcal{F}}=\tilde{G}_k$ and $\tilde{\mathcal{G}}=\tilde{G}_l$. By (\ref{fgr-ip}) we have
 \begin{equation}
 \Gamma_{k,l}\ =\  \pi\ 
 \left\langle\ \delta(-\Delta-(2E(\lambda)-\lambda)\ ) 
  \tilde{G}_{l,2},\tilde{G}_{k,2}\ \right\rangle.
  \end{equation}
  To see that $\Gamma_{k,l}$ is non-negative, observe that for any $s\in\mathbb{C}^N$, we have
  \begin{equation}
 s^*\ \Gamma\ s\ =\  \sum_{k,l=1}^{N}\ \Gamma_{k,l}\ s_k\ \bar{s_l}\ =\ 
   \pi\ 
 \left\langle\ \delta(-\Delta-(2E(\lambda)-\lambda)\ ) 
  \tilde{\mathcal{G}},\tilde{\mathcal{G}} \right\rangle\ \ge\ 0,
  \label{sstarGammas}
  \end{equation}
  where $\tilde{\mathcal{G}}=\sum_{k=1}^{N} s_k\ \tilde{G}_{k,2}$.\\ \\
 
 For the second statement we only sketch the proof. Recall the transformation of $L(\lambda)$ in \eqref{trans}. Then for any $2\times 1$ vector functions $\vec{F}$ and $\vec{G}$ we have $$
 \begin{array}{lll}
 \langle (L(\lambda)+2iE(\lambda)-0)^{-1}P_c\vec{F},P_c\vec{G}\rangle&=&-i\langle (H+2E(\lambda)+i0)^{-1}A^{*}P_c\vec{F},A^{*}P_c\vec{G}\rangle\\
 &=&-i\langle K(\lambda) (H_{0}+V_1+2E(\lambda)+i0)^{-1}A^{*}P_c\vec{F},A^{*}P_c\vec{G}\rangle
 \end{array}
$$ with $K(\lambda)$ is the operator defined as $[1+K_{small}]^{-1}$ with $K_{small}:=(H_{0}+V_1+2E(\lambda)+i0)^{-1}V_{small}.$ The operator $(H_0+V_1+2E(\lambda)+i0)^{-1}$ is well defined by the fact $-\lambda-2E(\lambda)\approx e_{0}-2(e_1-e_0)$ is not an eigenvalue of $-\Delta+V$ and the operator $-\Delta+V$ has no embedded eigenvalues in the essential spectrum.

Since the operator $K_{small}:\langle x\rangle^{2} {L}^{\infty}\rightarrow \langle x\rangle^2 {L}^{\infty}$ has a small norm and continuous in $\lambda$ we have $$[1+K_{small}]^{-1}=\sum_{n=0}^{\infty}(-K_{small})^n$$ is continuous in $\lambda.$ This together with the fact $(H_{0}+V_1+2E(\lambda)+i0)^{-1}A^{*}P_c\vec{F}\in \langle x\rangle^2 {L}^{\infty}$ is continuous in $\lambda$ implies that $\langle (L(\lambda)+2iE(\lambda)-0)^{-1}P_c\vec{F},P_c\vec{G}\rangle$ is continuous in $\lambda.$

The proof is complete.
\begin{flushright}
$\square$
\end{flushright}

\nit{\bf Proof of Proposition \ref{simplify-ip}:} The entries of $\Gamma$ are expressions of the form
\begin{equation}
-Re\ \left\langle
(L(\lambda)+2iE(\lambda)-0)^{-1}P_{c}\ \mathcal{F}, iJP_c\ \mathcal{G}\right\rangle,
\label{fgr-ip2}
\end{equation}
which we now proceed to simplify. 
Recall $L(\lambda)$ is of the form 
 \begin{equation}
 L(\lambda)\ =\ 
 (-\Delta+\lambda)
 \left(
\begin{array}{lll} 0&1\\-1&0 \end{array}
\right)+
\left(\begin{array}{lll}
0&V_{1}\\
V_{2}&0
\end{array}
\right)
\nn\end{equation}
where $V_{1}$ and $V_{2}$ are real-valued and exponentially decaying as $|x|$ tends to infinity. 
 Introduce the unitary matrix
 \begin{equation}
 U\ =\ \frac{1}{\sqrt{2}}\left(
\begin{array}{lll}
1&i\\ i&1 \end{array}
\right). \label{Udef}
\end{equation}
Note that 
\begin{equation}\label{Lsigma3H}
L(\lambda)\ =\ i\ U\ \sigma_3\ \mathcal{H}(\lambda)\ U^*,\ \ \mathcal{H}^*=\cH
\end{equation}
where
\begin{align}
\cH\ &:=\ \mathcal{H}_0+\  \tilde{V}\nn\\
\mathcal{H}_0\ &:=(-\Delta\ +\ \lambda)\ I
\end{align}
\begin{align}
\tilde{V}\ &:=\ \left(\begin{array}{lll}
V_1-V_2&-i(V_1+V_2)\\
i(V_1+V_2)&V_1-V_2
\end{array}\right)\nn\\
\sigma_{3} \ &:=\left(
\begin{array}{lll}
1&0\\ 0&-1 \end{array}
\right).
\nn\end{align}

We now use the unitary transformation, $U$, to obtain an expression in terms of the operator $\sigma_3H$:
\begin{align}
& -\langle (L(\lambda)+2iE(\lambda)-0)^{-1}P_{c}\ \mathcal{F}, i J\ P_c\ \mathcal{G}\rangle\nn\\
\ &=-\  
\left\langle \left(\ i\ U\ [\sigma_3\cH(\lambda)+2E(\lambda)+i0]\ U^*\ \right)^{-1}P_{c}\ \mathcal{F}, i J\ P_c\ \mathcal{G}\right\rangle\nn\\
\ &=\ \ \left\langle\ \left(\sigma_3\cH(\lambda)+2E(\lambda)+i0\right)^{-1}\  U^*P_{c}\ \mathcal{F},  U^* J\ P_c\ \mathcal{G}\right\rangle\nn\\
\ &=\ \ \left\langle\ \left(\sigma_3\cH(\lambda)+2E(\lambda)+i0\right)^{-1}\  U^*P_{c}\ \mathcal{F},  (U^* J\ U)\ U^* P_c\ \mathcal{G}\right\rangle\nn\\
\ &=\  -i\ \left\langle\ \left(\sigma_3\cH(\lambda)+2E(\lambda)+i0\right)^{-1}\  U^*P_{c}\ \mathcal{F},  \sigma_3\ U^* P_c\ \mathcal{G}\right\rangle,\label{Uchain}
\end{align}
where we have used that $U^*JU=i\sigma_3$.

Next, we introduce  $P_c(\sigma_3\cH)$, the projection onto the continuous spectral part of $\sigma_3H$ and wave operators  
 $W:L^2\to P_c(\sigma_3\cH))\ L^2$ and $Z:P_c(\sigma_3H)L^2\to L^2$, see \cite{CPV:05}, which satisfy
\begin{align}
P_c(\sigma_3\cH)^*\ \sigma_3\ =\ \sigma_3\ P_c(\sigma_3\cH),\nn\\
W^*\ \sigma_3\ =\ \sigma_3\ Z,\ \ \ Z^*\ \sigma_3\ =\ \sigma_3\ W\nn\\
Z\ \sigma_3 \cH\ =\ \sigma_3\ \cH_0\ Z\label{WZprops}
\end{align} 

Now we use the wave operators $W$ and $Z$ to transform the previous expression into one, in terms of the ``free operator'' $\sigma_3(-\Delta+\lambda)$. First, note that $U^*P_c\ \mathcal{F}$ lies in the
 range of $P_c(\sigma_3\cH)$, and therefore there exists $\tilde{\mathcal{F}}=(\tilde{\mathcal{F}}_1,\tilde{\mathcal{F}}_2)^T$ 
 such that $W\tilde{\mathcal{F}}=U^*P_c\ \mathcal{F}$.  Similarly, there exists $\tilde{\mathcal{G}}=(\tilde{\mathcal{G}}_1,\tilde{\mathcal{G}}_2)^T$, such that  $W\tilde{\mathcal{G}}=U^*P_c\ \mathcal{G}$. Substitution into the final expression in (\ref{Uchain}) and use of the properties (\ref{WZprops}) we have:
\begin{align}
& i\ \left\langle\ \left(\sigma_3\cH(\lambda)+2E(\lambda)+i0\right)^{-1}\  U^*P_{c}\ \mathcal{F},  \sigma_3\ U^* P_c\ \mathcal{G}\right\rangle\nn\\
 \ &=\ i\ \left\langle\ \left(\sigma_3\cH(\lambda)+2E(\lambda)+i0\right)^{-1}\ W\tilde{\mathcal{F}} ,  \sigma_3\ W\tilde{\mathcal{G}}\right\rangle\nn\\
\ &=\   i\ \left\langle\ \left(\sigma_3\cH(\lambda)+2E(\lambda)+i0\right)^{-1}\ W\tilde{\mathcal{F}} ,  Z^*\ \sigma_3\ \tilde{\mathcal{G}}\right\rangle\nn\\
\ &=\   i\ \left\langle\ Z\ \left(\sigma_3\cH(\lambda)+2E(\lambda)+i0\right)^{-1}\ W\tilde{\mathcal{F}} ,\ \sigma_3\ \tilde{\mathcal{G}}\right\rangle\nn\\
\ &=\  i\ \left\langle\ \left(\sigma_3(-\Delta+\lambda)+2E(\lambda)+i0\right)^{-1}\ Z\ W\tilde{\mathcal{F}} ,\ \sigma_3\ \tilde{\mathcal{G}}\right\rangle\nn\\
\ &=\  i\ \left\langle\ \left(\sigma_3
 (-\Delta+\lambda)+2E(\lambda)+i0\right)^ {-1}\ \tilde{\mathcal{F}} ,\ \sigma_3\ \tilde{\mathcal{G}}\right\rangle\label{WZchain}
 \end{align}
 Referring back to (\ref{fgr-ip2}), we recall that we are interested in the real part of this expression.
 \begin{align}
&-Re\  i\ \left\langle\ \left(\sigma_3
 (-\Delta+\lambda)+2E(\lambda)+i0\right)^ {-1}\ \tilde{\mathcal{F}} ,\ \sigma_3\ \tilde{\mathcal{G}}\right\rangle\nn\\
&=\  Im\ \left\langle\ \left(\sigma_3
 (-\Delta+\lambda)+2E(\lambda)+i0\right)^ {-1}\ \tilde{\mathcal{F}} ,\ \sigma_3\ \tilde{\mathcal{G}}\right\rangle\nn\\
 &=\ \ Im\ \left\langle\  
 \left(
 \begin{array}{ll}
  (-\Delta+\lambda+2E(\lambda)+i0)^{-1} & 0\\
  0 & -(-\Delta+\lambda-2E(\lambda)-i0)^{-1})  \end{array}
 \right)\ \tilde{\mathcal{F}} ,\ \sigma_3\tilde {\mathcal{G}}\ \right\rangle\nn\\
 &=\ \ Im\ \left\langle\  
 \left(
 \begin{array}{ll}
  (-\Delta+\lambda+2E(\lambda)+i0)^{-1} & 0\\
  0 & (-\Delta+\lambda-2E(\lambda)-i0)^{-1})  \end{array}
 \right)\ \tilde{\mathcal{F}} ,\ \tilde {\mathcal{G}}\ \right\rangle\nn\\
 &\ =\  \ Im\ \left\langle\ (-\Delta-(2E(\lambda)-\lambda)-i0)^{-1}\tilde{\mathcal{F}}_2,\tilde{\mathcal{G}}_2\ \right\rangle\nn\\
 & =\ \pi\ \left\langle\ \delta(-\Delta-(2E(\lambda)-\lambda)\ ) 
  \tilde{\mathcal{F}}_2,\tilde{\mathcal{G}}_2\ \right\rangle.\label{fgr-nonneg}
 \end{align}
The last equality uses that  $0<2E(\lambda)-\lambda\in\sigma_c(-\Delta)$, $-2E(\lambda)-\lambda\notin\sigma_c(-\Delta)$ and the distributional (Plemelj) identity:
\begin{equation}
Im\ (x-i0)^{-1}\ =\ \lim_{\varepsilon\downarrow0}\ Im\ (x-i\varepsilon)^{-1}\ =\ \pi\ \delta(x)
\label{plemelj}
\end{equation} \\ 
Summarizing, we have shown
\begin{equation}
-Re\ \left\langle
(L(\lambda)+2iE(\lambda)-0)^{-1}P_{c}\ \mathcal{F}, iJP_c\ \mathcal{G}\right\rangle\ =\ 
 \pi\ \left\langle\ \delta(-\Delta-(2E(\lambda)-\lambda)\ ) 
  \tilde{\mathcal{F}}_2,\tilde{\mathcal{G}}_2\ \right\rangle.
  \label{F2G2-id1}
  \end{equation}
  This completes the proof of Proposition \ref{simplify-ip}.
 
\subsection{FGR for symmetric potentials}\label{subsec:FGR}
In this section we derive the simpler form of the FGR matrix and condition for positivity in the case where   the potential $V(x)$ is a function of $|x|$. In fact, it is proved in Lemma ~\ref{mainLem2} that if the potential $V$, hence $\phi^{\lambda}$, is spherically
symmetric, then the functions $\xi_{n},\ \eta_{n}$ satisfy
\begin{equation}
\xi_{n}=\frac{x_{n}}{|x|}\xi(|x|),\
\eta_{n}=\frac{x_{n}}{|x|}\eta(|x|)\label{specialstruc}\end{equation}
for some functions $\xi(|x|)$ and
$\eta(|x|)$. 
 By the assumptions on $V$, $\phi^{\lambda},\ \xi_{k},\ \eta_{k},\
k=1,2,\cdot\cdot\cdot,N=d$ we have
\begin{equation}
G_{k}(z,x)~=~x_{k}(z\cdot x)G(|x|)\nn
\end{equation}
for some
radial vector function $G(|x|)$.
 \\ \\
Before stating the results we define two
constants
\begin{align}
Re\ Z_{0}^{(1,1)}\ &\ =\ 
- Re\ \left\langle (L(\lambda)+2iE(\lambda)-0)^{-1}P_{c}\ x_{1}^2\ G(|x|),
i\ J\ x_1^2\ G(|x|)\right\rangle\nn\\
Re\ Z_{0}^{(2,2)}\ &=\ 
- Re\  \left\langle (L(\lambda)+2iE(\lambda)-0)^{-1}P_{c}\ x_{1}x_2\ G(|x|),
i\ J\ x_1 x_2\ G(|x|)\right\rangle
\label{Z11Z22}
\end{align}
\begin{proposition}
(i)\ Suppose that $V$, $\xi_{n},\ \eta_{n}$ satisfy the conditions above.
Then the assumption (FGR) holds provided that 
\begin{equation}
 ReZ_{0}^{(1,1)}>0,\ \ 
ReZ_{0}^{(2,2)}>0.\label{ReZZjj}
\end{equation}
(ii)\ From Proposition \ref{simplify-ip}, it follows that 
\begin{equation}
 ReZ_{0}^{(1,1)}\ge0,\ \ 
ReZ_{0}^{(2,2)}\ge0.
\end{equation}
and, generically, that strict positively (\ref{ReZZjj}) holds.
\end{proposition}
\nit {\bf Proof:}\  For any vectors $s,\beta,z\in \mathbb{C}^{N}$, we
define 
\begin{equation}
\mathcal{Q}(s,\beta;z):=-Re\left\langle (L(\lambda)+2iE(\lambda)-0)^{-1}P_{c}(z\cdot
x)(s\cdot x)G(|x|), iJ(z\cdot x)(\beta\cdot x)G(|x|)\right\rangle,
\end{equation} 
Note that 
\begin{equation}
\mathcal{Q}(s,s;z)=\frac{1}{2}s^*\ [Z(z,\bar{z})+Z^{*}(z,\bar{z})]\ s\ =\ Re\ \  s^*Z(z,\bar{z}) s
\nn\end{equation}
Therefore, verifying (FGR) is equivalent to checking that there is a constant $C>0$, for which 
\begin{equation}
\mathcal{Q}(s,s;z)\ \ge\ C\ |s|^2\ |z|^2,\ \ s, z\in\mathbb{C}^d.
\label{FRG2}
\end{equation}

To simplify $\mathcal{Q}(s,s;z)$, first note that since operator $L(\lambda)$ and $G(|x|)$ are invariant under transformations $x\mapsto T^*x$, where $T$ is unitary, the value of $\mathcal{Q}(s,\beta;z)$ is unchanged when replacing $x$ by $T^*x$.\\ Therefore,
 \begin{equation}
 \mathcal{Q}(s,\beta;z)\ =\ \mathcal{Q}(Ts,T\beta;Tz)
 \label{QT}
 \end{equation}
Now choose $T$ to be a unitary matrix, such that 
\begin{equation}
 Tz\ =\ |z|\ e_1\ =\ |z|\ (1,0,\dots,0)^T.
 \nn\end{equation}
 With this choice of $T$, we have by (\ref{QT}) with $\beta=s$,
\begin{equation}
\mathcal{Q}(s,s;z)=-Re\left\langle (L(\lambda)+2iE(\lambda)-0)^{-1}P_{c}|z|x_{1}(T s\cdot x)G(|x|),
i|z|x_{1}(T s\cdot x)JG(|x|)\right\rangle.
\label{QTss}
\end{equation}
The following argument will show that
 $\mathcal{Q}(s,s;z)\ge C\ |Ts|\ |z|^2\ =\ C\ |s|^2\ |z|^2$, the latter holding since $T$ is unitary. Therefore, without any loss of 
 generality consider (\ref{QTss}) with $T$ set equal to the identity.
  Explicitly writing out the inner products and using bilinearity and symmetry, we have
\begin{align}
& \ \mathcal{Q}(s,s;z)\nn\\ 
&= \ -|z|^2\ 
 Re \ \sum_{p,q=1}^d  \left\langle (L(\lambda)+2iE(\lambda)-0)^{-1}P_{c}\ x_{1}x_p\ G(|x|),
i\ x_1 x_q\ JG(|x|)\right\rangle\ s_p\ \overline{s_q}\nn\\
&=\ -|z|^2\ 
 Re\ \sum_{p=1}^d  \left\langle (L(\lambda)+2iE(\lambda)-0)^{-1}P_{c}\ x_{1}x_p\ G(|x|),
i\ x_1 x_p\ JG(|x|)\right\rangle\ |s_p|^2\nn\\
&=\ -|z|^2\ Re\ \left\langle (L(\lambda)+2iE(\lambda)-0)^{-1}P_{c}\ x_{1}^2\ G(|x|),
i\ x_1^2\ JG(|x|)\right\rangle\ |s_1|^2\nn\\
&\ \ \ -\ |z|^2\ 
Re\  \left\langle (L(\lambda)+2iE(\lambda)-0)^{-1}P_{c}\ x_{1}x_2\ G(|x|),
i\ x_1 x_2\ JG(|x|)\right\rangle\ \sum_{q=2}^d |s_q|^2\nn\\
&=\ |z|^2\ \left(\  Re\ Z_0^{(1,1)}\  |s_1|^2\ +\ Re\ Z_0^{(2,2)}
\sum_{q=2}^d\ |s_q|^2\ \right)\nn\\
&\ \ \  \geq \ 
\min\{Re\ Z_0^{(1,1)},Re\ Z_0^{(2,2)}\ \}\ |s|^2\ |z|^2\ \equiv\ C\ |s|^2\ |z|^2\ >\  0.
\nn
\end{align}


The proof is complete.

\subsection{Choice of Basis for the Degenerate Subspace: Proof of Proposition ~\ref{prop:OrthonormalBasis}}\label{subsec:choice}

In the proof of the proposition we need the following lemma.
\begin{lemma}
If $u=\left(
\begin{array}{lll}
u_{1}\\
iu_{2}
\end{array}
\right)\not=0$ is an eigenfunction of $L(\lambda)$ with eigenvalue $iE(\lambda),\ E(\lambda)>0,$ then
\begin{equation}\label{eq:positivity}
\langle u_{1},u_{2}\rangle>0.
\end{equation}
\end{lemma}
\begin{proof}
The fact $L(\lambda)u=iE(\lambda)u$ yields
\begin{equation}\label{eq:OperatorOnU}
L_{-}(\lambda)u_{2}=E(\lambda)u_{1},\ L_{+}(\lambda)u_{1}=E(\lambda)u_{2}.\end{equation}
Therefore $$\langle u_{1},u_{2}\rangle=\frac{1}{E(\lambda)}\langle L_{-}(\lambda)u_{2},u_{2}\rangle.$$

Equation (~\ref{eq:positivity}) follows from the two claims that
$L_{-}(\lambda)$ is a positive definite self-adjoint operator on the
space $\{v|v\perp \phi^{\lambda}\}$ and
$u_{2}\not\in span\{\phi^{\lambda}\}.$ The first fact is well known (see
e.g. ~\cite{Wei86}). We prove the second by contradiction.
Suppose that $u_{2}=c\phi^{\lambda}$ for some constant $c$, then we
have $L_{-}(\lambda)u_{2}=0,$ which together with
(~\ref{eq:OperatorOnU}) and the fact $E(\lambda)\not=0$ implies
$u_{1}=u_{2}=0$, i.e. $u=0$. This contradicts to the fact $u\not=0.$
Thus $u_{2}\not\in span\{\phi^{\lambda}\}.$
\end{proof}
\textbf{Proof of Proposition ~\ref{prop:OrthonormalBasis}} We start
the proof by constructing $N$ independent vectors $u_{n}\in
span\{v_{1},\ v_{2},\cdot\cdot\cdot,v_{N}\},\
n=1,2,\cdot\cdot\cdot, N$ such that the vector 
$\left(
\begin{array}{lll} 1&0\\ 0&i \end{array} \right)u_{n}$
 is real.

Suppose that $v_{n}=\left(
\begin{array}{lll}
v_{1}^{(n)}\\
v_{2}^{(n)}
\end{array}
\right)$. Then the definition of $L(\lambda)$ in
(~\ref{eq:LinOpera}) implies $\left(
\begin{array}{lll}
Rev_{1}^{(n)}\\
iImv_{2}^{(n)}
\end{array}
\right)$ and $\left(
\begin{array}{lll}
Imv_{1}^{(n)}\\
-iRev_{2}^{(n)}
\end{array}
\right)$ are also eigenfunctions of $L(\lambda)$ with eigenvalues
$iE(\lambda)$. This together with the fact $$\{\left(
\begin{array}{lll}
Rev_{1}^{(n)}\\
iImv_{2}^{(n)}
\end{array}
\right),\ \left(
\begin{array}{lll}
Imv_{1}^{(n)}\\
-iRev_{2}^{(n)}
\end{array}
\right), n=1,2,\cdot\cdot\cdot, N\}=\{v_{n},\
n=1,2\cdot\cdot\cdot, N\}$$ enables us to choose $N$ independent
eigenfunctions for $iE(\lambda)$: $u_{n},\ n=1,2,\cdot\cdot\cdot, N,$ such that $\left(
\begin{array}{lll}
1&0\\
0&i
\end{array}
\right)u_{n}$ are real vectors.

Using (~\ref{eq:positivity}) and a standard Gram-Schmidt procedure in linear
algebra, one can find $N$ pairs of real functions $(\xi_{n},\
\eta_{n}),\ n=1,2\cdot\cdot\cdot, N,$ such that $span\{\left(
\begin{array}{lll}
\xi_{n}\\
i\eta_{n}
\end{array}
\right),\ n=1,2\cdot\cdot\cdot, N\}=span\{v_{n},n=1,2,\cdot\cdot, N\}$ and $\langle \xi_{n},\eta_{m}\rangle=\delta_{n,m}.$

We now turn to the verification of (~\ref{eq:UnexpectedFact}). The observations
\begin{align}
& L_{-}(\lambda)-L_{+}(\lambda)\ =\ 2f^{'}[(\phi^{\lambda})^{2}](\phi^{\lambda})^{2},\nn\\
&L_{-}(\lambda)\eta_{n}\ =\ E(\lambda)\xi_{n},\
\text{and}\ L_{+}(\lambda)\xi_{n}\ =\ E(\lambda)\eta_{n},\
n=1,2,\cdot\cdot\cdot, N,\nn
\end{align}
yield
$$
\begin{array}{lll}
& &\int f^{'}[(\phi^{\lambda})^{2}](\phi^{\lambda})^{2}(\xi_{m}\eta_{n}-\xi_{n}\eta_{m})dx\\
&=&\frac{1}{2}[\langle \xi_{m},L_{-}(\lambda)\eta_{n}\rangle-\langle L_{+}(\lambda)
\xi_{m},\eta_{n}\rangle-\langle \xi_{n},L_{-}(\lambda)\eta_{m}\rangle+\langle L_{+}(\lambda)\xi_{n},\eta_{m}\rangle]\\
&=&0.
\end{array}
$$

Finally,  (~\ref{eq:orthogonality}) is seen as follows:
\begin{align}
&\langle \phi^{\lambda}, \xi_{n}\rangle=\frac{1}{E(\lambda)}\langle
\phi^{\lambda},L_{-}(\lambda)\eta_{n}\rangle=\frac{1}{E(\lambda)}
\langle L_{-}(\lambda)\phi^{\lambda},\eta_{n}\rangle=0,\nn\\
&\langle
\partial_{\lambda}\phi^{\lambda},\eta_{n}
\rangle=\frac{1}{E(\lambda)}\langle
\partial_{\lambda}\phi^{\lambda},L_{+}(\lambda)\xi_{n}\rangle=-\frac{1}{E(\lambda)}
\langle\phi^{\lambda},\xi_{n}\rangle=0\nn
\end{align}
\subsection{The identity $P_c(JH)^*J\ =\ JP_c(JH)$}
\begin{proposition}\label{projection-prop}
 $L=JH, H=H^{*}\ \ \implies\ \  
 [P_{c}(L)]^{*}J\ =\ JP_{c}(L)$ .
 \end{proposition}
 \nit{\bf Proof:}
  Represent $P_c(L)$ as a Riesz  projection 
  \begin{equation}
  P_{c}(L)=\frac{1}{2\pi i}\oint (zI-JH)^{-1}dz,
  \label{RieszPc}
  \end{equation}
   where the integration is counter-clockwise, moreover the essential spectrum of $L$ is $[i\lambda,i\infty)\cup (-i\infty,-i\lambda]$. The spectrum associated with  the  upper branch, $[i\lambda,i\infty)$, is
   given by 
  \begin{align}
  &P^{+}(JH)\ =\ \frac{1}{2\pi}[A-B],\ \ {\rm where}
  \label{P+}\\
  &A\ =\ \int_{\lambda}^{\infty}\ (i\tau+0-JH)^{-1}\ d\tau\nn\\
  &B\ =\ \int_{\lambda}^{\infty}\ (i\tau-0-JH)^{-1}\ d\tau.
  \nn\end{align}
   We claim that
\begin{equation}\label{eq:transfo}
A^{*}J=-JB,\ B^{*}J=-JA.
\end{equation}
This implies $[P^{+}(JH)]^{*}J=JP^{+}(JH)$. Similarly, $[P_{c}(JH)]^{*}J=JP_{c}(JH)$.\\

To complete the proof of Proposition \ref{projection-prop}, we now prove ~\eqref{eq:transfo}. By direct computation, using $J^{*}=-J$, we have 
\begin{equation}
A^{*}=\int_{\lambda}^{\infty}(-i\tau+0+HJ)^{-1}\ d\tau\ 
\nn\end{equation}
Therefore,
\begin{equation}
A^{*}J\ =\ \int_{\lambda}^{\infty}(Ji\tau J-J0J-JJHJ)^{-1}\ d\tau\ J=
 \int_{\lambda}^{\infty}(-J)(i\tau-0-JH)^{-1}(-J)\ d\tau\ J=-JB.
\end{equation}
thus proving the first identity in ~\eqref{eq:transfo}. $B^{*}J=-JA$ can be proved similarly.

\subsection{Time Convolution Lemmas: Proof of Proposition ~\ref{prop:conv-est}}\label{subSEC:Time-Con}

\nit {\bf Proof:}\ In what follows we only prove the case $\sigma=1$ of (~\ref{conv-est}), the other cases and (~\ref{log-conv-est}) are similar.
\begin{align}
I(t):=\int_0^t\ \frac{1}{(1+t-s)^{3\over2}}\ \frac{1}{T_0+s}\ ds\ & \le\ 
\frac{1}{(1+{t\over2})^{3\over2}}\int_0^{t\over2}\  \frac{1}{T_0+s}\ ds
\ +\  \frac{1}{T_0+{t\over2}}\ \int_{t\over2}^t  \frac{1}{(1+t-s)^{3\over2}}\
 ds\nn\\
 &\ \le\ \frac{\log(1+{t\over 2T_0})}{(1+{t\over2})^{\frac{3}{2}} }\ +\ \frac{2}{T_0+{t\over2}}
 \nn\end{align}
 On the other hand, we also have
 \begin{equation}
 \int_0^t\ \frac{1}{(1+t-s)^{3\over2}}\ \frac{1}{T_0+s}\ ds\  \le\ \frac{2}{T_0}
 \nn\end{equation}
 Thus, 
 \begin{equation}
 I(t)\ \le\ c_1\ \min\left\{\frac{1}{T_0}\ ,\  \frac{1}{1+t}\ \right\}
 \nn\end{equation}
We now claim  that for some constant  $c>0$, that $I(t)\le c\ (T_0+t)^{-1}$. It sufficies to find a constant $c$, independent of $T_0$ and $t$,
 such that 
 \begin{equation}
 m(t)\ :=\ (T_0+t)\ \min\{\ \frac{1}{T_0}\ ,\ \frac{1}{1+t}\ \}\ \le\ c
 \nn\end{equation}
 If $t$ is such that the above $\min$ is $T_0^{-1}$, then 
 $T_0^{-1}\le (1+t)^{-1}$ or $t\le T_0-1$. Then, $m(t)\le (2T_0-1)T_0^{-1}\le {3\over2}$. If $t$ is such that the above $\min$ is $(1+t)^{-1}$, then 
  $t\ge T_0-1$. Therefore, $m(t)\le (2T_0-1)T_0^{-1}$ since $m(t)$ is decreasing with $t$. Since $T\ge2$, $m(t)\le 3/2$. This completes the proof. 
\subsection{Bounds on Solutions to a Weakly Perturbed ODE: Proof of Lemma ~\ref{LM:RiccatiEst}}\label{subSEC:Riccati}
\begin{proof}
Let $\beta$ denote the solution to the differential equation 
\begin{align}\label{eq:equationW}
\partial_{t}|\beta_\rho|^{2}\ &=\ -|\beta_\rho|^{4}+g\nn\\
|\beta_\rho|^2(0)\ &=\ |z(0)|^2\ -\ \rho,\ \ \rho>0.
\nn \end{align} 
Since 
\begin{align}
\partial_{t} \left(\ |z(t)|^2-|\beta_\rho(t)|^2\ \right)\ &\le\ -|z(t)|^4\ +\ |\beta_\rho(t)|^4\nn\\
& =\ -\left(\ |z(t)|^2\ +\ |\beta_\rho(t)|^2\ \right)\ \left(\ |z(t)|^2\ -\ |\beta_\rho(t)|^2\ \right)\nn
\end{align}
with the initial condition
$$|z(0)|^2-|\beta_\rho(0)|^2\ =\ \rho>0.$$

 Thus, $|z(t)|^2\ \le\ |\beta_\rho(t)|^2$ for all $t\ge0$ and letting $\rho$ tend to zero, we have
 \begin{equation} 
 |z(t)|^2\ \le\ |\beta(t)|^2
 \nn\end{equation}
 so it suffices to prove the bound:
 \begin{equation}
|\beta(t)|\ \leq\ (1+K\ c_\#\ T_0^{-\delta})\  (\kappa+t)^{-\frac{1}{2}},\ \ \ \kappa=\min\{T_{0},|w_{0}|^{-2}\}
\label{betabound}
\end{equation}
where  $\beta(t)$ solves the initial value problem 
\begin{align}
\partial_{t}|\beta|^{2}\ &=\ -|\beta|^{4}+g\nn\\
|\beta(0)|^2\ &=\ |w_0|^2.
\label{beta-ivp}\end{align}

\nit The proof of (\ref{wbound}) for $\beta$ is divided into two cases: 
  $|w_{0}|\geq T_{0}^{-1/2}$ and 
 $|w_{0}|~< ~T_{0}^{-1/2}.$

First consider case (i): $|w_{0}|\geq T_{0}^{-1/2}$. By local existence for the initial value problem (\ref{beta-ivp}), we have that for some   $t_1>0$, 
\begin{equation}\label{eq:Hypo}
 \frac{1}{2}\frac{1}{(T_{0}+t)^{\frac{1}{2}} }\ \le\ |\beta(t)|,\ \ t\in[0,t_1].
\end{equation} 
Then, using the assumed bound on $g(t)$, (\ref{gbound}), we have 
\begin{align}
|g(t)|\ &\le\ \frac{c_\#}{(T_0+t)^{2+\delta}}\ =\ \frac{c_\# }{(T_0+t)^{2}}\ \frac{1}{(T_0+t)^{\delta}}\nn\\
 &\ \le\ 2^4\ c_\#\ |\beta(t)|^4\cdot\frac{1}{T_0^\delta}\ =\ c_{1\#}\ \ T_0^{-\delta}\ |\beta(t)|^4,
 \nn\end{align}
 where $c_{1\#}:=2^4\ c_\#$.
 It follows from (\ref{beta-ivp}) that 
 \begin{equation}
 \partial_{t} |\beta(t)|^2\ \le\ -\left(\ 1-c_{1\#}\ T_0^{-\delta}\ \right)|\beta(t)|^4
\nn\end{equation}
or
\begin{equation}
 \partial_{t} |\beta(t)|^{-2}\ \ge\ \left(\ 1-c_{1\#}\ T_0^{-\delta}\ \right).
\nn\end{equation}
Integration over the interval $[0,t], \ t\le t_1$,  yields
\begin{equation}
  |\beta(t)|\ \le\
 \frac{1+c_{2\#}\ T_0^{-\delta}}{ (|w_0|^{-2}+t)^{{1\over2}} },
 \label{uplow}\end{equation}
 where $c_{2\#}\sim c_{1\#}\sim c_\#$,
 and we use that $c_\#\ T_0^{-\delta}$ is sufficiently small.
 Now set $\kappa=\min\{|w_0|^{-2}, T_0\}$ and we have
 \begin{equation}
 |z(t)|\ \le\ |\beta(t)|\ \le\
 \frac{1+c_{2\#}\ T_0^{-\delta}}{ (\kappa+t)^{{1\over2}} },\ \ 0\le t\le t_1
 \nn\end{equation}
 Now suppose that $[0,\Xi)$ denotes the maximal subset of $\R_+$, on which the upper bound in 
  (\ref{uplow}) holds.   If $\Xi<\infty$, then by continuity and the assumption that $|w_0|\ge T_0^{-{1\over2}}$, we have
  \begin{align}
  & |\beta(\Xi)|\ =\
 \frac{1+c_{2\#}\ T_0^{-\delta}}{ (|w_0|^{-2}+\Xi)^{{1\over2}} }\ \ge\
  \frac{1}{ (|w_0|^{-2}+\Xi)^{{1\over2}} }\ 
  \ \ge\ \frac{3}{4}\ \frac{1}{ (T_0+\Xi)^{{1\over2}} },\nn\\
& \label{upequality}\end{align}
 implying (see (\ref{eq:Hypo}) that  the above argument  can be applied beyond $t=\Xi$, contradicting its maximality.

\nit Finally, we consider {\bf case (ii):} $|\beta_{0}|< T_{0}^{-1/2}.$\\
 Denote by $\beta_{1}(t)$ the solution to
(~\ref{eq:Riccati}) with the initial condition $\beta_{1}(0)=T_{0}^{-1/2}.$ As shown in the previous case $|\beta_{1}(t)|\leq (1+K\ c_\# T_0^{-\delta})(T_{0}+t)^{-1/2}.$ Observing that $$
\begin{array}{lll}
\partial_{t}(|\beta|^{2}-|\beta_{1}|^{2})&=&-(|\beta|^{2}+|\beta_{1}|^{2})(|\beta|^{2}-|\beta_{1}|^{2}),\\
|\beta_{0}|^{2}- |\beta_{1}(0)|^{2}&<&0
\end{array}
$$ we have that for any time $t$, $|\beta(t)|^{2}\leq |\beta_{1}(t)|^{2}$. This together with the estimate of $\beta_{1}$ completes the proof of the second case.
\end{proof}

\end{document}